\documentclass[aps,prb,
twocolumn,
groupedaddress,showpacs,floatfix]{revtex4-1}

\usepackage[encapsulated]{CJK}
\usepackage{ucs}
\usepackage[utf8x]{inputenc}
\usepackage{multirow}
\usepackage{amsmath}
\usepackage{amssymb}
\usepackage{mathrsfs}
\usepackage{graphics}
\usepackage{graphicx}
\usepackage{bbold}
\usepackage{bm}

\usepackage{hyperref}

\begin{document}

\title{Spin-orbit interaction in the \texorpdfstring{$\bm{k\cdot p}$}~ theory for cubic crystals}


\author{Warren J. Elder}
\author{E. S. Tok}
\altaffiliation{On sabbatical leave from Department of Physics, National University of Singapore, Singapore}
\author{Jing Zhang}
\email{Corresponding email: jing.zhang@imperial.ac.uk}
\affiliation{Blackett Laboratory, Department of Physics, Imperial College London, Prince Consort Road, London SW7 2BW, UK}

\date{\today}

\begin{abstract}
Recent work by Elder, Ward and Zhang [Phys. Rev. B {\bf 83}. 165210 (2011)]  has shown need for correction and modification of current implementation of the $\bm{k\cdot p}$ method and operator ordering scheme using the interaction parameters defined under double group consideration. This manuscript examines the difference in treatment of spin-orbit interaction under the single and double group formulations. We show that the restriction to the adapted double group bases, brought about by the imposition of single group selection rule in calculating the $\bm{k\cdot\pi}$ interaction, is not appropriate. In addition, the unitary transformation employed in the literature to diagonalise the intra-band spin orbit interaction in semiconductors with diamond lattice can not remove the inter-band terms. It leads to a bases set for valence band ordered differently from $D_{\frac{3}{2}}^+$ in the $O(3)$ group thus invalidating any correlation of magnetic quantum number to the $z$ component of angular momentum. Under the double group consideration, spin-orbit interaction affects all the zone centre states and offers a mechanism for changing the relative positions of various zone centre states in the conduction band. In addition to the mixing caused by $\bm{k}$ independent spin orbit terms, formation of hybridised orbitals under double group rules also produce mixing. This can lead to inversion in materials such as $\alpha$-tin. The re-arrangement of conduction band zone centre states leads to a negative $\gamma_2$ in most materials with respect to the valence band bases having the same order as $D_{\frac{3}{2}}^+$ in the $O(3)$ group. The double group formulation is also required in the description of Zeeman interaction, including the $\hat{\bm{S}}\cdot \bm{B}$ term, for the mixed zone centre states. It provides a correspondence between second order interaction parameters and the Luttinger invariants. A form of the Hamiltonian generally used in the literature is obtained from the results of EWZ using a unitary transformation of the valence band basis and re-definition of some of the Luttinger invariants. Such a unitary transformation enables explanation of optical transition selection rules for this form of Hamiltonian and has the effect of inverting the signs of $\gamma_2$ in the valence band. The inter-band block of the transform Hamiltonian differs from that of EWZ but the associated change of sign of  $\gamma_2$ ensures the same coupling. Various single group formulations are compared to the double group formulation with the the critical difference identified as mixing caused by inter-band spin orbit interaction. The double group formulation is used to model the dispersion in Ge and Si. It overcomes deficiencies in single group formulation and is able to explain the conduction band/spin split off band effective mass concurrently using one set of parameters. 
\end{abstract}


\maketitle

\section{Introduction}
\label{sec:intro}
The $\bm{k\cdot p}$ method has been developed under the single group formulation, initially without consideration of the spin degree of freedom\cite{Shockley:1950fh,Dresselhaus:1955da} (referred to as the DKK model after Dresselhaus-Kip-Kittel\cite{Dresselhaus:1955da} in this manuscript). The effects of electron spin has been considered through the spin orbit interaction terms in the one electron Hamiltonian, obtained from Dirac equation using Foldy-Wouthuysen transform\cite{[][{. The sign in front of the term $e\phi$ in Eq.(36) is incorrect.}]Foldy:1950gg}. The treatment of spin orbit interaction differentiates the implementation of $\bm{k\cdot p}$ method into single and double group formulations. In this manuscript, a single group formulation of the $\bm{k\cdot p}$ method utilises single group selection rules and the group theoretical method in the evaluation of $\bm{k\cdot p/k\cdot\pi}$ perturbation, whereas the double group formulation make use of double group selection rules in the evaluation of the $\bm{k\cdot\pi}$ perturbation. In constructing models with a limited near set of zone centre bases, the method of L\"owdin\cite{LowdinPO:1951io} is frequently used to account for the effects of remote set of states as L\"owdin interactions. This method can be derived from quasi-degenerate perturbation theory\cite{Wagner:1986wm,fizik:2003wv}.

To incorporate the effects of electron spin, double group selection rules are required, as energy eigenfunctions of the unperturbed Hamiltonian form the bases of irreducible representations (IR) of the double group. Recent work by Elder, Ward, and Zhang\cite{Elder:2011kg} (referred to as EWZ in this manuscript) follows this prescription, both in terms of classification of zone centre energy eigenstates, and the use of selection rules in the evaluation of the matrix representation of the $\bm{k\cdot\pi}$ perturbation with respect to the zone centre states. The use of double group selection rules enables the work of EWZ to take into account mixing produced by the inter-band $\bm{k}$ independent spin orbit interaction under the one electron theory. From the perspective of a semi-empirical technique, it also accounts for mixing due to many electron effects such as hybridisation in the formation of covalent bond in cubic semiconductor crystals. 

The method of invariant featured prominently in the development of the $\bm{k\cdot p}$ method\cite{LuttingerJM:1956hi, BirPikus,Suzuki:1974gd,Cho:1976hh,Trebin:1979jl}. Without specifying the bases, the Hamiltonian is obtained by ensuring its invariance under symmetry operations of the lattice. It is expressed in terms irreducible perturbation components such as wavevector $\bm{k}$, corresponding generators, and a set of invariants commonly known as Luttinger invariants. This analysis can be performed using either single or double group selection rules. Correspondence has been established between the DKK, and EWZ models derived from perturbation theory and models derived under the method invariant using single, and double group selection rules, respectively. Under double group selection rules, there are five invariants describing the valence band, two for the spin orbit split off band and three for the inter-band block between the valence band and spin split off band. In contrast, there are only four invariants describing the corresponding zone centre states under the single group selection rules. These invariants are expected to be linearly independent of one another and relations formed between invariants and second order interaction parameters obtained under perturbation theory should preserve such linear independence.

Historically, spin-orbit interaction is considered as a perturbation under the single group formulation\cite{Dresselhaus:1955da,Kane:1956iz}. Various models have been developed to account for experimentally observed effects of spin. These include the relativistically corrected DKK model\cite{BirPikus,Suzuki:1974gd}, the Kane model\cite{Kane:1956iz,Kane:1966wa}, extended Kane model\cite{Rossler:1984fd,Zawadzki:1985je}, and the Weiler model\cite{Weiler:1978da}. They differ in the classification of the zone centre states used in the perturbative expansion, but all make use of the single group selection rules in the evaluation of $\bm{k\cdot p}$ or $\bm{k\cdot\pi}$ perturbations. Implicit in these models is the absence of mixing between zone centre states without spin during the process of constructing zone centre states with spin. The complexity of these models increases towards that of the double group formulation, with the Weiler model being able to explain the finite invariant $q$ and generally different Luttinger parameters in blocks coupled by the spin orbit interaction. However, all these models overestimate the zone centre conduction band effective mass in Ge, and underestimate the effective mass of spin-split off band in Si. There are similar issues in compound semiconductors where Hermann-Weisbuch\cite{Hermann:1977fc} parameters are introduced to attribute such discrepancies to unknown remote states. It is generally assumed that the method of invariant and perturbation theory approach refer to the same zone centre bases set when comparing them.

There are three elements in the treatment of spin-orbit interaction under the framework of single group formulation; 1) there is an assumption that the spin-orbit interaction is small allowing, a) the use of perturbation theory, b) the use of single group product bases as `zero' order un-perturbed wave functions, and c) the neglect of the $\bm{k}$ independent inter-band spin orbit interaction which causes mixing; 2) the spin orbit interaction matrix for the valence band referring to the product bases is obtained from `p' like atomic orbitals; 3) the unitary transformation for the degenerate valence band, which diagonalises the intra-band $\bm{k}$ independent spin-orbit interaction, is also derived from the `p' like atomic orbital. This manuscript focuses on each of these elements and shows that some of these assumptions are inappropriate for the consideration of spin-orbit interaction of electrons in cubic crystals. It is found that the single group selection rules imposes restrictions upon the bases. The existence of a finite spin orbit splitting is well known experimentally, and requires a finite intra-band $\bm{k}$ independent spin orbit interaction when evaluated against adapted double group bases of the single group formulation. This implies that the inter-band $\bm{k}$ independent spin orbit terms should also be finite, and poses a contradiction between the mixing caused by such terms, and the restriction placed upon adapted double group bases from adherence to the single group selection rules.

This manuscript is divided into twelve sections. Section \ref{sec:spin_orbit_product}-\ref{sec:incomplete_bases} examines the treatment of spin orbit interaction under current single group formulation, and discusses the limitation of the adapted double group bases when used in the single group formulation. Section \ref{sec:unitary_transform} examines the consequence of a unitary transformation which permits the derivation of the 6-band Hamiltonian widely used in the literature. The group theoretical background of the unitary transformation, both in terms of choice of representation matrices and order of basis, are discussed in appendix \ref{app:grp_background}. Section \ref{sec:double_group} discusses the different mechanisms of mixing arising from the construction of zone centre cell periodic wave functions, that subsequently require the use of double group formulation. This further illuminates how the spin orbit interaction affects the ordering of conduction band zone centre states derived from anti-bonding hybridised orbitals. Sections \ref{sec:magnetic_interaction}-\ref{sec:orientation} introduce the treatment of magnetic interactions under the double group formulation. The correspondence between the Luttinger invariants from both symmetric and anti-symmetric L\"owdin terms and second order interaction parameters in each block are established. The dependence of the Hamiltonian on the orientation of magnetic field is also derived. The following three sections then make a detailed examination of the treatment of spin orbit interaction under the single group formulation with the use of the group theoretical method. Section \ref{sec:relativistic_dkk} describes the treatment of spin orbit effect under the DKK model using relativistic corrections. The matrix representation of $H_{so1}^s$ and $H_{so2}^s$ are related to the matrix representation of the $\hat{\bm{p}}$ operator. Section \ref{sec:extended_kane} extends the Weiler model to incorporate remote states of all symmetries. Comparisons are then made between the Weiler model and double group formulation of EWZ, with the main distinction being the allowed mixing of single group states under EWZ. Section \ref{sec:lowdin_bases} re-examines the L\"owdin interaction obtained from quasi-degenerate perturbation theory. It shows that the double group bases, used in EWZ, are the logical choice in presence of inter-band $\bm{k}$ independent spin orbit interaction. Any form of Hamiltonian obtained from method of invariant refers to zone centre bases of the double group formulation. Section \ref{sec:double_group_parameters} extract double group interaction parameters for Si and Ge from experimentally derived Luttinger parameters, and compares dispersion relations obtained near the zone centre under the different models. The last section summarises the differences between the double group formulation and other models in the literature. The key difference between the approaches is mixing of zone centre states caused by $\bm{k}$ independent spin orbit interaction, and consideration of hybridised orbitals with spin degree of freedom. Other general points that have arisen from the development of double group formulation are also discussed.

In this manuscript, the notation of Koster et al.\cite{Koster:1963um} on group representations are used. The Condon Shortley phase convention is followed when spherical harmonics or irreducible spherical tensor operators are used. Most of the discussions focus on materials with a diamond lattice, though some remarks are made in relation to materials with a zincblende lattice. To facilitate discussions, the first order interaction between states belonging to various IRs are given in appendix \ref{app:first_order} together with the bases used in appendix \ref{app:bases}. The DKK Hamiltonian and relations between Luttinger invariants due to L\"owdin term and second order interaction parameters, are obtained under the single group formulation in appendix \ref{app:lowdin}. Similar relations between the Luttinger invariants and second order interaction parameters under the double group formulation are also obtained using perturbation theory for the symmetric L\"owdin interaction, and shown in appendix \ref{app:lowdin}. The relationship between some of the first order interaction parameters under single group formulation are obtained in appendix \ref{app:xis} by transforming the first order interaction with respect to the single group product bases. A distinction is made between Luttinger invariants obtained from method of invariant and experimentally determined Luttinger parameter. The latter is dependent on the model used to obtain the parameters from experimental observations. 

A remark is also appropriate here on the terms used in the description of zone centre basis functions for the perturbative expansion, and as bases of IR of the symmetry group of the unperturbed Hamiltonian. The solution of the DKK Hamiltonian at $\bm{k}=0$ forms the basis of single group IRs. Conversely, the transformation properties of eigenfunctions of the unperturbed DKK Hamiltonian, may be represented by the relevant {\em single group basis functions} of the IR. The {\em single group product bases}, obtained by direct product of spinor states and bases of {\em one} single group IR, form the bases of double group representation which is generally reducible. The unitary transformation of a single group product basis into {\em adapted double group basis} decompose such representation into IRs of the double group. It is able to represent the transformation properties of energy eigenfunctions of the unperturbed Hamiltonian, which does not contain mixed states derived from single group states with different energy. In other words, the unperturbed Hamiltonian does not contain $\bm{k}$ independent inter-band interactions. The use of single group product bases or adapted double group bases does not imply treatment of spin orbit interaction unless the relativistic corrections are included. The {\em double group bases} can represent the symmetry properties of eigenfunctions of Hamiltonian containing $\bm{k}$ independent inter-band interactions. Due to multiplicity in the decomposition of direct product of double group IR $\Gamma_8^\pm$, two set of {\em independent} bases in the $\Gamma_8^\pm$ IR are required to fully describe the angular dependence of relevant interactions/operators. For this purpose, the adapted double group bases from the same single group IR are {\em not} independent. The spin operator $\hat{\bm{S}}$ works on the spin degree of freedom and has matrix representation with respect to the zone centre bases. With respect to the single group product bases, the matrix representation of $\hat{\bm{S}}$ is simply $\frac{\hbar}{2}\sigma_\mu\otimes \mathbb{1}_{n\times n}$ where $\sigma_\mu$ are Pauli matrices, and $n$ is the dimension of the single group IR. The matrix representation of $\hat{\bm{S}}$ with respect to the adapted double group bases may then be obtained using the relevant unitary transformation. This is, however, not the case when the matrix representation of $\hat{\bm{S}}$ is formed between general double group bases. It can be obtained using group theoretical technique by considering the transformation properties of $\hat{\bm{S}}$.

\section{Spin-orbit interaction under the single group formulation}
\label{sec:spin_orbit_product}
The time independent one electron Schr\"odinger equation in absence of magnetic field may be cast into the following form after application of the Bloch's theorem,
\begin{widetext}
\begin{eqnarray}
\label{eqn:H0}
\underbrace{\underbrace{\left[\frac{\hat{\bm{p}}^2}{2m_0}+V(r)\right.}_{H_0^s}+\underbrace{\frac{1}{2m_0^2c^2}\hat{\bm{S}}\cdot(\nabla V(r)\times\hat{\bm{p}})}_{H_{so1}^s}}_{H_0}&+&\underbrace{\underbrace{\frac{\hbar}{m_0}\bm{k}\cdot \hat{\bm{p}}}_{H_{k\cdot p}^s}+\underbrace{\left.\frac{\hbar}{2m_0^2c^2}\bm{k}\cdot\left(\hat{\bm{S}}\times\nabla V(r)\right)\right]}_{H_{so2}^s}}_{H_1}u_{n,\bm{k}}(r) \notag\\
&=&\left(E_n(\bm{k})-\frac{\hbar^2k^2}{2m_0}\right)u_{n,\bm{k}}(r).
\end{eqnarray}
\end{widetext}
The zone centre states for the perturbative expansion are obtained from the solution of the Hamiltonian $H_0$, by setting $\bm{k}=\bm{k_0}$. This problem can then be treated in two ways\cite{Kane:1966wa,BirPikus}. The representation is formed at the point of expansion ($\bm{k_0}=\bm{0}$ in this case) with $H_{so1}^s$ incorporated as part of the unperturbed Hamiltonian $H_0$, which yields the double group formulation of EWZ. This requires the use of double group selection rules. The zone centre bases are therefore the eigenstates of $H_0$, and form the bases of IR of the double group. In the second approach, the product between the solution of $H_0^s$ and spinor state are used as bases, but the representation of $H_{so1}^s$ is not diagonal, and both intra-band and inter-band terms exist. An additional assumption is then made that single group selection rules are applicable when treating the $\bm{k\cdot\pi}$ interaction. From group theoretical viewpoint, the two approaches differ in the use of single or double group selection rules when dealing with the $\bm{k\cdot\pi}$ perturbation. They are formally equivalent under one electron theory, and should yield the same dispersion relation, if the bases for perturbation expansion are complete. The second approach thus assumes single group selection rules (for the $\bm{k\cdot\pi}$ perturbation), requires that the one electron Schr\"odinger equation is valid, and give rise to the single group formulation generally used in the literature. The advantage of using the second approach is the link between Luttinger invariants among bands which are degenerate under DKK\cite{Kane:1966wa}. 

In constructing the Hamiltonian with a limited set of bases, a unitary transformation of the single group product bases is made to remove the weak interaction between the near set and remote set to the desired order in $\bm{k}$ (see section \ref{sec:lowdin_bases}). The unitary transformation should diagonalise the unperturbed Hamiltonian $H_0$ which contains $H_{so1}^s$. However, the single group selection rule imposes the requirement of no `mixing' between different non-degenerate zone centre states of the DKK model\footnote{If mixing occurs, then the reduced tensor element and/or the angular dependence of the $\bm{k\cdot\pi}$ interaction would be different, contrary to the assumption of single group formulation. This effectively requires double group selection rules.}. Therefore, the unitary transformation occurs within the space spanned by each single group product bases, and it preserves the orthogonality between states derived from non-degenerate solutions of $H_0^s$. In absence of the inter-band $\bm{k}$ independent spin orbit interaction, the unitary transformation which diagonalises $H_{so1}^s$ also diagonalises $H_{0}$. Under these circumstances, the adapted double group bases under the restrictions of the single group formulation, can represent the angular dependence of the zone centre states. However, in the presence of inter-band $\bm{k}$ independent spin orbit interaction, there is a conflict between the use of single group selection rules and the need to diagonalise $H_{so1}^s$. This follows from the fact that unitary transformation, which occurs within the space spanned by single group product bases, can not diagonalise the inter-band $\bm{k}$ independent spin orbit interaction. The validity of single group formulation within the framework of one electron theory then rests on whether the inter-band $\bm{k}$ independent spin orbit interaction can be neglected.

Under the DKK model with relativistic corrections, the $H_{so1}^s$ term is treated as a perturbation in parallel with the $\bm{k\cdot p}$ perturbation. Using the single group product bases to represent zone centre states, this perturbation can be separated into the spatial and spin parts such that,
\begin{equation}
\label{eqn:kane}
\left<\phi_i|H_{so1}^s|\phi_j\right>=\bm{S}_{\alpha,\beta}\otimes \mathcal{L}_{mn}
\end{equation}
where, $\bm{S}_{\alpha,\beta}=\frac{\hbar}{2}\bm{\sigma}$ and $\mathcal{L}_{mn}$ are the matrix representation of $\hat{\bm{S}}$ and $\nabla V(r)\times\hat{\bm{p}}$ evaluated with respect to the spinor states and single group basis\cite{Kane:1956iz,Lew}, respectively. The operator $\nabla V(r)\times\hat{\bm{p}}$ transforms according to $\Gamma_4^+$ IR under the $O_h$ group symmetry operations. The matrix representation of the operator $H_{so1}^s$ with respect to the single group product bases can be constructed from a tensor product of $\bm{S}_{\alpha,\beta}$ and $\mathcal{L}_{mn}$. Selection rules derived from group theory are applied when evaluating this matrix using Wigner-Eckart theorem. They indicate that spin-orbit interaction can only affect zone centre states associated with single group states with $\Gamma_4^\pm$ and $\Gamma_5^\pm$ symmetry as `intra-band' terms, and states with $\Gamma_3^\pm, \Gamma_4^\pm$, and $\Gamma_5^\pm$ symmetry as `inter-band' terms. For zone centre states with $\Gamma_5^+$ and $\Gamma_4^-$ symmetry in the valence and conduction bands, the unitary transformation with single group restriction diagonalises the intra-band terms, and partitions the representation space as $\Gamma_6^+\otimes\Gamma_5^+=\Gamma_8^+\oplus\Gamma_7^+$ and $\Gamma_6^+\otimes\Gamma_4^-=\Gamma_8^-\oplus\Gamma_6^-$. It is important to note that while the intra-band $\bm{k}$ independent terms are diagonalised, the inter-band terms remain off-diagonal. Such inter-band terms, particularly those which mix single group states with different symmetry, contribute to L\"owdin terms via their modification of zone centre states. These particular terms are, however, neglected in all single group formulations of $\bm{k\cdot p}$ theory. 

\begin{table*}
\centering
\caption{Spin orbit perturbation calculated using Eq.(\ref{eqn:kane}).}\label{tbl:kane}
\begin{ruledtabular}
\begin{tabular}{ccc}
Single group bases & $\bm{S}_{\alpha\beta}\otimes \mathcal{L}_{mn}$ & Diagonalised $H_{so1}$ \\ \hline
$\Gamma_4^-:\:\{\left|x\right>,\left|y\right>,\left|z\right>\}$ & $\begin{pmatrix} 0 & -i &  0 & 0 & 0 & 1\\  i & 0 & 0 & 0 & 0 & -i \\ 0 & 0 & 0 &-1 & i & 0 \\ 0 & 0 & -1 & 0 & i & 0 \\ 0 & 0 & -i & -i & 0 & 0 \\ 1 & i & 0 & 0 & 0 & 0 \end{pmatrix}$ & $\Gamma_8^-\oplus\Gamma_6^-: \begin{pmatrix} 1 & 0 & 0 & 0 & 0 & 0 \\ 0 & 1 & 0 & 0 & 0 & 0 \\ 0 & 0 & 1 & 0 & 0 & 0 \\ 0 & 0 & 0 & 1 & 0 & 0 \\ 0 & 0 & 0 & 0 & -2 & 0 \\ 0 & 0 & 0 & 0 & 0 & -2\end{pmatrix}$ \\
$\Gamma_5^+:\:\{\left|yz\right>,\left|zx\right>,\left|xy\right>\}$ & $\begin{pmatrix} 0 & i &  0 & 0 & 0 & -1\\  -i & 0 & 0 & 0 & 0 & i \\ 0 & 0 & 0 & 1 & -i & 0 \\ 0 & 0 & 1 & 0 & -i & 0 \\ 0 & 0 & i & i & 0 & 0 \\ -1 & -i & 0 & 0 & 0 & 0 \end{pmatrix}$ & $\Gamma_8^+\oplus\Gamma_7^+: \begin{pmatrix} -1 & 0 & 0 & 0 & 0 & 0 \\ 0 & -1 & 0 & 0 & 0 & 0 \\ 0 & 0 & -1 & 0 & 0 & 0 \\ 0 & 0 & 0 & -1 & 0 & 0 \\ 0 & 0 & 0 & 0 & 2 & 0 \\ 0 & 0 & 0 & 0 & 0 & 2\end{pmatrix}$ 
\end{tabular}
\end{ruledtabular}
\end{table*}
The angular dependence of the representation matrices of $H_{so1}^s$ evaluated with respect to these bases are shown in Table.\ref{tbl:kane}. The unitary transformation which takes the single group product bases into the adapted double group bases for the $\Gamma_8^+\oplus\Gamma_7^+(\Gamma_5^+)$ or $\Gamma_8^-\oplus\Gamma_6^-(\Gamma_4^-)$ IRs, are listed in appendix \ref{app:unitary_transform}. They are obtained from relations between the product basis and the corresponding double group basis published by Onodera et al.\cite{[][{. The bases listed in this reference do not follow time reversal requirement stipulated by Eq.(2) of EWZ.}]Onodera:1966bj}, or those given in EWZ. Both similarity transformations can diagonalise the $\bm{k}$ independent spin orbit interactions evaluated with respect to the single group product bases that involve $\Gamma_4^-$ and $\Gamma_5^+$ IRs, as shown in Table \ref{tbl:kane}. 

There are two issues related to the matrix representation of $H_{so1}^s$ and the unitary transformation under the single group formulation. First is the reduced tensor elements of the matrix representation of $H_{so1}^s$. For the states with $\Gamma_5^+$ and $\Gamma_4^-$ symmetry in the valence and conduction band, the reduced tensor element of the intra-band term is generally equated to $\Delta/3$, where $\Delta$ is the relevant experimentally observed spin splitting between the related states under double group classification\cite{Kane:1956iz}. The spin splitting is clearly finite and the intra-band term then can not be neglected. This is examined further in section \ref{sec:relativistic_dkk}, where the matrix representation of $H_{so1}^s$ is related to matrix representation of momentum in the same way for both inter and intra-band terms. If one accepts a finite reduced tensor element for the intra-band term, one also expect that the reduced tensor elements of the symmetry permitted inter-band terms should also be finite. Therefore,  inter-band $\bm{k}$ independent spin orbit interaction may not be neglected. This raises the important issue of what the differences are between the properties of zone centre bases of the double group formulation, and those of the adapted double group bases under the restriction of the single group formulation. As we shall see in the next section, the adapted double group bases with the restriction of the single group formulation are incomplete. Mixing due to inter-band $\bm{k}$ independent spin orbit interaction is crucial to the difference between the double group formulation and single group formulation under the one electron theory. Many electron effects to be discussed in section \ref{sec:double_group}, also lead to mixing making the use of the double group formulation a necessity.

The second issue relates to the order of bases within the double group IR. Historically, the valence band and spin orbit split off band zone centre states are considered to form the $p_{\frac{3}{2}}$ and $p_\frac{1}{2}$ mutliplets\cite{Dresselhaus:1955da} even though the lowest basis of $\Gamma_5^+$ IR has $l=2$. The same unitary transformation is applied to states with $\Gamma_4^-$ and $\Gamma_5^+$ symmetry. Consequently, the historically chosen transformation takes the single group product basis from $\Gamma_4^-$ into basis of $\Gamma_8^-\oplus\Gamma_6^-(\Gamma_4^-)$ according to the $\left|J,M_j\right>$ notation. When the same transformation is applied to the single group product bases involving $\Gamma_5^+$ IR, such as zone centre states of valence band of Ge, the resulting adapted double group basis clearly form representation of $\Gamma_8^+\oplus\Gamma_7^+(\Gamma_5^+)$ but is {\em not} ordered according to the $\left|J,M_j\right>$ notation. This means that it is inappropriate to interpre the $M_j$ quantum number in terms of z-component of angular momentum. The impact of this discrepancy is discussed further in section \ref{sec:unitary_transform}.

A final observation is the energetic ordering in the valence band due to spin orbit interaction, obtained from the specific lowest single group basis and shown in Table \ref{tbl:kane}, is in contradiction to experimental observation. In fact, the sign of the reduced tensor element is dependent on the $l$ number of the single group bases based on the central field approximation. So the results shown in Table \ref{tbl:kane} remains compliant to group theoretical method. 

\section{Incompleteness of adapted double group bases under the single group formulation}
\label{sec:incomplete_bases}
In this section, the use of adapted double group bases within the single group formulation is examined to establish whether such basis are complete. The adapted double group basis, obtained by unitary transformation of single group product basis, clearly form the basis of representations of the double group. The matrix representation of a given operator, with well defined transformation properties, can be evaluated with respect to the bases of the IR. The number of linearly independent matrices required to describe such an operator is determined by the selection rules\cite{Koster:1958kea,Lax,Butler:1981ws}. Multiplicity in the decomposition of the direct product of $\Gamma_8^\pm$ IRs of the $O_h$ group is a key property that differentiates double group IR from the single group IR\cite{Koster:1958kea,Lax,Butler:1981ws,Damhus:1984uu}. For double group IR of $\Gamma_8^+$, it is $\Gamma_8^+\otimes\Gamma_8^+=[\Gamma_1^+\oplus\Gamma_3^+\oplus\Gamma_5^+]\oplus\{\Gamma_2^+\oplus 2\Gamma_4^+\oplus\Gamma_5^+\}$, while in contrast for single group IR of $\Gamma_5^+$ it is, $\Gamma_5^+\otimes\Gamma_5^+=\left\{\Gamma_1^+\oplus\Gamma_3^+\oplus\Gamma_5^+\right\}\oplus\left[\Gamma_4^+\right]$\footnote{IRs in $\{\}$ are from symmetric product and those in $[]$ are from anti-symmetric product. Time reversal properties of operator places additional constraint on the selection rules\cite{Lax}.}.  According to appropriate selection rules, operators such as total angular momentum $\hat{\bm{J}}$, which transforms according to the $\Gamma_4^+$ IR, are described by two linearly independent matrices when evaluated with respect to the double group bases of the $\Gamma_8^+$ IR, but only one linearly independent matrix when evaluated with respect to the single group bases of $\Gamma_5^+$ IR. For the total angular momentum operator $\hat{\bm{J}}$, one must be able to obtain two linearly independent matrices by means of enumeration of different bases of the $\Gamma_8^+$ IR. Thus, the double group bases are only complete if multiple set of bases yield the required number of linearly independent matrices. 

Since the $O_h$ group is a sub group of $O(3)$, the bases of IR of the $O(3)$ group also form bases of representation of the  $O_h$ group and are generally reducible. Similarly, single group bases may be constructed from bases of IR of the $SO(3)$ group\cite{Altmann:1965br,Wormer:2001ec}. The adapted double group bases can then be constructed from the bases of single group IR using the transformations described in appendix \ref{app:unitary_transform}. 
Letting $V_l^\Gamma$ define the relation between the bases $\left|\phi_i^\Gamma\right>$ of single group IR $\Gamma$, and bases $\left|\psi_j^l\right>$ of the $D_l$ IR of $SO(3)$ group, we have,
\[
\left|\phi_i^\Gamma\right>=\sum_{j=1}^{2l+1}{V_l^\Gamma}_{ji}\left|\psi_j^l\right>.
\]
where $l$ is the orbital angular momentum quantum number.
Then the adapted double group bases may be related to the product of spin states with bases $\left|\psi_j^l\right>$ of the $D_l$ IR of the $SO(3)$ group, by a matrix defined by,
\begin{equation}
U_l^\Gamma=\left(\mathbb{1}_2\otimes V^\Gamma_l\right)U^\Gamma.
\end{equation}
where $U^\Gamma$ is the unitary transformation which block diagonilses the representation matrices, and are given in appendix \ref{app:unitary_transform}. The matrix representation of the angular momentum operator evaluated with respect to the adapted double group bases can be obtained from,
\begin{equation}
\label{eqn:adapted}
\bm{J}^{adap}={U_l^\Gamma}^\dag\bm{J}^{sph}U_l^\Gamma
\end{equation}
where $\bm{J}^{sph}$ is the matrix representation of $\hat{\bm{J}}$ operator evaluated with respect to the product bases between the spinor states and bases of $D_l$ IR of the $SO(3)$ group. The matrix representation $\bm{J}^{sph}$ (in unit of $\hbar$) can be readily obtained from Pauli matrices and $\bm{L}_{(2l+1)\times(2l+1)}$, which is the matrix representation of the orbital angular momentum evaluated with respect to bases of $D_l$\cite{Merzbacher:1998vw},
\begin{equation}
\label{eqn:product}
\bm{J}^{sph}=\frac{\bm{\sigma}}{2}\otimes\mathbb{1}_{(2l+1)\times(2l+1)}+\mathbb{1}_{2\times 2}\otimes\bm{L}_{(2l+1)\times(2l+1)}.
\end{equation}

Considering the lowest basis of $\Gamma_5^+$ IR with $l=2$, the Cartesian components of the matrix representation for the total angular momentum (in units of $\hbar$) in the $\Gamma_8^+$ partition can be obtained using Eq.(\ref{eqn:adapted}), and are given as,
\begin{widetext}
\begin{equation}
\label{eqn:j_mu_odd}
\mathcal{J}_x^\prime=\frac{\sqrt{3}}{6}\begin{pmatrix}0 & 1 & 0 & -\frac{2}{\sqrt{3}} \\ 1 & 0 & 0 & 0 \\ 0 & 0 & 0 & 1 \\ -\frac{2}{\sqrt{3}} & 0 & 1 & 0\end{pmatrix}\quad
\mathcal{J}_y^\prime=\frac{\sqrt{3}i}{6}\begin{pmatrix}0 & -1 & 0 & -\frac{2}{\sqrt{3}} \\ 1 & 0 & 0 & 0 \\ 0 & 0 & 0 & -1 \\ \frac{2}{\sqrt{3}} & 0 & 1 & 0\end{pmatrix}
\mathcal{J}_z^\prime=\frac{1}{6}\begin{pmatrix}1 & 0 & 0 & 0 \\ 0 & 3 & 0 & 0 \\ 0 & 0 & -3 & 0 \\ 0 & 0 & 0 & -1\end{pmatrix}
\end{equation}
\end{widetext}
These matrices constitute one of the two linearly independent sets of matrices required to describe the angular dependent part of the matrix representation of the total angular momentum operator. If the adapted double group bases from {\em one} single group IR ($\Gamma_5^+$ in this case) are complete, the second set of linearly independent matrices should be obtainable from other bases with different $l$ number. However, results from using single group bases with $l=4$ (see Table \ref{tbl:basis_fn}) or any other bases transforming according to $\Gamma_5^+$ IR, returns a set of matrices linearly dependent on those given in Eq.(\ref{eqn:j_mu_odd}). This is not surprising given the way in which the adapted double group bases are constructed from bases of one single group IR. The single group selection rules stipulate that there is only one linearly independent matrix in the representation of $\hat{\bm{J}}$ with respect to the single group bases. The direct product of spinor states with such bases  followed by the unitary transformation can not generate any linear independence among the matrix representation of $\hat{\bm{J}}$ with respect to single group product bases. One may also note that the matrix representation $\mathcal{J}_\mu^\prime$, shown in Eq.(\ref{eqn:j_mu_odd}), does not satisfy the condition $\exp[-i\mathcal{J}^\prime_\mu4\pi]=\mathbb{1}$\footnote{If one consider the matrix representation of $\hat{\bm{J}}$ in the $\Gamma_8^+\oplus\Gamma_7^+$ space, then the condition $\exp[-i\mathcal{J}^\prime_\mu4\pi]=\mathbb{1}$ is satisfied.}. This reflects the fact the adapted double group bases of $\Gamma_8^+$ and $\Gamma_7^+$ remain coupled. Since  zone centre states form bases of IR of the double group, $\mathcal{J}_\mu^\prime$ must be scaled accordingly to satisfy the condition $\exp[-i\mathcal{J}^\prime_\mu4\pi]=\mathbb{1}$. Under the single group formulation, the matrix representation of the angular momentum operator evaluated with respect to the adapted double group bases of the $\Gamma_8^+$ valence band may always be represented by,
\begin{equation}
\label{eqn:j_mu_single}
J_\mu^{adapt}=c_1\mathcal{J}_\mu
\end{equation}
where $\mathcal{J}_\mu=-3\mathcal{J}_\mu^\prime$ and $\exp\left(-iJ_\mu^{adapt}4\pi\right)=\mathbb{1}$ or $c_1\in\mathbb{Z}$.  Therefore the adapted double group bases with restriction to one single group IR parentage are incomplete due to their inability to generate the required number of linearly independent matrices through enumeration.

The representation of $\Gamma_8^+$ may also be generated from the direct product $\Gamma_6^+\otimes\Gamma_3^+=\Gamma_8^+$ or $\Gamma_6^+\otimes\Gamma_4^+=\Gamma_8^+\oplus\Gamma_6^+$. The second set of linearly independent matrices can be easily generated using either the general double group bases with well defined $j$ number, or adapted double group bases from different single group parentage. To illustrate this point, the angular dependent part of matrix representation of $\hat{\bm{J}}_\mu$ is evaluated with respect to the basis of $D_{\frac{3}{2}}^+$ IR of the $O(3)$, which also form basis of $\Gamma_8^+$ IR of the $O_h$ group, are given by,
\begin{widetext}
\begin{equation}
\label{eqn:j_mu_even}
J_x=\frac{\sqrt{3}}{2}\begin{pmatrix}0 & 1 & 0 & 0 \\ 1 & 0 & \frac{2}{\sqrt{3}} & 0 \\ 0 & \frac{2}{\sqrt{3}} & 0 & 1 \\ 0 & 0 & 1 & 0\end{pmatrix}\quad
J_y=\frac{\sqrt{3}i}{2}\begin{pmatrix}0 & -1 & 0 & 0 \\ 1 & 0 & -\frac{2}{\sqrt{3}} & 0 \\ 0 & \frac{2}{\sqrt{3}} & 0 & -1 \\ 0 & 0 & 1 & 0\end{pmatrix}
J_z=\frac{1}{2}\begin{pmatrix}3 & 0 & 0 & 0 \\ 0 & 1 & 0 & 0 \\ 0 & 0 & -1 & 0 \\ 0 & 0 & 0 & -3\end{pmatrix}.
\end{equation}
\end{widetext}
Alternatively, they may be obtained from adapted double group bases derived from single group $\Gamma_4^+$ IR. This set of matrices are linearly independent from those obtained in Eq.(\ref{eqn:j_mu_odd}), but both are obtained from basis that transform according the $\Gamma_8^+$ IR. The matrix representation of the total angular momentum operator, evaluated with respect to the general double group bases under the double group formulation, is given by,
\begin{equation}
\label{eqn:j_mu_double}
J_\mu^{double}=c_1\mathcal{J}_\mu+c_2J_\mu .
\end{equation}
The coefficient $c_1$ and $c_2$ are subject to the condition $\exp\left(-iJ_\mu^{double}4\pi\right)=\mathbb{1}$. 
This requirement yield 
\begin{subequations}
\label{eqn:J_constraint}
\begin{eqnarray}
c_1&=&\frac{1}{8}\left(3n-m\right)\\
c_2&=&\frac{1}{8}\left(3m-n\right)
\end{eqnarray}
\end{subequations}
where $m,n \in \mathbb{Z}$. It is clear that the ratio of the eigenvalue of $J_z^{double}$ for the $\left|\frac{3}{2},\frac{3}{2}\right>$ and $\left|\frac{3}{2},\frac{1}{2}\right>$ states is $(3c_2-c_1)/(c_2-3c_1)=m/n$ for the general double group bases but a constant $1/3$ for the adapted double group basis, as indicated in Eq.(\ref{eqn:j_mu_single}). This particular feature of the zone centre energy eigenfunctions under the double group formulation, namely requiring two linearly independent matrices to represent the angular dependence of $\hat{\bm{J}}$, can never be represented by the product bases from one single group IR or the corresponding adapted double group bases. Thus the difference between the single, and double group formulations can be attributed to {\em mixing} of different non-degenerate single group states in the construction of compatible zone centre states under the double group classification.

The matrix representation of $\hat{\bm{J}}^2$ with respect the zone centre states, as bases of $\Gamma_8^+$ IR, is also diagonal. Hence, the zone centre energy eigenstates are always eigenstates of total angular momentum, but the eigenvalues of $\hat{\bm{J}}^2$ and $\hat{\bm{J}}_\mu$ are not generally integer multiples of $\hbar^2/4$ and $\hbar/2$ respectively. They are material dependent, and the eigenvalue of $\hat{\bm{J}}^2$ is proportional to $\frac{1}{4}\left(15c_1^2-9c_1c_2+15c_2^2\right)$. It is not then possible to equate the difference in $z$ components of angular momentum to integer multiples of $\hbar$.

In obtaining generators for irreducible perturbation transforming according to $\Gamma_4^+$ symmetry, Luttinger used the two linearly independent matrices which form the matrix representation from $\hat{\bm{J}}_\mu$ and $\hat{\bm{J}}_\mu^3$ operators using a single set of basis\cite{LuttingerJM:1956hi}. The reason that $\hat{\bm{J}}_\mu$ and $\hat{\bm{J}}_\mu^3$ operators return different linearly independent matrices, even when evaluated with respect to the same set of basis, is that they are different operators that have the same transformation properties. The two different linearly independent matrices obtained from a single set of adapted double group basis reflect the properties of two different operators, not the properties of the basis. Hence, the total angular momentum operator, is more appropriately described by Eq.(\ref{eqn:j_mu_odd}) or Eq.(\ref{eqn:j_mu_double}) depending on the use of single or double group formulation. If the zone centre states can be described by adapted double group bases under the single group restriction, then the second invariant $c_2$ for angular momentum operator would always be zero. The Zeeman interactions are different from the total angular momentum operator, and they are  discussed in section \ref{sec:magnetic_interaction}.

The same issues arise when derving the matrix representation of momentum operator $\hat{\bm{p}}$, or perturbation operator $\hat{\bm{\pi}}$, between states with $\Gamma_8^\pm$ symmetry. As shown in Eq.(\ref{eqn:double_first_order_G8pm}), it is described by two linearly independent matrices under the double group selection rules, but only one matrix if adapted double group bases are used under the single group restriction\cite{Rossler:1984fd,Zawadzki:1985je}. Given the finite $q$ Luttinger parameter observed in experiment, the use of adapted double group bases under the single group formulation is not appropriate.

The cause of the incompleteness of the adapted double group bases under the single group restriction arises formally from the absence, or neglect, of the inter-band $\bm{k}$ independent spin orbit interaction. It is well known that spin orbit interaction will cause mixing between $\Gamma_8^\pm$ IRs of the same parity\cite{BirPikus,Pollak}. To properly consider the effects of perturbations to second order in $\bm{k}$, such mixing should be included. Then the zone centre states of $\Gamma_8^-$ symmetry in the conduction band would contain a mixture of bases derived from $\Gamma_6^+\otimes\Gamma_4^-$ and $\Gamma_6^+\otimes\Gamma_3^-$ IRs. The restriction to one single group IR is in principle at least, incorrect. One may argue that spin orbit effect is a small perturbation and such mixing would be minimal. However, additional mechanisms of mixing exist if one considers the way in which hybridised orbitals are constructed with consideration of spin (see section \ref{sec:double_group}). This mixing  effect can be readily seen in the analysis shown in sections \ref{sec:extended_kane}, and \ref{sec:double_group_parameters}.

The incompleteness of adapted double group bases when used under the single group formulation implies that the treatment of spin-orbit interaction is inappropriate from a more fundamental view point. Mixing may arise as a result of many electron effects/configuration in addition to the effects of the k independent interband spin orbit interaction. A consequence of this is that zone center states with $\Gamma_7^-$ and $\Gamma_6^+$ symmetry may also experience the effects of spin orbit interaction as a shift in their energy, though the degeneracy of such states would not be lifted. While the matrix element of $H_{so1}^s$ is always zero between states derived from single group states of $\Gamma_2^-, \Gamma_1^+$ and $\Gamma_3^\pm$ IRs as dictated by symmetry, selection rules permit their existence under double group formulation between states associated with $\Gamma_7^-, \Gamma_6^+$ and $\Gamma_8^\pm$ IRs. The consequences of this observation are discussed in section \ref{sec:double_group} under the double group formulation.

\section{Comparison with 6-band Hamiltonian obtained under single group formulation}
\label{sec:unitary_transform}
The purpose of this section is to establish the relations between the form of the 6-band Hamiltonian expressed in EWZ, and those in the literature obtained under a single group approach using either perturbation theory or method of invariant. This should ease re-interpretation of any experimental data that relies on labelling of $M_j$ numbers, its correlation to $z$ component of angular momentum, and its use in designation of heavy and light hole states. It should also highlight the difference in sign between the $\gamma_2$ Luttinger invariant in the Hamiltonian of EWZ, and those used in the literature. 

It is first important to clearly define the concept of {\em heavy} and {\em light} hole bands. The heavy hole (light hole) band is generally interpreted as the set of Kramer degenerate pairs in the valence band that has the larger (smaller) effective mass at the zone centre in the $\left<001\right>$ principle axes directions. The $M_j$ number, serving as an index to the degenerate states at the zone centre, defines the order of the basis in the IR, and the form of the representation matrices of the $O_h$ group. It is frequently correlated to $z$ components of angular momentum. As stated in appendix \ref{app:grp_background}, a direct link to angular momentum can not be made for point groups and any such link has to be based on the fact that $O_h$ group is a subgroup of $O(3)$, and the bases of $D_{\frac{1}{2}}^\pm$ and $D_{\frac{3}{2}}^\pm$ IRs of the $O(3)$ group also form the bases of $\Gamma_6^\pm$ and $\Gamma_8^\pm$ IRs of the $O_h$ group. Thus, the order of $\left|J,M_j\right>$ bases of IRs of the $O_h$ group must be the same as those in the $\left|j,m_j\right>$ bases of $D_j^\pm$ IR of the $O(3)$ group, for the link between $M_j$ and $z$ component of angular momentum to exist. For this to occur, the representation matrices of the $O(3)$ group differ in sign between proper and improper rotations. If the correlation between the $M_j$ number to $z$ component angular momentum is to be made, the bases of $O_h$ group must be ordered in such as way that its representation matrices follows the same rule. The double group bases used by EWZ and those given in appendix \ref{app:bases} follows this convention.

Within the 4-band model, the $\left|J,M_j\right>=\left|\frac{3}{2},\pm\frac{3}{2}\right>$ pairs describe states with an effective mass along the $\left<001\right>$ directions given by $\frac{m_0}{\gamma_1-2\gamma_2}$, while the $\left|J,M_j\right>=\left|\frac{3}{2},\pm\frac{1}{2}\right>$ pairs describe states with an effective mass of $\frac{m_0}{\gamma_1+2\gamma_2}$. The $\left|J,M_j\right>$ quantum numbers have the significance defined in section III of EWZ, and determine the time reversal and spatial transformation properties of the specific basis function within the IR. Clearly, the $M_j$ index of the heavy and light hole bands exchange with each other if the sign of $\gamma_2$ changes. Once the sign of $\gamma_2$ is determined by the material parameters (see Eq.(\ref{eqn:g2_single}) and Eq.(\ref{eqn:g2_double}) in appendix \ref{app:lowdin} or Eq.(18b) of Ref.\onlinecite{Elder:2011kg}), the association of heavy or light hole bands with appropriate Kramer pairs is determined. In appendix \ref{app:lowdin}, a condensed version of the derivation of relations between the Luttinger invariants and second order interaction parameters is detailed, with a different choice of bases for the $\Gamma_8^\pm$ IR as compared with those in EWZ. Since the definition of a second order interaction parameter ensures that $F, G, H_1, H_2$ are negative and $\zeta$'s are positive (except $\zeta_{\Gamma_8^+,\Gamma_8^+}^{\Gamma_8^-,3}$)\footnote{These arguments on signs does not hold for material such as $\alpha-$Sn where inversion of the band structure occurs.},  the sign of $\gamma_2$ is mostly determined by the relative dominance between $F$ and $H_1$ in the single group formulation (see Eq.(\ref{eqn:g2_single})), or $\zeta_{\Gamma_8^+,\Gamma_8^+}^{\Gamma_6^-}$ and  $\zeta_{\Gamma_8^+,\Gamma_8^+}^{\Gamma_7^-}$ in the double group formulation (see Eq.(\ref{eqn:g2_double})). The sign of $\gamma_2$ is therefore mostly determined by the symmetry of the lowest zone centre conduction band states. In Ge and most other III-V compound semiconductors, a negative $\gamma_2$ may be expected due to the symmetry of lowest conduction band zone centre states being $\Gamma_7^-(\Gamma_6)$, and the bases in $\Gamma_8^+$ ordered in the same way as $D_{\frac{3}{2}}^+$.

The designation of valence states does not always correlate with numerical value of experimentally measured effective mass in different crystallographic directions. For example, when the in-plane mass is measured using cyclotron resonance with the magnetic field applied in the $[001]$ direction, the 4-band Hamiltonian yields a mass of $\frac{1}{\gamma_1\pm\gamma_2}$ where the $+$ ($-$) sign refers to the $\left|\frac{3}{2},\pm\frac{3}{2}\right>$ ($\left|\frac{3}{2},\pm\frac{1}{2}\right>$) pair. The relation between these measured masses and $M_j$ numbers would be in opposition to the heavy and light hole band assignment\cite{Hensel:1974da}. The energy eigenstates with an arbitrary $\bm{k}$ are also generally a mixture of zone centre states due to presence of off-diagonal terms in 4-, or higher-band models. In addition, one can not make an association between the mass measured in an arbitrary orientation, with zone centre states defined with respect to the principal axes.

The association of the heavy and light hole bands with the bases of the 6-band Hamiltonian, derivable from appendix D of EWZ, depends on the sign of Luttinger invariant $\gamma_2$. The Hamiltonian has been obtained from perturbation theory (see appendix \ref{app:lowdin}) and the method of invariant, and may be written as,
\begin{widetext}
\begin{equation}
\label{EWZ}
L=-\frac{\hbar^2}{2m_0}\begin{pmatrix}
P+Q & -S & R & 0 & \sqrt{\frac{3}{2}}S_{vs}^* & \sqrt{2}Q_{vs} \\
-S^* & P-Q & 0 & R & \sqrt{2}R_{vs}^* & \frac{1}{\sqrt{2}}S_{vs}^*\\
R^* & 0 & P-Q & S & \frac{1}{\sqrt{2}}S_{vs} & -\sqrt{2}R_{vs} \\
0 & R^* & S^* & P+Q & -\sqrt{2}Q_{vs} & \sqrt{\frac{3}{2}}S_{vs} \\
\sqrt{\frac{3}{2}}S_{vs}  & \sqrt{2}R_{vs} & \frac{1}{\sqrt{2}}S_{vs}^* & -\sqrt{2}Q_{vs} & P_{s} & 0 \\
\sqrt{2}Q_{vs} & \frac{1}{\sqrt{2}}S_{vs} & -\sqrt{2}R_{vs}^* & \sqrt{\frac{3}{2}}S_{vs}^* & 0 & P_{s}
\end{pmatrix}
\end{equation}
\end{widetext}
where,
\begin{eqnarray*}
P&=&\gamma_1k^2,\quad Q=\gamma_2(k_\|^2-2k_z^2),\quad S=2\sqrt{3}\gamma_3k_- k_z, \\
R&=&-\sqrt{3}\gamma_2(k_x^2-k_y^2)+2\sqrt{3}i\gamma_3k_xk_y,\\
Q_{vs}&=&\gamma_2^{vs}(k_\|^2-2k_z^2),\quad S_{vs}=2\sqrt{3}\gamma_3^{vs} k_- k_z,\\
R_{vs}&=&-\sqrt{3}\gamma_2^{vs}(k_x^2-k_y^2)+2\sqrt{3}i\gamma_3^{vs} k_xk_y,\quad P_{s}=\gamma_1^sk^2,\\
k_\pm&=&k_x\pm ik_y,\quad k_\|^2=k_x^2+k_y^2,\quad k^2=k_x^2+k_y^2+k_z^2.
\end{eqnarray*} 
If $\gamma_2>0$, the heavy hole states near the zone centre are associated with $\left|J,M_j\right>=\left|\frac{3}{2},\pm\frac{3}{2}\right>$ basis functions. If $\gamma_2<0$, then the heavy hole states near the zone centre are associated with the $\left|\frac{3}{2},\pm\frac{1}{2}\right>$ basis functions.

Any unitary transformation can be made to the basis transforming as a $\Gamma_8^+$ IR without affecting the dispersion relations of the valence bands. Thus, a unitary transformation can be considered which permutes the $M_j=\pm\frac{3}{2}$ and $M_j=\mp\frac{1}{2}$ states. This is, in some way, equivalent to using the bases of $\Gamma_8^-$ IR in place of the $\Gamma_8^+$ IR. Most results in the literature were obtained using the ordering of the odd bases, even though the zone centre valence states of Ge must have positive parity, and those in most III-V compound semiconductors are dominated by the spatially even part of the wave functions. The group theoretical reason behind this unitary transformation is given in appendix \ref{app:grp_background}. The transformation of the basis functions of the valence states are then describe by the relation,
\begin{eqnarray}
\label{eq:u_matrix}
\left(\left|\phi_1^\prime\right>, \left|\phi_2^\prime\right>, \left|\phi_3^\prime\right>, \left|\phi_4^\prime\right>\right)&=&\left(\left|\phi_1\right>, \left|\phi_2\right>, \left|\phi_3\right>, \left|\phi_4\right>\right)U \notag \\
U&=&\begin{pmatrix}
0 &0& -1& 0\\ 0 &0& 0& 1\\ 1 & 0 & 0 & 0 \\ 0 & -1 & 0 & 0
\end{pmatrix} 
\end{eqnarray}
where the basis transform as a row vector. The $U$ matrix has the same form as the $T$ matrix originally identified in EWZ and corrected in Eq.(2) of the erratum\cite{Elder2}. The $T$ matrix transforms the basis of the $\Gamma_8^-$ IR to ensure that the representation matrices for elements of the $\Gamma_8$ IR of the $T_d$ group are obtained under compatibility relations. However, the $U$ matrix is applied to the bases of $\Gamma_8^+$ IR transforming as a row vector, so that $U$ is actually the inverse of $T$ defined by EWZ. The unitary transformation of the valance band and spin split off band is then given by,
\begin{equation}
\label{eq:u_transform}
M=\begin{pmatrix}U & 0_{4\times 2} \\ 0_{2\times 4} & \mathbb{1}_{2\times 2} \end{pmatrix}
\end{equation}
The 6-band Hamiltonian {\em referring to the transformed bases} may be obtained from Eq.(\ref{EWZ}) using the transformation in Eq.(\ref{eq:u_matrix}) and is written as,
\begin{widetext}
\begin{equation}
\label{transformed}
L=-\frac{\hbar^2}{2m_0}\begin{pmatrix}
P-Q & -S & -R^* & 0 & \frac{1}{\sqrt{2}}S_{vs} & -\sqrt{2}R_{vs} \\
-S^* & P+Q & 0 & -R^* & \sqrt{2}Q_{vs} & -\sqrt{\frac{3}{2}}S_{vs}\\
-R & 0 & P+Q & S & -\sqrt{\frac{3}{2}}S_{vs}^* & -\sqrt{2}Q_{vs}\\
0 & -R & S^* & P-Q & \sqrt{2}R_{vs}^* & \frac{1}{\sqrt{2}}S_{vs}^*\\
\frac{1}{\sqrt{2}}S_{vs}^* & \sqrt{2}Q_{vs} & -\sqrt{\frac{3}{2}}S_{vs} & \sqrt{2}R_{vs} & P_{s} & 0 \\
-\sqrt{2}R_{vs}^* & -\sqrt{\frac{3}{2}}S_{vs}^* & -\sqrt{2}Q_{vs} & \frac{1}{\sqrt{2}}S_{vs} & 0 & P_{s}
\end{pmatrix}
\end{equation}
where terms have the same definition as in Eq.(\ref{EWZ}).

An effective Luttinger invariant can be defined as $\gamma_2^{eff}=-\gamma_2$. This leads to the inversion of the sign in front of the $Q, R $ and $R^*$ terms, replaces $R$ with $R^*$, and $R^*$ with $R$.  When a similar definition of effective Luttinger invariants, ${\gamma_2^{eff}}^\prime=-\gamma_2^{vs}$ and ${\gamma_3^{eff}}^\prime=-\gamma_3^{vs}$ in the interband block is made, the sign on all terms in the interband block inverts. The form of the L\"owdin contribution to the Hamiltonian is then the same as the one obtained by Luttinger and Kohn\cite{LuttingerJM:1955ee}, Suzuki and Hensel\cite{Suzuki:1974gd} and Chao and Chuang\cite{Chao:1992fv}, and given by,
\begin{equation}
\label{Chuang}
L=-\frac{\hbar^2}{2m_0}\begin{pmatrix}
P+Q & -S & R & 0 & -\frac{1}{\sqrt{2}}S_{vs} & \sqrt{2}R_{vs} \\
-S^* & P-Q & 0 & R & -\sqrt{2}Q_{vs} & \sqrt{\frac{3}{2}}S_{vs}\\
R^* & 0 & P-Q & S & \sqrt{\frac{3}{2}}S_{vs}^* & \sqrt{2}Q_{vs}\\
0 & R^* & S^* & P+Q & -\sqrt{2}R_{vs}^* & -\frac{1}{\sqrt{2}}S_{vs}^*\\
-\frac{1}{\sqrt{2}}S_{vs}^* & -\sqrt{2}Q_{vs} & \sqrt{\frac{3}{2}}S_{vs} & -\sqrt{2}R_{vs} & P_{s} & 0 \\
\sqrt{2}R_{vs}^* & \sqrt{\frac{3}{2}}S_{vs}^* & \sqrt{2}Q_{vs} & -\frac{1}{\sqrt{2}}S_{vs} & 0 & P_{s}
\end{pmatrix}
\end{equation}
where, the definition of the terms differs from those in Eq.(\ref{EWZ},\ref{transformed}) as,
\begin{eqnarray*}
Q &=&\gamma_2^{eff}(k_\|^2-2k_z^2)=-\gamma_2(k_\|^2-2k_z^2)\\ 
R&=&-\sqrt{3}\gamma_2^{eff}(k_x^2-k_y^2)+2\sqrt{3}i\gamma_3k_xk_y=\sqrt{3}\gamma_2(k_x^2-k_y^2)+2\sqrt{3}i\gamma_3k_xk_y  \\
Q_{vs}&=&{\gamma_2^{eff}}^\prime(k_\|^2-2k_z^2)=-\gamma_2^{vs}(k_\|^2-2k_z^2) \\
S_{vs}&=&2\sqrt{3}{\gamma_3^{eff}}^\prime k_- k_z=-2\sqrt{3}\gamma_3^{vs} k_- k_z  \\
R_{vs}&=&-\sqrt{3}{\gamma_2^{eff}}^\prime(k_x^2-k_y^2)+2\sqrt{3}i{\gamma_3^{eff}}^\prime k_xk_y=\sqrt{3}\gamma_2^{vs}(k_x^2-k_y^2)-2\sqrt{3}i\gamma_3^{vs} k_xk_y.
\end{eqnarray*} 
\end{widetext}

For the purpose of comparison, an approximation is made to restore the single group selection rules and ignore the spin orbit splitting in the definition of L\"owdin terms. This approximation is consistent with those considered under the single group formulation\cite{Kane:1966wa}. Thus, utilising the single group $\Gamma_5^+$ IR heritage of the 6-dimensional $\Gamma_8^+ \oplus \Gamma_7^+$ subspace, one can write the interband Luttinger invariants as,
\begin{subequations}
\label{single_grp_relation}
\begin{eqnarray}
{\gamma_2^{eff}}^\prime&=&-\gamma_2^{vs}=-\gamma_2=\gamma_2^{eff}\\
{\gamma_3^{eff}}^\prime&=&-\gamma_3^{vs}=\gamma_3\\
\gamma_1^s&=&\gamma_1
\end{eqnarray}
\end{subequations}
This leads to equality among effective Luttinger invariants in the interband block, and their counter parts in the main valence band block in Eq.(\ref{Chuang}). In this case, all the newly defined effective Luttinger invariants are positive, for most semiconductors with diamond and zincblende lattices, except diamond and possibly Si as indicated by Lawaetz\cite{[{}][{. The order of bases used correspond to those used in Eq.(\ref{Chuang}). Hence $\gamma_2>0$ for Ge and GaAs.}]Lawaetz:1971cm}. 

The form of Hamiltonian in Eq.(\ref{Chuang}) is similar to those obtained by Dresselhaus\cite{Dresselhaus:1955da}, Kane\cite{Kane:1956iz} and Luttinger and Kohn\cite{LuttingerJM:1955ee} using the either the alternative diagonalisation matrix or bases functions of the $\Gamma_8^-$ and $\Gamma_6^-$ IRs. Any difference in the signs of the terms between expressions in the literature and Eq.(\ref{Chuang}), is due to different choice of phase in the bases. The form of the Hamiltonians shown in Eq.(\ref{EWZ}) and Eq.(\ref{Chuang}) are related to each other by the unitary transformation given by Eq.(\ref{eq:u_transform}), and are therefore both correct as they give rise to the same dispersion relation, for a given set of Luttinger invariants. However, the the methodology employed to obtain Eq.(\ref{Chuang}) using the basis of $\{\Gamma_6^+,\Gamma_8^-,\Gamma_6^-\}$\cite{LuttingerJM:1955ee,Suzuki:1974gd} can only be justified under the method of invariant\cite{Trebin:1979jl} as shown in appendix \ref{app:grp_background}. The basis of Hamiltonian shown in Eq.(\ref{Chuang}), remain as those of the $\Gamma_8^+\oplus\Gamma_7^+$ IRs. As shown in appendix \ref{app:grp_background}, the association of $M_j$ with the $z$ component of angular momentum is not correct for the form of Hamiltonian in Eq.(\ref{Chuang}), and the subsequent association with heavy and light hole band (accepting $z$ components of angular momentum as $M_j\hbar$) is not valid. 

The zone centre energy eigenstates in the crystal are the eigenstates of the angular momentum operator but the eigenvalues for $\hat{\bm{J}}^2$ and $\hat{\bm{J}}_z$ are not generally integer multiples of $\hbar^2/4$ and $\hbar/2$ respectively. Any link between the $M_j$ quantum number and $z$ component is, at best, a correlation. In dealing with physical effects such as optically induced transitions, it is important to evaluate the relevant matrix elements in the Fermi Golden rule that determines  transition probability. For example, the relevant perturbation operators, for transition induced by light polarised as $\sigma_\pm$ in the Faraday configuration, are $\hat{\bm{p}}_x\pm i\hat{\bm{p}}_y$\cite{Groves:1970jm}. The permitted transition can easily be obtained using the form of $M_{\alpha^u,\alpha^v}^x\pm iM_{\alpha^u,\alpha^v}^y$ matrices given in appendix \ref{app:first_order}. For light linearly polarised in the $\left<001\right>$ direction, the corresponding transition matrix between states of the valence band ($\Gamma_8^+$), and lowest conduction band ($\Gamma_7^-)$ is given by,
\[
M_{\Gamma_8^+,\Gamma_7^-}^z=\frac{\hbar}{m}\begin{pmatrix}0 & 0 & 0 & 2 \\ 2 & 0 & 0 & 0\end{pmatrix}^T.
\]
It is clear that only states $\left|\frac{3}{2},\pm \frac{3}{2}\right>$ in the valence band are coupled to the conduction band states with $\Gamma_7^-$ symmetry by light polarised in the $\left<001\right>$ direction. As pointed out in the erratum to EWZ and earlier, the nature of $\left|J,M_j\right>=\left|\frac{3}{2},\pm \frac{3}{2}\right>$ (heavy or light) depends on the sign of $\gamma_2$. For materials with negative $\gamma_2$, these states corresponds to the light hole band as observed in Ge and many other compound semiconductors.

The unitary transformation indicated in Eq.(\ref{eq:u_matrix}) must also be applied to other interaction matrices, such as the strain dependent perturbation, and as a result the sign of the deformation potentials should be carefully considered. However, an important consequence of this implicit unitary transformation is the incorrect association of coupling provided by $Q_{vs}$ term between the $\left|\frac{3}{2},\pm\frac{1}{2}\right>$ states in the valence band and $\left|\frac{1}{2},\mp\frac{1}{2}\right>$ spin split off states, based on the form of Hamiltonian in Eq.(\ref{Chuang}). Such an association has been used in the interpretation of experimental data of materials under uniaxial stress, and underpins some of our understanding of the the electronic structure in these crystals\cite{Hensel:1963dd,Hasegawa:1963gr,Pollak:1968iw,Laude:1971kf}. 

In the rest of the manuscript, any reference to the 6-band Hamiltonian refers to Eq.(\ref{EWZ}). Reference to Eq.(\ref{transformed}-\ref{Chuang}) are stated explicitly.

\section{Double group formulation}
\label{sec:double_group}
It appears that all the three elements in the treatment of spin-orbit interaction under the single group formulation had some deficiencies. The use of double group formulation is then a natural way of dealing with spin-orbit interaction, by treating the $\bm{k}$ independent spin orbit term in Eq.(\ref{eqn:H0}) directly in the un-perturbed Hamiltonian. The discussions in section \ref{sec:incomplete_bases} indicated that the adapted double group basis were incomplete when utilised under a single group approach. This was attributed to a restriction placed upon such bases by the lack of mixing under the single group formulation. In this section, mixing is considered under both the single and double group formulations to establish the key difference between the two approaches. Other impacts of the double group formulation on the analysis of material parameters is also considered.

One of the main distinctions between double group formulation of EWZ and traditional single group formulations, is the use of hybridised orbital-derived zone centre states as bases for the expansion of Hamiltonian $H(\bm{k})$, instead of those derived from atomic orbitals. The action of crystal field on the atomic orbitals leads to zone centre states with symmetry given in Table \ref{tbl:atomic_symmetry}. In constructing zone centre states from atomic orbitals, a degree of mixing of atomic orbitals is permitted under the single group formulation, provided that they are compatible under the single group\cite{Dresselhaus:2008gx}. The compatibility between atomic orbitals and single group IR are shown in Table \ref{tbl:single_mixing}. We will consider these mixing with the condition that they do not alter the angular dependence of first order $\bm{k\cdot\pi}$ interaction under the single group formulation. In other words, mixing of atomic orbitals is subject to compatibility of single group IRs. Thus, states with $\Gamma_8^+ \oplus \Gamma_7^+$ symmetry derived from the single group $\Gamma_5^+$ IR contain contributions from $p, d_{xy}$ and $f_{z^3}$ orbitals, but no contribution from $d_{z^2}$ or $f_{z(x^2-y^2)}$ orbitals.
\begin{table*}[htdp]
\caption{Atomic orbitals symmetry and their contribution to bonding and anti-bonding states}
\begin{center}
\begin{ruledtabular}
\begin{tabular}{clcccc}
Atomic& \multirow{2}{*}{Symmetry}& \multicolumn{2}{c}{Bonding states} & \multicolumn{2}{c}{Antibonding states} \\ \cline{3-6}
orbital & & No spin~~ & With spin~~ & No spin~~ & With spin~~ \\ \hline
\multirow{1}{*}{s}& \multirow{1}{*}{$\Gamma_1^+ (s)$} & $\Gamma_1^+$ & $\Gamma_6^+$ & $\Gamma_2^-$ & $\Gamma_7^-$ \\ \cline{2-6}
\multirow{1}{*}{p} & \multirow{1}{*}{$\Gamma_4^- (p)$}& $\Gamma_5^+$ & $\Gamma_8^+\oplus\Gamma_7^+$ & $\Gamma_4^-$ &$\Gamma_8^-\oplus\Gamma_6^-$ \\ \cline{2-6}
\multirow{1}{*}{d}& $\left\{\begin{matrix}\Gamma_5^+ (d_{xy}) \\ \Gamma_3^+(d_{z^2})\end{matrix}\right.$& $\begin{matrix}\Gamma_5^+\\ \Gamma_3^+\end{matrix}$ & $\begin{matrix}\Gamma_8^+\oplus\Gamma_7^+\\ \Gamma_8^+\end{matrix}$ & $\begin{matrix}\Gamma_4^-\\ \Gamma_3^-\end{matrix}$ &$\begin{matrix}\Gamma_8^-\oplus\Gamma_6^-\\ \Gamma_8^-\end{matrix}$ \\ \cline{2-6}
\multirow{1}{*}{f}& $\left\{\begin{array}{l}\Gamma_2^- (f_{xyz})\\ \Gamma_4^- (f_{z^3})\\\Gamma_5^- (f_{z(x^2-y^2)})\end{array}\right.$& $\begin{matrix}\Gamma_1^+\\\Gamma_5^+\\\Gamma_4^+\end{matrix}$ & $\begin{matrix}\Gamma_6^+\\\Gamma_8^+\oplus\Gamma_7^+\\\Gamma_8^+\oplus\Gamma_6^+\end{matrix}$ & $\begin{matrix}\Gamma_2^-\\\Gamma_4^-\\\Gamma_5^-\end{matrix}$ &$\begin{matrix}\Gamma_7^-\\\Gamma_8^-\oplus\Gamma_6^-\\\Gamma_8^-\oplus\Gamma_7^-\end{matrix}$ \\ 
\end{tabular}
\end{ruledtabular}
\end{center}
\label{tbl:atomic_symmetry}
\end{table*}
\begin{table}[htdp]
\caption{Mixing of atomic orbitals permitted under single group formulation}
\begin{center}
\begin{ruledtabular}
\begin{tabular}{ll} 
Symmetry (zone & \multirow{2}{*}{Contributing atomic orbitals}\\
centre state) &   \\ \hline
$\Gamma_6^+$  & $s,\:f_{xyz}$ \\ 
$\Gamma_8^+\oplus\Gamma_7^+$  & $p,\:d_{xy},\:f_{z^3}$ \\ 
$\Gamma_8^+$  & $d_{z^2}$ \\ 
$\Gamma_8^+\oplus\Gamma_6^+$  & $f_{z(x^2-y^2)}$ \\ 
$\Gamma_7^-$ & $s^*,\:f_{xyz}^*$ \\
$\Gamma_8^-\oplus\Gamma_6^-$ & $p^*,\:d_{xy}^*,\:f_{z^3}^*$ \\ 
$\Gamma_8^-$ & $d_{z^2}^*$ \\ 
$\Gamma_8^-\oplus\Gamma_7^-$ & $f_{z(x^2-y^2)}^*$ \\ 
\end{tabular}
\end{ruledtabular}
\end{center}
\label{tbl:single_mixing}
\end{table}%
\begin{table}[htdp]
\caption{Mixing of atomic orbitals permitted under double group formulation}
\begin{center}
\begin{ruledtabular}
\begin{tabular}{lcl} 
Symmetry (zone &~~~& \multirow{2}{*}{Contributing atomic orbitals}\\
centre state) &  \\ \hline
$\Gamma_6^+$ & & $s,\:f_{xyz},\:f_{z(x^2-y^2)}$ \\ 
$\Gamma_7^+$ & & $p,\:d_{xy},\:f_{z^3}$ \\ 
$\Gamma_8^+$ & & $p,\:d_{xy},\:d_{z^2},\:f_{z^3},\:f_{z(x^2-y^2)}$ \\
$\Gamma_6^-$ & & $p^*,\:d_{xy}^*,\:f_{z^3}^*$ \\ 
$\Gamma_8^-$ & & $p^*,\:d_{xy}^*,\:d_{z^2}^*,\:f_{z^3}^*,\:f_{z(x^2-y^2)}^*$ \\ 
$\Gamma_7^-$ & & $s^*,\:f_{xyz}^*,\:f_{z(x^2-y^2)}^*$ \\ 
\end{tabular}
\end{ruledtabular}
\end{center}
\label{tbl:double_mixing}
\end{table}%
Under the single group formulation, incorporation of zone centre states with $\Gamma_8^+$ ($\Gamma_3^+$) and $\Gamma_8^+ \oplus \Gamma_6^+$ ($\Gamma_4^+$) symmetry containing $d_{z^2}$ and $f_{z(x^2-y^2)}$ orbitals, respectively, would lead to new bonding states which are not generally observed in experiment. They therefore must be considered as non-participating in the formation of the covalent bond. Likewise, under the single group formulation participation of $f_{z(x^2-y^2)}^*$ orbital in anti-bonding states with $\Gamma_7^-$ symmetry is not possible, as this orbital is derived from a single group IR with $\Gamma_5^-$ symmetry instead of $\Gamma_2^-$ symmetry. It is clear that under the single group formulation, restrictions are placed upon the mixing of different symmetry compliant orbitals into the bonding and anti-bonding zone centre states.

Under the double group formulation, the action of crystal field and spin orbit interaction is considered together as part of the un-perturbed Hamiltonian. Full mixing of orbitals within individual double group IRs is permitted. This can be explained under the effect of hybridisation under the action of both the crystal field and spin orbit interaction, as indicated in Table \ref{tbl:double_mixing}. Bonding states with $\Gamma_8^+$ symmetry and anti-bonding state with $\Gamma_7^-$ symmetry, may then arise from  $p, d_{xy}, d_{z^2} f_{z^3}, f_{z(x^2-y^2)}$ and $s^*, f_{xyz}^*, f_{z(x^2-y^2)}^*$ orbitals, respectively. The $k$-independent spin orbit interaction leads to further mixing between some states with the same parity in crystals with the diamond lattice, and among all states in crystals with zincblende lattice. All the atomic orbitals are permitted to participate in the formation of bonding and anti-bonding zone centre states. In other words, the double group formulation calls for symmetrization of linear combination of wave functions to be performed at double group level, and the Hamiltonian of the unperturbed system includes the $k$ independent spin orbit interaction. 

The energetic ordering of the zone centre states has a strong impact on the sign of the $\gamma_2$ Luttinger invariant in the valence band, as stated in section \ref{sec:unitary_transform} and shown in appendix \ref{app:lowdin}. This ordering also has a strong impact upon the bandstructure of the semiconducting material. From the point of view of a semi-empirical technique, the energies of zone centre states are generally determined from experimental observation. As a result, mixing allowed under the double group formulation would not alter the the assignment of zone centre state energies in practical implementation of the $\bm{k\cdot p}$ method as they are determined from experiment. However, under the double group treatment of spin-orbit interaction, the effects of spin can shift the energy levels of states with $\Gamma_7^-$ and $\Gamma_6^+$ symmetry. This provides an additional mechanism through which the ordering of zone centre states in the conduction band can be modified. Without solving the unperturbed Schr\"odinger equation under the double group formulation, group theoretical methods indicate that spin can effect states with $\Gamma_7^-$ and $\Gamma_6^+$ symmetry in addition to the lifting of degeneracy between states with $\Gamma_8^+ \oplus \Gamma_7^+$ and $\Gamma_8^- \oplus \Gamma_6^-$ symmetries. Since $\nabla V(r)\times\hat{\bm{p}}$ and $\hat{\bm{S}}$ both transform like an axial vector ($\Gamma_4^+$) under the $O_h$ group, the $H_{so1}^s$ term transforms according to $\Gamma_4^+\otimes\Gamma_4^+=\Gamma_1^+ \oplus\Gamma_3^+ \oplus\Gamma_5^+\oplus\Gamma_4^+$, and the interaction is time reversal even. The product decomposition of states with either $\Gamma_7^-$ or $\Gamma_6^+$ symmetry both give $[\Gamma_1^+] \oplus \{\Gamma_4^+\}$ IR in the decomposed direct sum. Thus, selection rules indicate that states $\Gamma_7^-$ and $\Gamma_6^+$ symmetry (and states with $\Gamma_7^+$, $\Gamma_6^-$, $\Gamma_8^{\pm}$) contain at least a $[\Gamma_1^+]$ IR in their decomposed direct sum from their direct product, permitting spin orbit interaction to shift the energy levels of such states at the zone centre. Consequently, many electron effects should be included in the determination of zone centre state energies or treated as a fitting parameter under the semi-empirical scheme. 

It is particularly interesting to examine the trend from diamond to $\alpha$-Sn, which indicates a reduction in energy of the $\Gamma_7^-$ state relative to other states. In materials with small spin splitting effects (diamond and Si) the $\Gamma_7^-$ state resides at energies above the $\Gamma_8^- \oplus \Gamma_6^-$ states. In Ge, with large spin splitting, the $\Gamma_7^-$ state has shifted below the $\Gamma_8^- \oplus \Gamma_6^-$ states and in $\alpha$-Sn with the largest spin splitting inversion has occurred. The double group formulation may then provide an explanation why states with $\Gamma_7^-$ symmetry reside at lower energies than states with  $\Gamma_8^-\oplus\Gamma_6^-$ symmetry in materials such as Ge or GaAs but not in materials such as diamond and Si. Without spin-orbit interaction, the lowest $\Gamma_7^-$ basis has $l=3$, whereas the lowest $ \Gamma_8^-\oplus\Gamma_6^-$ bases have $l=1$. While this may explain why the $\Gamma_7^-$ state should have higher energy among the degenerate anti-bonding hybridised orbitals (higher number of nodes associated with larger $l$ number indicate higher kinetic energy), as in the case of diamond and Si, spin orbit interaction should be taken into account to decide the final ordering of the conduction states. With the large orbital angular momentum quantum number associated with the bases of $\Gamma_7^-$ IR, one could expect a larger downward shift of $\Gamma_7^-$ states due to spin orbit interaction compared with that of the $\Gamma_8^-\oplus\Gamma_6^-$ states. This is, of course, dependent upon the deviation of zone centre states with $\Gamma_7^-$ symmetry away from those prescribe by the single group formulation (participation from $s^*, f_{xyz}^*$ orbitals {\em only}, as shown in Table \ref{tbl:single_mixing}). The mixing, or introduction of $f_{z(x^2-y^2)}^*$ orbitals into states with $\Gamma_7^-$ symmetry under the double group formulation, provides the required deviation from a single group prescription. Thus, in the case of semiconductors comprising of heavier elements, such as Ge and $\alpha$-Sn, one can therefore expect a stronger spin orbit effect and a larger degree of mixing in the zone centre states. Spin-orbit interaction and mixing may then have a deciding effect on the ordering of the conduction band zone centre states in these materials, even though no such effect is expected under the single group formulation of the $\bm{k\cdot p}$ method\cite{Cardona:2010gl}. 

The effects of mixing in zone centre state as described under the double group formulation may be illuminated further by making a comparison between crystals with a diamond and zincblende lattice. In crystals such as SiC or AlP, the conduction band state with $\Gamma_6$ symmetry is below states with $\Gamma_8$ and $\Gamma_7$ symmetry. By considering the compatibility relations between the $T_d$ and $O_h$ groups, these materials behave like Ge or GaAs, in terms of the ordering of zone centre states in the conduction band, even though the constituent elements belong to the same row as, or above that of Si in the periodic table. The distinction between them is the lack of inversion symmetry in crystals with zincblende lattice. This allows the $k$ independent spin orbit interaction to couple the valence band zone centre states to those of the conduction band under the double group selection rules. This additional coupling, over what was allowed in crystals with diamond lattice, enhances the deviation of the states in the conduction band with $\Gamma_6$ symmetry from those prescribed by the single group formulation. Therefore, spin orbit interaction and the additional coupling expected in zincblende crystals such as SiC and AlP, can enhance the downward shift of states with $\Gamma_6$ symmetry towards the valence band in comparison to those with $\Gamma_7$ and $\Gamma_8$ symmetry.

In the latter part of this section we shall examine the impact of double group formulation on the analysis and interpretation of material parameters. The sign of $\gamma_2$ for Si\cite{Hensel:1963dd} is generally considered negative within the double group framework of EWZ, as in the case of Ge, GaAs and many other semiconductors. To determine if this is the case for Si, one could measure the ratio of heavy and light hole masses in SiGe alloys. If the Luttiger invariants in Ge and Si have the same sign one could expect the heavy/light hole mass ratio to be a monotonic function of composition from Ge to Si, if the Luttinger invariants are smooth monotonic functions of composition. This is indeed the conclusion of simulation based on Kane's model reported by Braunstein\cite{Braunstein:1963fu}. However, the experimental data shown in Fig.10 of Ref.\onlinecite{Braunstein:1963fu} indicates that the mass ratio approaches unity, but would then clearly have to increase back to the value of Si. This indicate a point of inflection exists in the experimental data. It can only occur if the value of $\gamma_2$ for Si is positive and the point of inflection corresponds to $\gamma_2=0$ at the given composition. Therefore the experimental data here suggests that Si, like diamond, has a positive $\gamma_2$, while Ge has negative $\gamma_2$. In the SiGe material system, one must therefore deal with possibility of $\gamma_2$ changing sign depending on composition. The scaling of various material parameters under envelope function approximation should take this into account.

The question is then raised about the interpretation of data by Hensel and Feher\cite{Hensel:1963dd}, and Hasegawa\cite{Hasegawa:1963gr}. Taking into account the difference in the order of basis as detailed in section \ref{sec:unitary_transform}, the determination of sign of $\gamma_2$ relies on application of specific second order perturbative term which leads to linear stress dependence on effective mass of hole states coupled to the spin orbit band. However, it is easy to see that the effective mass of all states in the valence band would be dependent on the applied stress (products of strain and quadratic $\bm{k}$ terms in the Hamiltonian contains $\Gamma_1^+$ irreducible component) from method of invariant. This term has the same order as those considered by Hasegawa\cite{Hasegawa:1963gr}. The conclusion that stress affects the effective mass of only one of the valence band states may not be correct. One aspect of the measurement of Hensel and Feher shows very similar behaviour of Si compared with Ge and most other semiconductors with diamond or zincblende lattice. Under uniaxial compression in the $\left<001\right>$ direction, the states with lighter in-plane mass (normally referred to as heavy hole band) are depopulated. Thus, the energy of the heavy hole states decreased relative to the energy of the light hole states under uniaxial compression. This would be consistent with Si having a negative $\gamma_2$ as in the case of Ge and GaAs, if the deformation potential $b$ has the same sign in both these materials. This result would then be in contradiction to results of Braunstein\cite{Braunstein:1963fu}. 

Similar optical reflectance measurements under uniaxial stress exist in the literature and should be able to resolve this issue, at least in principle. The experimental data for GaAs and Ge\cite{Pollak:1968iw,Chandrasekhar:1977bi} are relatively easy to interpret. There is clear evidence that the light hole states move upwards in energy under the uniaxial compression in the $\left<001\right>$ direction and its value has a non-linear dependence on the stress. This leads to the conclusion that $\gamma_2$ is negative and heavy hole states are associated with $\left|J,M_j\right>=\left|\frac{3}{2},\pm\frac{1}{2}\right>$ quantum numbers. Data for Si is inconclusive\cite{Pollak:1968iw,Laude:1971kf}.  The quadratic dependence associated with  $\left|J,M_j\right>=\left|\frac{3}{2},\pm\frac{3}{2}\right>$ states can not be identified clearly from Fig.4 of Ref.\onlinecite{Laude:1971kf}. This may be due to the small spin orbit splitting in the valence band of Si compared with the large strain perturbation actually applied over the range of stress used. Judging from data of Hensel et al\cite{Hensel:1963dd}, decoupling of states occurs at stress of approximately $4\times 10^7$ Pascal. This is smaller than the smallest none zero stress of $10^8$ Pascal used by Laude et al.\cite{Laude:1971kf} Therefore, the sign of $\gamma_2$ for Si remains undetermined.

The relations between Luttinger invariants and second order interaction parameters are important when obtaining material parameters for the $\bm{k\cdot p}$ method. Thus it is important to relate {\em all} the Luttinger invariants to the second order interaction parameters, under both the single and double group selection rules for appropriate comparisons of the two formulations. In absence of magnetic field, there are only three Luttinger invariants $\gamma_1, \gamma_2$ and $\gamma_3$. To extract the five second order $\zeta$ parameters defined in EWZ from the experimentally measured Luttiger parameters, would then be an under determined problem. With magnetic field applied, two further invariants $\kappa$ and $q$ are available in addition to $\gamma_1, \gamma_2$ and $\gamma_3$, ensuring that there is a one to one correspondence between invariants and second order interaction parameters. Under the DKK model, there are a total of four invariants, $L, M$, $N$ and the magnetic invariant $P$, which can be related to a set of four second order interaction paramters, $F$, $G$, $H_1$ and $H_2$. Efforts to relate the five invariants obtained under the double group selection rules with second order interaction parameters in other models based on single group formulation lead to linear dependence between the invariants\cite{Hensel:1969ko,Groves:1970jm,Lawaetz:1971cm,Trebin:1979jl}, contrary to the symmetry arguments. In the following section, a full set of relations are extracted between invariants and second order interaction parameters under the double group selection rules by considering the magnetic perturbation. 

\section{Anti-symmetric L\"owdin term}
\label{sec:magnetic_interaction}
In this section the anti symmetric L\"owdin term is derived from perturbation theory and group theoretical method under the double group formulation. The relations between all the Luttinger invariants and second order interaction parameters are derived. This enables a one to one mapping between the Luttinger invariants and independent second order interaction parameters in each block of the Hamiltonian. Representation of Zeeman interactions are then obtained from relevant terms in Eq.(\ref{eqn:H0}).

The matrix representation of Hamiltonian between zone centre states associated with representation $\alpha^u$ and $\alpha^v$ (classified according to double group representations) in the presence of magnetic field may be written in block form as,
\begin{subequations}
\begin{widetext}
\label{eqn:Lowdin}
\begin{eqnarray}
H_{\alpha^u,\alpha^v}&=&H_0^{\alpha^u,\alpha^v}+H_{\bm{k\cdot\pi}}^{\alpha^u,\alpha^v}+H_{L\ddot{o}wdin:S}^{\alpha^u,\alpha^v}+H_{L\ddot{o}wdin:A}^{\alpha^u,\alpha^v}+ H_{so3}\\
H_{\bm{k\cdot\pi}}^{\alpha^u,\alpha^v}&=&\sum_\mu k_\mu \bm{M}_{\alpha^u,\alpha^v}^\mu\\
H_{L\ddot{o}wdin:S}^{\alpha^u,\alpha^v}&=&\sum_{\alpha^\beta}\sum_{\mu}^{x,y,z}\sum_{\nu}^{x,y,z}\frac{k_\mu \left[\bm{M}^\mu_{\alpha^u,\alpha^\beta}\bm{M}^\nu_{\alpha^\beta,\alpha^v}+\bm{M}^\nu_{\alpha^u,\alpha^\beta}\bm{M}^\mu_{\alpha^\beta,\alpha^v}\right]k_\nu(E_{\alpha^u}+E_{\alpha^v}-2E_{\alpha^\beta})}{4(E_{\alpha^u}-E_{\alpha^\beta})(E_{\alpha^v}-E_{\alpha^\beta})}  \label{eqn:Lowdin_S} \\
H_{L\ddot{o}wdin:A}^{\alpha^u,\alpha^v}&=&\sum_{\alpha^\beta}\sum_{\mu}^{x,y,z}\sum_{\nu}^{x,y,z}\frac{k_\mu \left[\bm{M}^\mu_{\alpha^u,\alpha^\beta}\bm{M}^\nu_{\alpha^\beta,\alpha^v}-\bm{M}^\nu_{\alpha^u,\alpha^\beta}\bm{M}^\mu_{\alpha^\beta,\alpha^v}\right]k_\nu(E_{\alpha^u}+E_{\alpha^v}-2E_{\alpha^\beta})}{4(E_{\alpha^u}-E_{\alpha^\beta})(E_{\alpha^v}-E_{\alpha^\beta})}  \label{eqn:Lowdin_A} \\
H_{so3}&=&g_0\mu_B\bm{S}_{\alpha^u,\alpha^v}\cdot \bm{B}
\end{eqnarray}
\end{widetext}
\end{subequations}
where $\bm{M}_{\alpha^u,\alpha^v}$ are defined in appendix \ref{app:first_order}, $H_0^{\alpha^u,\alpha^v}$ is diagonal and contains zone centre state energies when $\alpha^u=\alpha^v$, $H_{\bm{k\cdot\pi}}^{\alpha^u,\alpha^v}$ is the first order $\bm{k\cdot\pi}$ interaction, and $H_{L\ddot{o}wdin}^{\alpha^u,\alpha^v}$ is the interaction between states belonging to $\alpha^u,\:\alpha^v$ IRs mediated by remote states in $\alpha^\beta$ IR\cite{LowdinPO:1951io}, and is partitioned into symmetric and anti-symmetric parts. The Zeeman interaction due to electron spin ($H_{so3}$) arises directly from the Foldy-Wouthysen transform. The anti-symmetric L\"owdin term arises from lack of commutability between components of wave vectors due to presence of magnetic field. The total Zeeman interaction consists of contributions from the anti-symmetric L\"owdin terms and those due to electron spin ($H_{so3}$).

With second order interaction parameters as defined by Eq.(11-13) in EWZ, the symmetric part of the L\"owdin term is obtained from Eq.(\ref{eqn:Lowdin_S}) using $\bm{M}$ matrices shown in appendix \ref{app:first_order}. The form of this symmetric L\"owdin term has been given in Eq.(\ref{EWZ}), and relations between Luttinger invariants and second order interaction parameters are given in Eq.(\ref{eqn:luttinger_inv}).

For the anti-symmetric part of the L\"owdin term, we first consider interactions mediated by the remote states with $\Gamma_6^-$ and $\Gamma_7^-$ symmetry. Constructing the anti-symmetric L\"owdin term from Eq.(\ref{eqn:Lowdin_A}) using $\bm{M}$ matrices obtained in appendix \ref{app:first_order} yields two distinct terms,
\begin{subequations} 
\label{eqn:zeeman_g67m}
\begin{eqnarray}
H_{L\ddot{o}wdin:A}^{\Gamma_6^-}&=&-2\mu_B\zeta_{\Gamma_8^+,\Gamma_8^+}^{\Gamma_6^-}\frac{i\hbar}{e}\left(J_x[k_y,k_z] +c.p.\right) \\
H_{L\ddot{o}wdin:A}^{\Gamma_7^-}&=&-2\mu_B\zeta_{\Gamma_8^+,\Gamma_8^+}^{\Gamma_7^-}\frac{i\hbar}{e}\left(\mathcal{J}_x[k_y,k_z] +c.p.\right)
\end{eqnarray}
\end{subequations} 
where $c.p.$ stands for cyclic permutation of the preceding term, $[k_\mu, k_\nu]=k_\mu k_\nu-k_\nu k_\mu$, and the two generator matrices $J$ and $\mathcal{J}$ are the same as those used in Eq.(\ref{eqn:j_mu_single},\ref{eqn:j_mu_double}). The form of these matrices and additional phase factor ``i'', are obtained from constructing the anti-symmetric L\"owdin terms using perturbation theory. Both sets of generator matrices transform according to $\Gamma_4^+$ IR of the $O_h$ group and are clearly linearly independent of each other. As in the case of symmetric L\"owdin term shown in EWZ, the interaction between states with $\Gamma_8^\pm$ symmetry contribute three distinct interaction parameters to the anti-symmetric L\"owdin term,
\begin{widetext}
\begin{subequations}
\label{eqn:zeeman_g8m}
\begin{eqnarray}
H_{L\ddot{o}wdin:A}^{\Gamma_8^-,1}&=&\mu_B\zeta_{\Gamma_8^+,\Gamma_8^+}^{\Gamma_8^-,1}\frac{i\hbar}{e}\left(\mathcal{J}_x[k_y,k_z] +c.p.\right)+\mu_B\zeta_{\Gamma_8^+,\Gamma_8^+}^{\Gamma_8^-,1}\frac{i\hbar}{e}\left(J_x[k_y,k_z] +c.p.\right) \\
H_{L\ddot{o}wdin:A}^{\Gamma_8^-,2}&=&4\mu_B\zeta_{\Gamma_8^+,\Gamma_8^+}^{\Gamma_8^-,2}\frac{i\hbar}{e}\left(\mathcal{J}_x[k_y,k_z] +c.p.\right)\\
H_{L\ddot{o}wdin:A}^{\Gamma_8^-,3}&=&5\mu_B\zeta_{\Gamma_8^+,\Gamma_8^+}^{\Gamma_8^-,3}\frac{i\hbar}{e}\left(\mathcal{J}_x[k_y,k_z] +c.p.\right)+3\mu_B\zeta_{\Gamma_8^+,\Gamma_8^+}^{\Gamma_8^-,3}\frac{i\hbar}{e}\left(J_x[k_y,k_z] +c.p.\right).
\end{eqnarray}
\end{subequations} 
\end{widetext}

Application of a magnetic field in any of the principal axes direction leads to the consideration of the problem under  envelope function theory as prescribed by Luttinger and Kohn\cite{LuttingerJM:1955ee}. The coupled differential equation is obtained by replacing only the components of wave vector in the Hamiltonian perpendicular to the magnetic field with $k_\alpha\rightarrow \hat{k}_\alpha=\frac{1}{i}\frac{\partial}{\partial r_\alpha}+\frac{e}{\hbar}A_\alpha$, where $e$ is the positive universal constant of electronic charge. For magnetic field aligned in the $\hat{z}$  direction, the commutation relation between the corresponding operators is,
\begin{equation}
[\hat{k}_x,\hat{k}_y]=\frac{eB}{i\hbar}
\end{equation}
under the symmetric gauge ($A_x=-\frac{By}{2}, A_y=\frac{Bx}{2}, A_z=0$). The wave vector component $k_z$ remains a good quantum number as translational symmetry along the direction of magnetic field is maintained. 

The matrix representation of $H_{so3}$ with respect to the single group product bases is well defined since the spin and orbital parts are separated. A unitary transformation into the adapted double group bases yield a contribution to the Zeeman interaction with modification to the $\kappa$ parameter by a factor $-g_0/6$ as shown in Eq.(\ref{eqn:lutt_sdpf}). The problem is more complicated given the mixing of the zone centre state. However, $\hat{\bm{S}}$ transforms according to $\Gamma_4^+$ IR of the $O_h$ group. Therefore, the matrix representation of $\hat{\bm{S}}$ with respect to the double group bases of $\Gamma_8^+$ can also be expressed in a similar way to Eq.(\ref{eqn:j_mu_double}), but with different set of reduced tensor elements, 
\begin{equation}
\label{eqn:sigma}
S_\mu=\hbar\left(\kappa_\sigma\mathcal{J}_\mu+q_\sigma{J}_\mu\right)
\end{equation}
where $\kappa_\sigma$ and $q_\sigma$ reflects the properties of zone centre states with $\Gamma_8^+$ bases due to mixing. As generators of transformation matrices for spinor states under rotation operation, $S_\mu$ are subject to the same constraint as the angular momentum operator shown in Eq.(\ref{eqn:J_constraint}). It is not clear how $\kappa_\sigma$ and $q_\sigma$ cab be determined from theoretical or experimental perspective.

Writing the Zeeman interaction as, 
\begin{eqnarray}
\label{eqn:zeeman_form}
H_{Zeeman}&=&H_{L\ddot{o}wdin:A}+H_{so3}\notag\\
&=&-2\mu_B\left[\kappa\left(\mathcal{J}_xB_x +c.p.\right)+q\left(J_xB_x +c.p.\right)\right]
\end{eqnarray}
and combining results from the symmetric part of L\"owdin term, the associated Luttinger invariants are then given by,
\begin{subequations}
\label{eqn:invariantzeta}
\begin{eqnarray}
\gamma_1&=&2\zeta_{\Gamma_8^+,\Gamma_8^+}^{\Gamma_7^-}+2\zeta_{\Gamma_8^+,\Gamma_8^+}^{\Gamma_6^-}+\zeta_{\Gamma_8^+,\Gamma_8^+}^{\Gamma_8^-,1}+8\zeta_{\Gamma_8^+,\Gamma_8^+}^{\Gamma_8^-,2}+4\zeta_{\Gamma_8^+,\Gamma_8^+}^{\Gamma_8^-,3}-1\\
\gamma_2&=&\zeta_{\Gamma_8^+,\Gamma_8^+}^{\Gamma_6^-}-\zeta_{\Gamma_8^+,\Gamma_8^+}^{\Gamma_7^-}-4\zeta_{\Gamma_8^+,\Gamma_8^+}^{\Gamma_8^-,2}-2\zeta_{\Gamma_8^+,\Gamma_8^+}^{\Gamma_8^-,3} \\
\gamma_3&=&\zeta_{\Gamma_8^+,\Gamma_8^+}^{\Gamma_6^-}+\zeta_{\Gamma_8^+,\Gamma_8^+}^{\Gamma_7^-}-2\zeta_{\Gamma_8^+,\Gamma_8^+}^{\Gamma_8^-,2}\\
\kappa&=&\zeta_{\Gamma_8^+.\Gamma_8^+}^{\Gamma_7^-}-\frac{1}{2}\zeta_{\Gamma_8^+.\Gamma_8^+}^{\Gamma_8^-,1}-2\zeta_{\Gamma_8^+.\Gamma_8^+}^{\Gamma_8^-,2}-\frac{5}{2}\zeta_{\Gamma_8^+.\Gamma_8^+}^{\Gamma_8^-,3}-\frac{g_0\kappa_\sigma}{2}\\
q&=&\zeta_{\Gamma_8^+.\Gamma_8^+}^{\Gamma_6^-}-\frac{1}{2}\zeta_{\Gamma_8^+.\Gamma_8^+}^{\Gamma_8^-,1}-\frac{3}{2}\zeta_{\Gamma_8^+.\Gamma_8^+}^{\Gamma_8^-,3}-\frac{g_0q_\sigma}{2}
\end{eqnarray}
The anti symmetric L\"owdin terms can be obtained for the conduction and spin-orbit split off bands, as well as the interband block between the valence and spin-orbit split off bands using the same methodology. For conduction\footnote{Hole energy is used and hence $\gamma_1^c<0$.} and spin-split off bands, the Zeeman interaction is written as,  $H_{Zeeman}=H_{L\ddot{o}wdin,A}+H_{so3}=-\frac{1}{2}{\mu_B\hbar}\kappa\left(\sigma_xB_x +c.p.\right)$. The Luttinger invariants for these blocks are related to corresponding double group second order interaction parameters by,
\label{eqn:invariantzeta_ccss}
\begin{eqnarray}
\gamma_1^s&=&\zeta_{\Gamma_7^+,\Gamma_7^+}^{\Gamma_7^-}+4\zeta_{\Gamma_7^+,\Gamma_7^+}^{\Gamma_8^-}-1\\
\kappa^s&=&2\zeta_{\Gamma_7^+,\Gamma_7^+}^{\Gamma_7^-}-4\zeta_{\Gamma_7^+,\Gamma_7^+}^{\Gamma_8^-}-g_0\kappa_\sigma^s\\
\gamma_1^c(\Gamma_7^-)&=&\zeta_{\Gamma_7^-,\Gamma_7^-}^{\Gamma_7^+}+4\zeta_{\Gamma_7^-,\Gamma_7^-}^{\Gamma_8^+}-1\\
g_c(\Gamma_7^-)&=&2\zeta_{\Gamma_7^-,\Gamma_7^-}^{\Gamma_7^+}-4\zeta_{\Gamma_7^-,\Gamma_7^-}^{\Gamma_8^+}-g_0\kappa_{\sigma}^c
\end{eqnarray}
For the interband block between the valence and spin-split off band, the Zeeman interaction may be written as, $H_{L\ddot{o}wdin,A}=-{\mu_B}\hbar\kappa^{vs}\left(J_x^{vs}B_x +c.p.\right)$, where $\bm{J}^{vs}$ is written as,
\begin{widetext}
\begin{equation*}
J_x^{vs}=-\frac{1}{\sqrt{2}}\begin{pmatrix}-1 & 0 \\ 0 & \sqrt{3} \\ \sqrt{3} & 0 \\ 0 & -1\end{pmatrix}\quad 
J_y^{vs}=-\frac{1}{\sqrt{2}}\begin{pmatrix}1 & 0 \\ 0 & -\sqrt{3} \\ \sqrt{3} & 0 \\ 0 & -1\end{pmatrix} \quad
J_z^{vs}=-\frac{1}{\sqrt{2}}\begin{pmatrix}0 & 2 \\ 0 & 0 \\ 0 & 0 \\ 2 & 0\end{pmatrix} \label{eqn:J_vs}
\end{equation*}
\end{widetext}
The form of these matrices were obtained from perturbation theory using Eq.(\ref{eqn:Lowdin_A}). They may also be obtained using the method of invariants by selecting the appropriate generating operator (irreducible tensor operator) along with the appropriate bases of $\Gamma_8^+$ and $\Gamma_7^+$ IRs of the $O_h$ group, as detailed in EWZ. Alternatively, one may consider the total angular momentum operator $\hat{\bm{J}}$ evaluated with respect to the $j=5/2$ bases of $\Gamma_8^+$ and $\Gamma_7^+$ IRs of the $O_h$ group. The invariants are then related to the second order interaction parameter by,
\label{eqn:invariantzeta_vs}
\begin{eqnarray}
\gamma_{2}^{vs}&=&\frac{1}{\sqrt{2}}\left(\zeta_{\Gamma_8^+,\Gamma_7^+}^{\Gamma_7^-}-\zeta_{\Gamma_8^+,\Gamma_7^+}^{\Gamma_8^-,1}-4\zeta_{\Gamma_8^+,\Gamma_7^+}^{\Gamma_8^-,2}\right)\\
\gamma_{3}^{vs}&=& \frac{1}{\sqrt{2}}\left(\zeta_{\Gamma_8^+,\Gamma_7^+}^{\Gamma_7^-}+\zeta_{\Gamma_8^+,\Gamma_7^+}^{\Gamma_8^-,1}+2\zeta_{\Gamma_8^+,\Gamma_7^+}^{\Gamma_8^-,2}\right) \\
\kappa^{vs}&=&\frac{1}{\sqrt{2}}\left(\zeta_{\Gamma_8^+,\Gamma_7^+}^{\Gamma_7^-}-\zeta_{\Gamma_8^+,\Gamma_7^+}^{\Gamma_8^-,1}+2\zeta_{\Gamma_8^+,\Gamma_7^+}^{\Gamma_8^-,2}\right)-g_0\kappa_\sigma^{vs}
\end{eqnarray}
\end{subequations}
Thus, there is a correspondence between the Luttinger invariants and second order interaction parameters in each block of the Hamiltonian. These relations may be compared with those of the DKK Hamiltonian with the $\bm{k}$ independent relativistic correction detailed in section \ref{sec:relativistic_dkk}, as shown in Eq.(\ref{eqn:lutt_sdpf}), or Table I of Ref.\onlinecite{Roth:1959kd}. Under the single group formulation, the following relations are obtained between Luttinger invariants in different blocks of the Hamiltonian,
\begin{subequations}
\begin{eqnarray}
q&=&0,\quad\kappa^c=g_0\\
\gamma_1^s&=&\gamma_1,\quad\kappa^s-\frac{g_0}{3}=4\left(\kappa+\frac{g_0}{6}\right)\\
\gamma_2^{vs}&=&\gamma_2,\quad\gamma_3^{vs}=-\gamma_3,\quad\kappa^{vs}+\frac{g_0}{3}=\kappa+\frac{g_0}{6}
\end{eqnarray}
\end{subequations}
With the adapted double group bases under the single group formulation, the contribution from Zeeman interaction due to electron spin is a constant matrix with $\kappa_\sigma=\frac{1}{3};\:\:q_\sigma=0$, $\kappa_\sigma^s=-\frac{1}{3}$, and $\kappa_\sigma^{vs}=\frac{1}{3}$. Thus the contribution from Zeeman interaction due to electron spin makes the Luttinger invariants associated with magnetic interaction different among the different blocks even under the single group formulation.

It should be noted that there is freedom in choosing the form and the scaling of the generator matrices shown in Eq.(\ref{eqn:zeeman_g67m}). Thus, the specific relations between Luttinger invariants and second order interaction parameters  are dependent on the choice of linearly independent generator matrices. The choice of generators made here is guided by perturbation theory, and naturally leads to simpler relations between the Luttinger invariants and second order interaction parameters. In the literature, the $J_\mu^3$ generator is generally used as a second linearly independent generator and is associated with the $q$ invariant. Considering the double group bases used in this manuscript, $J_\mu^3$ is equivalent to $\mathcal{J}_\mu^3=\frac{5}{2}\mathcal{J}_\mu+\frac{3}{4}J_\mu$. This means that the $\kappa$ and $q$ invariants defined in this manuscript are related to those generally quoted in the literature $(\kappa^\prime, q^\prime)$ by the following relation,
\begin{subequations}
\label{eqn:kappaqprime}
\begin{eqnarray}
\kappa&=&\kappa^\prime+\frac{5}{2}q^\prime\\
q&=&\frac{3}{4}q^\prime
\end{eqnarray}
\end{subequations}

The treatment of magnetic interaction for $B$ along the $[001]$ direction yields results that are no different from those of Luttinger\cite{LuttingerJM:1956hi}. The distinction between $\mathcal{J}_\mu$ and $J_\mu^3$, can easily be explained in terms of linear combination of generators. However, there is a difference in the process of deriving Luttinger invariants from experimental data. If the magnetic field is applied the the $[001]$ direction, the $g$ factors for the valence band are given by,
\begin{subequations}
\begin{eqnarray}
g_\|(M_j=\pm\frac{3}{2})&=&-2\kappa+6q \\
g_\| (M_j=\pm\frac{1}{2})&=&-6\kappa+2q
\end{eqnarray}
\end{subequations}
These relations may be compared with the expression given in Table I of Ref.\onlinecite{Hensel:1969ko} as, $g_\|=2\kappa^\prime+\frac{q^\prime}{2}=2\kappa-6q$. 
In the literature, the magnetic invariants $\kappa^\prime$ and $q^\prime$ are related to second order interaction parameters obtained under single group formulation using single group selection rules\cite{Groves:1970jm}. The measured parameters are thus linearly related to one another, contravening double group selection rules and introducing errors in their values. The mixing of zone centre states also requires a different treatment of contribution to Zeeman interaction from the electron spin compared to single group formulation. 

A point to note is the difference between matrix representations of total angular momentum shown in Eq.(\ref{eqn:j_mu_double}) and Zeeman interaction shown in Eq.(\ref{eqn:zeeman_form}). While the two expressions share the same generator matrices, the reduced tensor elements are not necessarily related. The coefficient $c_1$ and $c_2$ for angular momentum are subject to constraints discussed in section \ref{sec:incomplete_bases} whereas $\kappa$ and $q$ of Zeeman interaction are not. Hence, the general relation between Zeeman interaction and total angular momentum is lost in the crystal. This distinction arises from the dependence of anti-symmetric L\"owdin interaction on both the remote states and near states in contrast to the sole dependence on near state for angular momentum. This point is demonstrated by the results of anti-symmetric L\"owdin term and angular momentum obtained from bases of $\Gamma_8^+$ IR derived from $\Gamma_5^+$ IR. There is only one generator for $\hat{\bm{J}}$, but two for the anti-symmetric L\"owdin term evaluated from perturbation theory. As a consequence, the unitary transformation which diagonalise the Zeeman interaction in a given orientation defined by external field differs from that generated by the total angular momentum operator $\hat{\bm{J}}$ through rotation of axes.

In other orientations, the perturbation approach illuminates the origin of the anisotropic nature of the interaction associated with $q$ invariant, and allows the derivation of an analytic expression for its matrix representation for the valence band for arbitrary orientations. Magnetic interactions with the $\boldsymbol{B}$-field directed in other orientations is discussed in the following section.

\section{Magnetic interactions with B-field in other orientations}
\label{sec:orientation}
With magnetic field applied in directions other than the $\left<001\right>$ directions, appropriate changes must be made to construct the relevant Hamiltonian. In the context of method of invariant, the procedure has been described by Luttinger\cite{LuttingerJM:1956hi}, and is used widely\cite{Xia:1991dz,Ikonic:cq}. An alternative method based on the transformation of wave vector only, was used by Eppenga and Schuurmans\cite{Eppenga:1988kk}. The method of Eppenga and Schuurmans differs from others in terms of the bases used. A perturbation theory based approach has been applied to study dispersion of crystals under uniaxial stress\cite{Laude:1971kf} by projecting the bases states of $\{\left|X\right>, \left|Y\right>, \left|Z\right>\}$ in appropriate directions. An operator ordered Hamiltonian for treatment of spatially confined systems was also obtained using the same approach\cite{vanDalen:1998tp,LASSEN:2004dk}. In this section, we derive the necessary unitary transformation induced by a change of the coordinate axes, and derive the L\"owdin interaction from the perturbation theory approach. The difference between methods of Luttinger and Eppenga and Schuurmans are clearly identified.

It is important to note that there are two coordinate systems, and the associated orthonormal bases of vector space, spanned by the degenerate zone centre energy eigenstates, form the representation the symmetry group. The $\{x,y,z\}$ axes are oriented along the $\left<001\right>$ crystallographic directions whereby the $z$ axis is chosen as the quantisation direction of the third component of angular momentum. The point group symmetry of the crystal defines this coordinate system. The orthonormal bases are simply the unperturbed zone centre energy eigenstates referring to the $\{x,y,z\}$ coordinate system $\left|\phi^{xyz}_{\alpha^u,i}\right>$. The second coordinate system with $\{123\}$ axes, is defined by the externally applied magnetic field $\mathbf{B}$, which is parallel to the $3$ direction and defines the quantisation direction of the third component of angular momentum. The corresponding zone centre energy eigenstates refer to the $\{123\}$ coordinate system $\left|\phi^{123}_{\alpha^u,i}\right>$. In principle, the Zeeman interaction should be diagonal with respect to such bases. In applying the envelope function theory, as prescribed by Luttinger and Kohn\cite{LuttingerJM:1955ee}, the replacement of wave vector components with operator occurs in the $\{123\}$ coordinate system. Using the symmetric gauge, we have,
\begin{subequations}
\begin{eqnarray}
k_1\rightarrow \hat{k}_1&=&\frac{1}{i}\frac{\partial}{\partial r_1}-\frac{eBr_2}{2\hbar}\\
k_2\rightarrow \hat{k}_2&=&\frac{1}{i}\frac{\partial}{\partial r_2}+\frac{eBr_1}{2\hbar}\\
\left[\hat{k}_1, \hat{k}_2\right] &=&\frac{eB}{i\hbar} \label{eqn:B_commutation}
\end{eqnarray}
\end{subequations}
The third wave vector component $k_3$ remains as a good quantum number. It is not possible to define the corresponding commutator in the $\{xyz\}$ coordinate system, since the commutation relation defined by Eq.(\ref{eqn:B_commutation}) only holds in the $\{123\}$ coordinate system. The bases of the vector space in the two coordinate system are related by a similarity transformation,
\begin{equation}
\label{eqn:bases_transform}
\left|\phi^{123}_{\alpha^u,j}\right>=\sum_iU^{\alpha^u}_{ij}\left|\phi^{xyz}_{\alpha^u,i}\right>.
\end{equation}
In principle, the action of this unitary transformation should diagonalise the Zeeman interaction. 

Prior to the procedure of formulating the envelope function theory, and subsequent second quantisation of Landau levels, one must obtain the un-perturbed Hamiltonian referring to the $\{123\}$ coordinate system in terms of wave vector components and optionally associated bases. The L\"owdin interaction defined in Eq.(\ref{eqn:Lowdin}) contains a typical term,
\[
k_m\mathcal{M}^m_{\alpha^u,\alpha^\beta}\mathcal{M}^n_{\alpha^\beta,\alpha^v}k_n
\]
where $m$ and $n$ index the directions in the $\{123\}$ coordinate system, and $\mathcal{M}^m_{\alpha^u,\alpha^\beta}$ are evaluated with respect to the $\left|\phi^{123}_{\alpha^u}\right>$ bases. This typical term may be related to $\bm{M}^m_{\alpha^u,\alpha^\beta}$ referring to the $\left|\phi^{xyz}_{\alpha^u}\right>$ bases, through the unitary transformation defined in Eq.(\ref{eqn:bases_transform}),
\begin{eqnarray*}
k_m\mathcal{M}^m_{\alpha^u,\alpha^\beta}\mathcal{M}^n_{\alpha^\beta,\alpha^v}k_n&=&k_m{U^{\alpha^u}}^\dag\bm{M}^m_{\alpha^u,\alpha^\beta}U^{\alpha^\beta}{U^{\alpha^\beta}}^\dag\bm{M}^n_{\alpha^\beta,\alpha^v}U^{\alpha^v}k_n \\
&=&{U^{\alpha^u}}^\dag k_m\bm{M}^m_{\alpha^u,\alpha^\beta}\bm{M}^n_{\alpha^\beta,\alpha^v}k_n U^{\alpha^v}
\end{eqnarray*}
Thus, one can evaluate $\bm{M}^m_{\alpha^u,\alpha^\beta}$ with respect to the $\left|\phi^{xyz}_{\alpha^u}\right>$ bases, then construct the symmetric and anti-symmetric L\"owdin terms, and finally make the unitary transformation into the bases of $\left|\phi^{123}_{\alpha^u}\right>$. The interaction matrix, $\bm{M}^m_{\alpha^u,\alpha^\beta}$ can be calculated using a projection of the momentum operator $\hat{p}_1,\hat{p}_2,\hat{p}_3$ along the direction $\{x,y,z\}$.  Thus, if the components of a spatial vector in the second coordinate system are related by,
\begin{equation}
\begin{pmatrix}
r_1\\r_2\\r_3
\end{pmatrix}=\begin{pmatrix}
R_{11} & R_{12} & R_{13} \\
R_{21} & R_{22} & R_{23} \\
R_{31} & R_{32} & R_{33} 
\end{pmatrix}\begin{pmatrix} x\\y\\z\end{pmatrix}
\end{equation}
where the matrix $\bm{R}$ is real and orthogonal, then the relation between interaction matrices is given by,
\begin{equation}
\bm{M}^m_{\alpha^u,\alpha^\beta}=\sum_{\mu\in\{x,y,z\}} R_{m,\mu}\bm{M}^\mu_{\alpha^u,\alpha^\beta}
\end{equation}
A typical term in the L\"owdin interaction is then given by,
\begin{equation}
k_m\bm{M}^m_{\alpha^u,\alpha^\beta}\bm{M}^n_{\alpha^u,\alpha^\beta}k_n=\sum_{\mu,\nu\in\{x,y,z\}} R_{m,\mu}R_{n,\nu}\bm{M}^\mu_{\alpha^u,\alpha^\beta} \bm{M}^\nu_{\alpha^u,\alpha^\beta}k_mk_n \label{eqn:rotate_M}
\end{equation}
Therefore a general term in the L\"owdin interaction transforms according to the direct product of two $l=1$ representations of the $SO(3)$ group, which is the symmetry group of proper rotations in 3-dimensions. 

Let us now focus on the magnetic interaction and anti-symmetric L\"owdin term. In applying these relations, we obtain exactly the same results as in Eq.(\ref{eqn:zeeman_g67m}-\ref{eqn:invariantzeta}) with the wave vector component referring to the $\{123\}$ coordinate system, but the corresponding generator matrices defined still referring to the bases $\left|\phi^{xyz}_{\alpha^u}\right>$. Thus, the Zeeman interaction can be written as,
\begin{equation}
\label{eqn:zeeman_form_other}
H_{Zeeman}=-2\mu_B\left(\kappa\mathcal{J}_3^{xyz}+qJ_3^{xyz}\right)B
\end{equation}
The invariants, as one may expect, are un-affected by the coordinate transformation, and their relations to the interaction parameters remain as described by Eq.(\ref{eqn:invariantzeta}). The generator matrices, on the other hand, also transform according to the $l=1$ representation of the $SO(3)$ group. This arises from the fact that the anti-symmetric part of the product representation also corresponds to the $l=1$ representation of the $SO(3)$ group. For the valence band, we then obtain two sets of linearly independent generators,
\begin{subequations}
\label{eqn:jz_transform}
\begin{eqnarray}
J_m^{xyz}&=&R_{m1}J_x+R_{m2}J_y+R_{m3}J_z \\
\mathcal{J}_m^{xyz}&=&R_{m1}\mathcal{J}_x+R_{m2}\mathcal{J}_y+R_{m3}\mathcal{J}_z
\end{eqnarray}
\end{subequations}
where the superscript $xyz$ indicates that these generators are referring to the basis $\left|\phi^{xyz}_{\alpha^u}\right>$, and the unitary transformation is still to be performed. They can be obtained either from these direct projections using Eq.(\ref{eqn:jz_transform}), or constructed directly from Eq.(\ref{eqn:Lowdin_A}) with aid of Eq.(\ref{eqn:rotate_M}). The only relevant terms in the anti-symmetric L\"owdin term are those related to $[\hat{k}_1,\hat{k}_2]$, as the magnetic field is along the $3$ axis. The other two terms are zero since $k_3$ remains a well defined quantum number and commutes with $\hat{k}_1$ and $\hat{k}_2$. 

To re-quantize the zone centre states along the 3 axis, one needs to diagonalise the Zeeman interaction,  shown in Eq.(\ref{eqn:zeeman_form_other}) for the valence band. In general, the two linearly independent matrices can not be diagonalised simultaneously, except when the 3 axis is in the $\left<001\right>$ directions. The required unitary transformation of bases, given in Eq.(\ref{eqn:bases_transform}) would become dependent on the material parameters $\kappa$ and $q$ and leave ($\kappa\mathcal{J}_3^{123}+qJ_3^{123})$ as a whole diagonal, but both $\mathcal{J}_3^{123}$ and $J_3^{123}$ individually non-diagonal. It is inconvenient and a unitary transformation which diagonalise $\mathcal{J}$ is used. It was mistakenly suggested that this alternative transformation quantises the degenerate states split by Zeeman interaction in the 3 axis\cite{LuttingerJM:1956hi}, but this is not true since $J_3$ remains non-diagonal. The unitary transformation which diagonalises the $\mathcal{J}$ generator and is used in place of Eq.(\ref{eqn:bases_transform}) is given by,
\begin{equation}
\label{eqn:J_diagonalisation}
U^{\Gamma_8^+}=\exp[-\mathrm{i}\bm{\mathcal{J}}\cdot \bm{\theta}]
\end{equation} 
where $\bm{\theta}$ defines the rotation of coordinate systems from $\{xyz\}$ to $\{123\}$. The transformation matrices for magnetic field aligned in $[110]$ and $[111]$ directions have been obtained in appendix \ref{app:orientation}. The diagonalised $\mathcal{J}_3^{123}$ generator has the same form as $\mathcal{J}_z$. The form of the $J_3^{123}$ generator for the magnetic field aligned in $[110]$ and $[111]$ directions are given below,
\begin{subequations}
\begin{equation}
J^{123}_3(\bm{\theta})=\exp[\mathrm{i}\bm{\mathcal{J}}\cdot \bm{\theta}]\left(R_{31}J_x+R_{32}J_y+R_{33}J_z\right)\exp[-\mathrm{i}\bm{\mathcal{J}}\cdot \bm{\theta}] \label{eqn:J_3_123}\\
\end{equation}
\begin{eqnarray}
J^{123}_3([111])&=&\frac{1}{2}\begin{pmatrix}-1 & 0 & 0 & 0\\ 0 & \frac{7}{3} & \frac{4}{3}\sqrt{2}i & 0\\ 0 & -\frac{4}{3}\sqrt{2}i &-\frac{7}{3} & 0\\ 0 & 0 & 0 & 1\end{pmatrix} \\
J^{123}_3([110])&=&\frac{1}{2}\begin{pmatrix}0 & 0 & \sqrt{3} & 0\\ 0 & 2 & 0 & - \sqrt{3}\\ \sqrt{3} & 0 & -2 & 0\\ 0 & - \sqrt{3} & 0 & 0 \end{pmatrix} 
\end{eqnarray}
\end{subequations}

Hence, the Zeeman interaction can always be written as,
\begin{equation}
H_{Zeeman}=-2\mu_B(\kappa\mathcal{J}_3^{123}+qJ_3^{123})B
\end{equation}
where $\mathcal{J}_3$ always has the diagonal form, but $J_3$ is defined by Eq.(\ref{eqn:J_3_123}). The $J_3$ generator is equivalent to the $S_3$ matrix in Ref.\onlinecite{LuttingerJM:1956hi}, and agrees with the form used in the paper (see Eq.(58)) once the differences between generators is taken into account. 

For the $\Gamma_8^-$ IR, the corresponding matrices can be obtained by using the unitary transformation defined by the matrix $T$ in Eq.(\ref{eqn:tmatrix}). In other words $J_3^{123}$ will be diagonal and $\mathcal{J}_3^{123}$ will be anisotropic. In the other interaction blocks of the Hamiltonian, the Zeeman terms are diagonalised into the same form as in the $\{xyz\}$ coordinate system.

We note that the unitary transformations defined in Eq.(\ref{eqn:bases_transform}), or those used in its place and defined by Eq.(\ref{eqn:J_diagonalisation}) are not necessary for the description of the problem since $\left|\phi^{xyz}_{\alpha^u}\right>$ are complete. They are performed to obtain more convenient form of Hamiltonian. Examination of method of Eppenga and Schruumans  shows that they have simply performed the transformation described by Eq.(\ref{eqn:rotate_M}) and followed by summation over the Cartesian axis $\{1,2,3\}$, but without the transformation of the bases to $\left|\phi^{123}_{\alpha^u,j}\right>$. Since they are only dealing with symmetric L\"owdin term without operator ordering, the result is linked to that of method of Luttinger by the unitary transformation of the bases describe by Eq.(\ref{eqn:J_diagonalisation}).

\section{Comparison of double group formulation with relativistically corrected DKK Hamiltonian}
\label{sec:relativistic_dkk}
Within the framework of one electron theory, the approaches employed to solve the Schr\"odinger equation under the double or single group formulation should be equivalent (giving the same dispersion relations), provided the bases for perturbative expansion are {\em complete}. The primary difference between the two approaches can be identified as the inclusion or exclusion of $H_{so1}^s$ in the unperturbed Hamiltonian. As a result, the bases in the perturbative expansion when solving Eq.(\ref{eqn:H0}) differ under the two approaches.

In this section, we first examine relativistic corrections to the DKK model by method of invariants. The matrix representation $H_{so1}^s$ and $H_{so2}^s$ are then evaluated with respect to a complete set of single group product bases. The validity of DKK model with relativistic corrections are then examined in terms of its ability to explain experimental observations.

\begin{table*}[htdp]
\caption{Relativistic irreducible perturbations containing first order in Pauli matrices ($\sigma_\mu$), and zero or second order in wave vector components ($k_\mu$). Terms with $\Gamma_4^+$ symmetry are time reversal odd, and terms with $\Gamma_1^+, \Gamma_3^+$, and $\Gamma_5^+$ symmetry are time reversal even.}
\begin{center}
\begin{ruledtabular}
\begin{tabular}{ccl}
IR & & Irreducible perturbation \\ \hline
$\Gamma_1^+$ && $\sigma_xB_x+\sigma_yB_y+\sigma_zB_z$\\
$\Gamma_3^+$ && $\left\{\begin{array}{l}\frac{1}{\sqrt{3}}(2\sigma_zB_z-\sigma_xB_x-\sigma_yB_y)\\\sigma_xB_x-\sigma_yB_y \end{array}\right.$ \\
$\Gamma_4^+$ && $\left\{\begin{matrix}\sigma_x\\ \sigma_y\\ \sigma_z\end{matrix}\right.$;\quad $\left\{\begin{matrix}\sigma_xk^2\\ \sigma_yk^2\\ \sigma_zk^2\end{matrix}\right. $;\quad $\left\{\begin{matrix}\sigma_xk_x^2\\ \sigma_yk_y^2\\ \sigma_zk_z^2\end{matrix}\right.$;\quad $\left\{\begin{matrix}\sigma_y\{k_x,k_y\}+\sigma_z\{k_z,k_x\}\\ \sigma_z\{k_y,k_z\}+\sigma_x\{k_x,k_y\}\\ \sigma_x\{k_z,k_x\}+\sigma_y\{k_y,k_z\}\end{matrix}\right.$\\
$\Gamma_5^+$ && $\left\{\begin{matrix}\sigma_yB_z+\sigma_zB_y\\ \sigma_zB_x+\sigma_xB_z\\ \sigma_xB_y+\sigma_yB_x\end{matrix}\right.$
\end{tabular}
\end{ruledtabular}
\end{center}
\label{tbl:rel_purturb}
\end{table*}
The effects of spin orbit interaction can be incorporated into a limited band model using the method of invariant\cite{BirPikus} and perturbation theory\cite{Suzuki:1974gd}. Here we only consider relativistic corrections to the 6-band model. Under the method of invariant,  possible perturbations due to the relativistic corrections are constructed from products of the matrix representation of the spin operator $\hat{\bm{S}}$, and components of wave vector $\bm{k}$. With respect to single group product bases, the matrix representation of spin operator $\hat{\bm{S}}$ is simply $\frac{\hbar}{2}\bm{\sigma}\otimes\mathbb{1}_{n\times n}$, where $\bm{\sigma}$ are Pauli matrices, and $n$ is the dimension of the single group IR. Utilising the character table of the $O_h$ group, it can be shown that $\Gamma_5^+\otimes\Gamma_5^+=\{\Gamma_1^+\oplus\Gamma_3^+\oplus\Gamma_5^+\}\oplus[\Gamma_4^+]$. Hence, the symmetry allowed intra-band irreducible perturbations must have $\Gamma_4^+$ symmetry, if it is time reversal odd, and $\Gamma_1^+, \Gamma_3^+$, or $\Gamma_5^+$ symmetry if it is time reversal even. The permitted irreducible perturbations, which are first order in $\sigma$ but zero or second order in $k$, are listed in Table \ref{tbl:rel_purturb}.

The 6-band Hamiltonian with relativistic corrections consists of a non-relativistic term $H_{nr}(\bm{k})$, the $\bm{k}$ independent relativistic correction $H_r(0)$, and the $\bm{k}$ dependent relativistic correction $H_r(\bm{k})$,
\begin{widetext}
\begin{subequations}
\begin{eqnarray}
H(\bm{k})&=&H_{nr}(\bm{k})+H_{r}(0)+H_{r}(\bm{k})\\
H_{nr}(\bm{k})&=&U_{\Gamma_5^+}^\dag\left(\mathbb{1}_{2\times 2}\otimes H_{DKK}(\bm{k})\right)U_{\Gamma_5^+} \\
H_{r}(0)&=&-\frac{\Delta}{3}U_{\Gamma_5^+}^\dag\left(\sigma_x\otimes I_x+c.p.\right)U_{\Gamma_5^+}=-\frac{\Delta}{3}U_{\Gamma_5^+}^\dag\left(\bm{\sigma}\cdot\bm{I}\right)U_{\Gamma_5^+} \\
H_{r}(\bm{k})&=&-\frac{\hbar^2}{2m_0}U_{\Gamma_5^+}^\dag\left(-a_1k^2\left(\bm{\sigma}\cdot\bm{I}\right)-3a_2\left[(\sigma_x\otimes I_x-\bm{\sigma}\cdot\bm{I}/3)k_x^2+c.p.\right]\right.\notag \\
&&\quad\quad\left.+\frac{3}{2}a_3[(\sigma_x\otimes I_y+\sigma_y\otimes I_x)\{k_x,k_y\}+c.p.]\right)U_{\Gamma_5^+}  \notag \\
&&-\mu_BU_{\Gamma_5^+}^\dag\left\{3b_1(\sigma_xB_x+c.p.)\otimes\mathbb{1}_{3\times 3}\right.+36b_2\left((\sigma_xB_x)\otimes(I_x^2-I^2/3)+c.p.\right) \notag \\
&&\quad\quad\left.+6b_3\left((\sigma_xB_y+\sigma_yB_x)\{I_x,I_y\}+c.p.\right)\right\}U_{\Gamma_5^+}
\end{eqnarray}
\end{subequations}
\end{widetext}
where $H_{DKK}$ is given by Eq.(\ref{eqn:H_dkk}), $U_{\Gamma_5^+}$ defines the unitary transformation from single group product basis to adapted double group basis, $c.p.$ denotes cyclic permutation of the preceding term, $\sigma_\mu$ are the Pauli matrices, and $I_\mu$ is the matrix representation of angular momentum (in units of $\hbar$) with respect to the $l=2$ real bases of $\Gamma_5^+$ IR shown in appendix \ref{app:bases}. They are given by,
\begin{equation}
I_x=\begin{pmatrix} 0 & 0 & 0 \\ 0 & 0 & i \\ 0 & -i & 0 \end{pmatrix}\quad
I_y=\begin{pmatrix} 0 & 0 & -i \\ 0 & 0 & 0 \\ i & 0 & 0 \end{pmatrix}\quad
I_z=\begin{pmatrix} 0 & i & 0 \\ -i & 0 & 0 \\ 0 & 0 & 0 \end{pmatrix}\quad
\end{equation}
The relativistically corrected Luttinger invariants are then expressed as,
\begin{subequations}
\begin{eqnarray}
\gamma_1&=&\gamma_1^{nr}+a_1\\
\gamma_2&=&\gamma_2^{nr}+a_2\\
\gamma_3&=&\gamma_3^{nr}+a_3\\
\kappa&=&\kappa^{nr}+b_1+7b_2-b_3\\
q&=&9b_2-3b_3\\
\gamma_{1}^{ss}&=&\gamma_1^{nr}-2a_1\\
\kappa^{ss}&=&4\kappa^{nr}-2b_1-32b_2-16b_3\\
\gamma_{2}^{vs}&=&\gamma_2^{nr}-\frac{a_2}{2}\\
\gamma_3^{vs}&=&-\gamma_3^{nr}+\frac{a_3}{2}\\
\kappa^{vs}&=&\kappa^{nr}+2b_1-4b_2-2b_3
\end{eqnarray}
\end{subequations}
where the non-relativistic parameters are determined by the second order interaction parameters $F, G, H_1$ and $H_2$, as shown in Eq.(\ref{eqn:lutt_sdpf}). Under this particular treatment of spin orbit interaction, it is clear that the existence of the $q$ invariant is purely due to the relativistic correction. Moreover,  the Luttinger invariants describing L\"owdin interactions in different blocks, now differ from the non-relativistic values, and from each other. 

In general, most implementations of the single group formulation neglect the relativistic $\bm{k}$ dependent correction to the Luttinger invariants.  Inclusion of $H_r(\bm{k})$ in the Hamiltonian modifies the Luttinger parameters, and removes the link between them in different blocks under the adapted double group bases, as originally envisaged by Kane\cite{Kane:1966wa}. It is therefore important to establish the numerical impact of both $H_r(\bm{k})$ and $H_r(0)$.

Let us now examine the matrix representation of $H_{so1}^s$, and $H_1=H_{\bm{k\cdot p}}^s+H_{so2}^s$ with respect to the single group product bases. The matrix representation of $H_{\bm{k\cdot p}}^s$ with respect to single group product bases of IRs $\alpha^u, \alpha^v$ may be written as,
\begin{eqnarray}
\left<\psi_{\alpha^u}^i\right|H_{\bm{k\cdot p}}^s\left|\psi_{\alpha^v}^j\right>&=&\sum_\mu k_\mu U_{\alpha^u,\alpha^v}^\mu \notag \\
&=&\sum_\mu\frac{\hbar k_\mu}{m_0}\left<\psi_{\alpha^u}^i\right|\hat{\bm{p}}_\mu\left|\psi_{\alpha^v}^j\right> \notag \\
\bm{U}_{\alpha^u,\alpha^v}&=& \mathbb{1}_{2\times 2}\otimes \bm{M}_{\alpha^u,\alpha^v} \label{eqn:kp_matrix}
\end{eqnarray}
where $M_{\alpha^u,\alpha^v}^\mu$ is defined by Eq.(\ref{eqn:M_matrix}) for the momentum operator. The matrix representation of $H_{so2}^s$ with respect to the same bases may be written as,
\[
\sum_\mu k_\mu\bm{V}_{\alpha^u,\alpha^v}^\mu=\sum_\mu \frac{\hbar^2k_\mu}{4m_0^2c^2}\left<\psi_{\alpha^u}^i\right|\left(\hat{\bm{\sigma}}\times \nabla V(r)\right)_\mu\left|\psi_{\alpha^v}^j\right>
\]
Under the single group formulation, in which the unperturbed Hamiltonian is $H_0^s$, the gradient of the potential may be expressed as commutator of the momentum operator and $H_0^s$,
\[
\bm{\nabla}V(r)=\frac{i}{\hbar}[\hat{\bm{p}},H_0^s]=\frac{i}{\hbar}(\hat{\bm{p}}H_0^s-H_0^s\hat{\bm{p}})
\]
Expressing the single group product bases in terms of their spinor and orbital part $\left|\psi_{\alpha^u}^i\right>=\left|\phi_{\alpha^u}^q\right>\left|\chi^i\right>$, $\bm{V}$ may be evaluated as,
\begin{widetext}
\begin{eqnarray}
\frac{\hbar^2}{4m_0^2c^2}\left<\psi_{\alpha^u}^i\right|(\hat{\bm{\sigma}}\times\nabla V(r))\left|\psi_{\alpha^v}^j\right>
&=&\frac{i\hbar}{4m_0^2c^2}\left<\chi^i\right|\hat{\bm{\sigma}}\left|\chi^j\right>\times\left<\phi_{\alpha^u}^q\right|(\hat{\bm{p}}H_0^s-H_0^s\hat{\bm{p}})\left|\phi_{\alpha^v}^r\right> \notag \\
&=&\frac{i(E_{\alpha^v}-E_{\alpha^u})}{4m_0c^2}\left<\chi^i\right|\hat{\bm{\sigma}}\left|\chi^j\right>\times\frac{\hbar}{m_0}\left<\phi_{\alpha^u}^q\right|\hat{\bm{p}}\left|\phi_{\alpha^v}^r\right> \notag \\
\bm{V}_{\alpha^u,\alpha^v}&=&\frac{i(E_{\alpha^v}-E_{\alpha^u})}{4m_0c^2}\bm{\sigma}\times\bm{M}_{\alpha^u,\alpha^v} \label{eqn:sigmagradv_matrix}
\end{eqnarray}
\end{widetext}
Hence, the matrix representation of $H_1$ may be written as,
\begin{equation}
\label{eqn:kdotpi_single}
H_1^{\alpha^u,\alpha^v}=\sum_\mu k_\mu\left(U^\mu_{\alpha^u,\alpha^v}+V^\mu_{\alpha^u,\alpha^v}\right)
\end{equation}
Comparing Eq.(\ref{eqn:kp_matrix}) and Eq.(\ref{eqn:sigmagradv_matrix}), one can see that the reduced tensor matrix element of $\bm{V}$ and $\bm{U}$ differs by a factor $\frac{i(E_{\alpha^v}-E_{\alpha^u})}{4m_0c^2}$, which is inversely proportional to the electron rest mass and is of the order of $10^{-5}$. This means that $H_{so2}^s$ is relativistically small compared to $H_{\bm{k\cdot p}}^s$, and may be neglected when the two terms occur together as part of $H_1$.

The same can not be said about the $\bm{k}$ independent spin orbit interaction $H_{so1}^s$. If all the zone centre states constitute a complete set of bases and $\hat{\mathbb{1}}=\sum_{\alpha^\beta,n}\left|\phi_{\alpha^\beta}^n\right>\left<\phi_{\alpha^\beta}^n\right|$, then the relevant matrix element with respect to the single group product bases is given by,
\begin{widetext}
\begin{eqnarray*}
\frac{\hbar}{4m_0^2c^2}\left<\psi_{\alpha^u}^i\right|\hat{\bm{\sigma}}\cdot (\nabla V(r)\times\hat{\bm{p}})\left|\psi_{\alpha^v}^j\right>&=&\frac{i}{4m_0^2c^2}\left<\chi^i\right|\hat{\bm{\sigma}}\left|\chi^j\right>\cdot\left<\phi_{\alpha^u}^q\right|[\hat{\bm{p}},H_0^s]\times\hat{\bm{p}}\left|\phi_{\alpha^v}^r\right> \notag \\
&=&\frac{i}{4m_0^2c^2}\left<\chi^i\right|\hat{\bm{\sigma}}\left|\chi^j\right>\cdot\left<\phi_{\alpha^u}^q\right|\hat{\bm{p}}H_0^s\hat{\mathbb{1}}\times\hat{\bm{p}} -H_0^s\hat{\bm{p}}\hat{\mathbb{1}}\times\hat{\bm{p}}\left|\phi_{\alpha^v}^r\right> \notag\\
&=&\frac{i}{4m_0^2c^2}\left<\chi^i\right|\hat{\bm{\sigma}}\left|\chi^j\right>\cdot\left\{\sum_{\alpha^\beta,n}\left<\phi_{\alpha^u}^q\right|\hat{\bm{p}}\left|\phi_{\alpha^\beta}^n\right>\times\left<\phi_{\alpha^\beta}^n\right|\hat{\bm{p}}\left|\phi_{\alpha^v}^r\right>E_{\alpha^\beta} \right. \notag\\
&&-\left.\sum_{\alpha^\beta,n}\left<\phi_{\alpha^u}^q\right|\hat{\bm{p}}\left|\phi_{\alpha^\beta}^n\right>\times\left<\phi_{\alpha^\beta}^n\right|\hat{\bm{p}}\left|\phi_{\alpha^v}^r\right>E_{\alpha^u}\right\} \notag
\end{eqnarray*}
\end{widetext}
Then the matrix representation of $H_{so1}^s$ with respect to the single group product bases, denoted as $W_{\alpha^u,\alpha^v}$ may be written as,
\begin{eqnarray}
\label{eqn:so1_single_grp}
W_{\alpha^u,\alpha^v}&=&\frac{i}{4m_0^2c^2}\bm{\sigma\cdot}\sum_{\alpha^\beta}\bm{P}_{\alpha^u,\alpha^\beta}\times\bm{P}_{\alpha^\beta,\alpha^v}(E_{\alpha^\beta}-E_{\alpha^u}) \notag \\
&=&\frac{i}{4\hbar^2c^2}\bm{\sigma\cdot}\sum_{\alpha^\beta}\bm{M}_{\alpha^u,\alpha^\beta}\times\bm{M}_{\alpha^\beta,\alpha^v}(E_{\alpha^\beta}-E_{\alpha^u}) 
\end{eqnarray}
where $\bm{P}$ is the matrix representation of momentum operator. The inverse rest mass dependence of the matrix representation of $H_{so1}^s$ is absorbed into known, finite reduced tensor elements of $\bm{M}$ matrices. It is clear that this term is not relativistically small from this particular view point. The angular dependence from $\bm{\sigma}\cdot \left(\bm{M}_{\alpha^u;\alpha^\beta}\times\bm{M}_{\alpha^\beta;\alpha^v}\right)$ is readily diagonalised as a block, by a unitary transformation such as $U^{\Gamma_5^+}$, given in appendix \ref{app:unitary_transform}. It is important to recognise that while intra-band terms become diagonal as part of the Hamiltonian after the unitary transformation, the inter-band $\bm{k}$ independent spin orbit interaction terms remain on the off-diagonal blocks when evaluated with respect to the adapted double group bases. These inter-band terms are generally ignored in models derived under the single group formulation. Since $\bm{\nabla}V(r)\times\hat{\bm{p}}$ transforms according to $\Gamma_4^+$ IR of the $O_h$ group, single group selection rules readily permits this $\bm{k}$ independent interaction as intra-band terms between states with $\Gamma_4^\pm$ or $\Gamma_5^\pm$ symmetries, or {\em inter-band} terms between states with $\Gamma_3^\pm, \Gamma_4^\pm$ or $\Gamma_5^\pm$ symmetries with the same spatial parity. Hence, one can not have a finite reduced tensor elements for the intra-band terms, without having a finite reduced tensor elements for inter-band terms.

The $\bm{k}$ dependent relativistic correction arises from construction of L\"owdin terms using the first order $\bm{k\cdot\pi}$ perturbation. Using the matrix representation of the $H_1$ given in Eq.(\ref{eqn:kdotpi_single}) in place of $\bm{M}_{\alpha^u,\alpha^v}$ in Eq.(\ref{eqn:Lowdin_S},\ref{eqn:Lowdin_A}) to evaluate the L\"owdin interaction with respect to the single group product bases, we obtain,
\begin{widetext}
\begin{eqnarray}
L_{L\ddot{o}wdin}^{\alpha^u,\alpha^v}&=&\sum_{\alpha^\beta}\sum_{\mu}^{x,y,z}\sum_{\nu}^{x,y,z}\frac{(E_{\alpha^u}+E_{\alpha^v}-2E_{\alpha^\beta})}{4(E_{\alpha^u}-E_{\alpha^\beta})(E_{\alpha^v}-E_{\alpha^\beta})}k_\mu\left[\underbrace{\left(U^\mu_{\alpha^u,\alpha^\beta}U^\nu_{\alpha^\beta,\alpha^v}\pm U^\nu_{\alpha^u,\alpha^\beta}U^\mu_{\alpha^\beta,\alpha^v}\right)}_{\mbox{non-relativistic}}\right. \notag \\
&&+\underbrace{\left(U^\mu_{\alpha^u,\alpha^\beta}V^\nu_{\alpha^\beta,\alpha^v}+V^\mu_{\alpha^u,\alpha^\beta}U^\nu_{\alpha^\beta,\alpha^v}\right)\pm \left(U^\nu_{\alpha^u,\alpha^\beta}V^\mu_{\alpha^\beta,\alpha^v}+V^\nu_{\alpha^u,\alpha^\beta}U^\mu_{\alpha^\beta,\alpha^v}\right)}_{\mbox{relativistic correction: first order}} \notag \\
&&\left.+\underbrace{\left(V^\mu_{\alpha^u,\alpha^\beta}V^\nu_{\alpha^\beta,\alpha^v}\pm V^\nu_{\alpha^u,\alpha^\beta}V^\mu_{\alpha^\beta,\alpha^v}\right)}_{\mbox{relativistic correction: second order}}\right]k_\nu  \label{eqn:rel_Lowdin} 
\end{eqnarray}
\end{widetext}
where $\bm{U}$ and $\bm{V}$ are those defined in Eq.(\ref{eqn:kp_matrix}, \ref{eqn:sigmagradv_matrix}), and the $\pm$ signs refers to symmetric and antisymmetric L\"owdin interaction. It is apparent from Eq.(\ref{eqn:sigmagradv_matrix}) that $\bm{V}$ is zero when $\alpha^u=\alpha^v$ as demonstrated by Bir and Pikus\cite{BirPikus}. Hence, there is no intra-band contribution to the relativistic correction to L\"owdin term. The first order relativistic corrections to the L\"owdin interactions, shown in Eq.(\ref{eqn:rel_Lowdin}), have reduced tensor elements which are a factor of $\frac{(E_{\alpha^v}-E_{\alpha^u})}{4m_0c^2}$ smaller compared with the corresponding non-relativistic L\"owdin interaction. This indicates that the $\bm{k}$ dependent relativistic correction to the L\"owdin interaction can be neglected when treating the problem using single group product bases or adapted double group bases.

While the method of invariant shows that the Luttinger invariants differ between blocks in the 6 band model, the relativistic  corrections that lead to these differences are negligible. Thus, the equality among these Luttinger invariants in different blocks remains valid. The $q$ invariant is permitted to exist by symmetry, but is very close to zero. Therefore, the  DKK model with relativistic corrections still can not explain phenomenon readily observed experimentally. For example, experiments indicate a non-zero $q$ invariant, and an effective mass for the spin orbit band that differs from the value obtained from $\gamma_1$. In addition, the perturbation method could not account for the mixing introduced among states in the remote set by the $\bm{k}$ independent spin orbit interaction. In terms of $\bm{k}$ dependence, there are four independent second order interaction parameters in the valence band, but five Luttinger invariants obtained from method of invariant using double group selection rules. 

Within the single group formulation, it has been assumed that the $\bm{k}$ independent spin orbit interaction with respect to the single group product bases is finite for intra-band terms in order to explain the experimentally observed splitting ($\Delta$) due to spin orbit interaction. The relation shown in Eq.(\ref{eqn:so1_single_grp}) relates the reduced tensor elements of matrix representation of $H_{so1}^s$ to reduced tensor elements of momentum operators. Thus, a finite intra-band term would imply that the symmetry permitted inter-band term is also finite. It is then important to include such terms in any model. However, the inter-band terms are generally ignored in single group formulations. Furthermore, its effect could not be included in the treatment of spin orbit interaction under limited set of zone centre states, as shown in the following paragraph. 

Inter-band $\bm{k}$ independent spin orbit interaction terms cause mixing between states, particularly those with $\Gamma_8^-$ symmetry but derived from $\Gamma_3^-$ and $\Gamma_4^-$ single group states. Unless these terms are included explicitly in the single group formulation of multiband models, the implicit effect of mixing leads to departure of zone centre states from their single group heritage of one specific single group IR. In other words, it calls for the double group selection rules and formulation. If states with $\Gamma_8^-$ symmetry are not included explicitly in the near set, then such an effect may manifest itself in the absence of equality between Luttinger invariants in different blocks according to double group classification but otherwise associated via its single group heritage. In anycase, the universal constants at the front of the Eq.(\ref{eqn:so1_single_grp}) still give a scaling factor of $6.4\times 10^{-8}eV^2\mbox{\AA}^{-2}$. Thus, contributions from remote states beyond the normal anti-bonding zone centre states are required for the summation in Eq.(\ref{eqn:so1_single_grp}) to yield the level of spin splitting observed in experiments. Such remote states would have energies which invalidate the assumptions of the Foldy-Wouthuysen transform. This raises a question on the validity of the general assumption that the reduced tensor element in Eq.(\ref{eqn:so1_single_grp}) is finite for the single group product bases. Alternatively, one may question the validity of single group product bases as representations of the zone centre states.  A mixed bases, wether caused by inter-band $\bm{k}$ independent spin orbit interactions or many electron effects in the construction of zone centre states, would support a finite spin splitting and $\bm{k}$ dependent relativistic corrections. Such mixed bases would then requires the double group selection rules and double group formulation.

The Kane model\cite{Kane:1957ig,Kane:1966wa} and extended Kane model\cite{Rossler:1984fd,Pfeffer:1990cq} were developed to account for finite spin orbit splitting, and to incorporate three or more zone centre energy levels in the multiband model. It assumes the zone centre states in the near set have energies of those obtained experimentally, and are therefore classified according to double group IRs. The model retains single group selection rules and reduced tensor elements of the single group first order interactions. Implicitly, no mixing is permitted upon the formation of zone centre states classified according to double group IRs. The effect of the remote set of states is included by using experimentally measured Luttinger parameters, which are then modified by the removal of contribution from  states newly included in the near set according to their double group classified energetic positions. The relations between Luttinger invariants shown in Eq.(\ref{eqn:single_grp_relations}) are assumed to be true, but the relation shown in Eq.(\ref{eqn:single_grp_qrelations}) is not. While it is clear from the procedure of modifying Luttinger parameters for the 8- or 14-band models that the inter-band parameters would be different from the valence band due to different energies, this distinction is not made under either the Kane or extended Kane models. On the other hand, the distinction was made by Weiler et al.\cite{Weiler:1978da} under the same single group selection rules, but with double group classification of remote as well as near set of states. The impact of this distinction is explored in the following section by extending the Weiler model to include additional bands, and comparing the results with those of experiment.

\section{Comparison with Weiler model and 30 band model}
\label{sec:extended_kane}
\begin{figure*}
\begin{center}
\includegraphics[width=12cm]{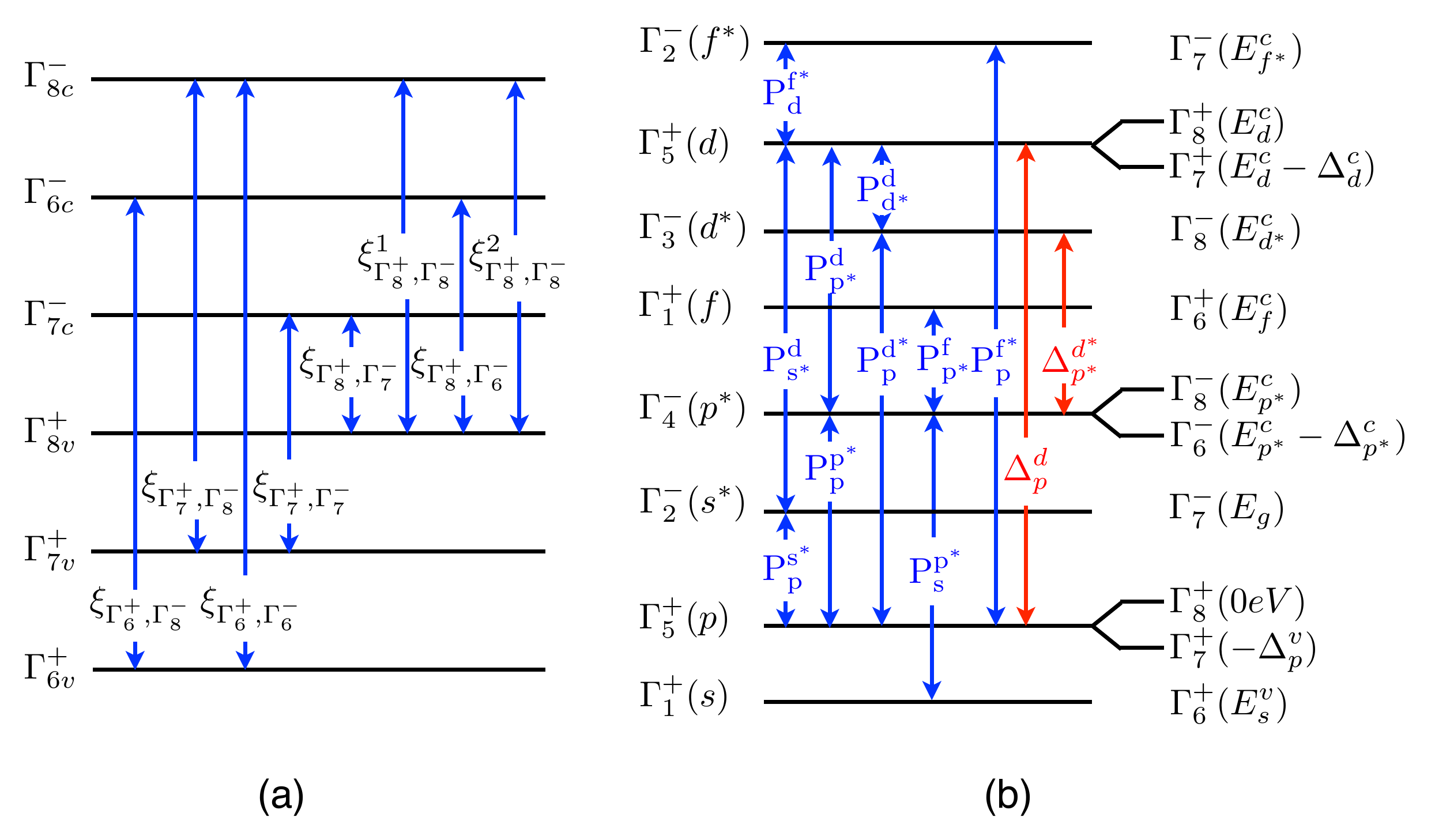}
\caption{Schematics of ordering of energies of zone centre states, allowed first order interactions and their interaction parameters under (a) double group formulation and (b) 30 band single group formulation for Ge.}
\label{fig:schematics}
\end{center}
\end{figure*}
In models with a limited number of bands, the magnitude of relativistic corrections to Luttinger invariants from the DKK model were found to be negligible. As a result, the relations between Luttinger invariants and second order interaction parameters differ between the single group formulation with relativistic corrections, and double group formulation. Specifically the lack of a $q$ invariant was problematic in the description of valence band structure. The Weiler model\cite{Weiler:1978da} was introduced to account for some of these experimental facts. The Weiler model assumes that the zone centre states have energies of those obtained experimentally, and are classified according to double group IRs. It however retains single group selection rules, and the reduced tensor elements of first order interaction in related blocks derived from that of the corresponding single group first order interaction. This treatment of the reduced tensor elements is identical to the approach employed in the extended Kane model for states in the near set. Thus, the first order interaction matrices are the same as those obtained by transforming the first order interaction under the single group into those under the adapted double group, as described in appendix \ref{app:xis}. In this respect, the Weiler model is still based on single group formulation. The model is ideally suited to explain the finite $q$ invariant in the valence band block through exactly the same mechanism proposed by Hensel and Suzuki\cite{Hensel:1969ko}. However, the links between the double group classified zone centre states, and single group selection rules or reduced tensor elements of first order interactions, are ambiguous. While it recognises that Luttinger invariants in different blocks of Hamiltonian are different (as opposed to Eq.(\ref{eqn:single_grp_relations})), no attempt was made to define independent second order interaction parameters and relate them to the relevant Luttinger invariants. In the Weiler model, remote states derived from single group states with $\Gamma_5^-$ symmetry were excluded. An implicit assumption of the model is the absence of inter-band $\bm{k}$ independent spin orbit interaction terms, and its effect of mixing of zone centre states. 

In this section, the Weiler model is extended (referred to as extended Weiler model in the rest of the manuscript) to include states with $\Gamma_8^-(\Gamma_3^-)$ and $\Gamma_8^-\oplus\Gamma_7^-(\Gamma_5^-)$ symmetry in the remote set of states. Relevant independent second order interaction parameters are defined taking into account the double group classification of zone centre states in terms of their energies, but retaining the single group first order interaction parameters which adhere to single group selection rules. Relations are then derived between Luttinger invariants and second order interaction parameters so that material parameters can be extracted under the assumptions of Weiler model, and their values can be compared with those of the experimentally determined parameters. Issues relating to the assumptions made in the Weiler model and other single group formulations, are addressed in the next section.

It is the case that many of the effects of the double group formulation pertaining to the energetic positions of the zone centre states, are already incorporated into the Weiler model through the use of empirical fitting parameters. While the energy of zone centre states with $\Gamma_7^-$ symmetry are unaffected by spin orbit interaction in the single group formulation, the use of empirical parameters allows any such effect to be incorporated in a practical implementation of the Weiler model. As shown in previous sections, the angular dependent form of the Hamiltonian for models up to, and including 8-bands are equivalent to those of the double group formulation. With parametric fitting, there then appears to be little room for differences to arise between the results of the double group formulation and the extended Weiler model. 

However, the main distinction between the two formulations is the allowance (or exclusion) of mixing between zone centre states derived from different single group symmetries. The effects of mixing in the single and double group formulations are exemplified in Fig.\ref{fig:schematics}, where the energetic positions of zone centre states and allowed first order interactions are shown schematically. Under the single group formulation, there is a one-to-one correspondence between the single group zone centre bases and the adapted double group IR (or pairs of IRs where spin orbit splitting exist). Allowed interactions are those permitted by single group selection rules. There are two inter-band $\bm{k}$ independent spin orbit interaction terms identified as $\Delta_{p^*}^{d^*}$ and $\Delta_p^d$. The term $\Delta_{p^*}^{d^*}$ in particular allows for mixing discussed at end of section \ref{sec:incomplete_bases}, and their incorporation would ensure the equivalence between single group and double group formulations based on one electron theory. However, this term is either set to zero\cite{Richard:2004wo} or not included at all\cite{Rideau:2006gu} in the 30-band model, and is absent in any other implementation of single group formulation. Consequently, mixing caused by inter-band $\bm{k}$ independent spin orbit interaction is ignored. If this term is included, it would introduce approximately 4\% of $\Gamma_8^-(\Gamma_3^-)$ state into the $\Gamma_8^-(\Gamma_4^-)$ state, assuming a energy separation $E_{d^\ast}^c-E_{p^\ast}^c=7.27eV$\cite{Richard:2004wo} and $\Delta_{p^*}^{d^*}=0.29eV$ in Ge. The effect of the inter-band $\bm{k}$ independent spin orbit interaction amounts to lifting of restrictions in the adapted double group bases and necessitates the double group selection rules. 

In the double group formulation, the zone centre states are solutions of unperturbed Hamiltonian $H_0$ in Eq.(\ref{eqn:H0}), which already contains the $\bm{k}$ independent spin orbit interaction term. Therefore, the zone centre states are already mixed in terms of adapted double group bases. Consequently, first order interactions are those permitted under the double group selection rules. Specifically, the interaction between states with $\Gamma_8^\pm$ IRs is characterised by two interaction parameters and two linearly independent matrices as described in Eq.(\ref{eqn:double_first_order_G8pm}). To adhere to double group selection rules, the two linearly independent matrices which form the first order interaction must be combined when the L\"owdin interaction is constructed using Eq.(\ref{eqn:Lowdin_A}). This introduces the three linearly independent second order interaction parameters, $\zeta_{\Gamma_8^+,\Gamma_8^+}^{\Gamma_8^-,1}, \zeta_{\Gamma_8^+,\Gamma_8^+}^{\Gamma_8^-,2}$ and $\zeta_{\Gamma_8^+,\Gamma_8^+}^{\Gamma_8^-,3}$ for L\"owdin interactions between states with $\Gamma_8^\pm$ symmetry. In contrast, any single group formulation only permits zone centre states to be associated with {\em one} single group IR. In this case, the first order interaction between states of $\Gamma_8^\pm$ IR can always be described by one interaction matrix if states derived from $\Gamma_5^-, \Gamma_3^+, \Gamma_4^+$ symmetry are excluded. The contributions to the L\"owdin term can be added at 2nd order level since each occurrence of these interactions is always associated with a different set of remote states. 

The angular dependent parts of the first order interaction associated with $\xi_{\Gamma_8^+,\Gamma_8^-}^1$ and $\xi_{\Gamma_8^+,\Gamma_8^-}^2$ double group parameters, are the same as the angular dependent parts of the interaction associated with $P_p^{p^*}$ and $P_p^{d^*}$ single group parameters, respectively. The distinction between single, and double group formulations is evident when constructing a L\"owdin term. The interaction described by $\xi_{\Gamma_8^+,\Gamma_8^-}^1$ and $\xi_{\Gamma_8^+,\Gamma_8^-}^2$ for a given pair of $\Gamma_8^\pm$ IR must be added at the first order level, whereas the corresponding terms $P_p^{p^*}$ and $P_p^{d^*}$ must be added at the second order level. Thus, there is no equivalent of the $\zeta_{\Gamma_8^+,\Gamma_8^+}^{\Gamma_8^-,3}$ parameter under the single group formulation, unless states with $\Gamma_8^-\oplus\Gamma_7^-(\Gamma_5^-)$ symmetry are included as remote states. The main difference between EWZ model and the extended Weiler model with 14-, or more bands is explicit in the number of linearly independent matrices required to describe first order interactions between states with $\Gamma_8^\pm$ symmetry\footnote{Compared with single group formulation\cite{Mayer:1991ef}, the inter-band $k$ independent spin orbit interaction such as $\Delta_-$ is not present in the double group formulation since these are included in the unperturbed Hamiltonian.}. The differences for models with fewer bands are implicit in the relations between second order interaction parameters and Luttinger invariants. The main impacts are then to the relations between the Luttinger invariants in various inter-, and intra-band blocks in the Hamiltonian, the fitting of experimentally derived Luttinger parameters in these different blocks, and the dispersion relations away from the zone centre. 

Under the single group selection rules, the first order interactions between states are given by Eq.(\ref{eqn:single_first_order}). The corresponding first order interaction between adapted double group bases may then be obtained using the procedure described in appendix \ref{app:xis}. While it is not possible to determine the sign of the first order interaction parameters ($\xi$) from symmetry argument, the sign of $\xi$s between states with double IRs derived from the same parent single group IR, are related. The following relations can then be extracted,
\begin{widetext}
\begin{subequations}
\label{eqn:xi_relations}
\begin{eqnarray}
P_p^{s^*}=\xi_{\Gamma_5^+(p),\Gamma_2^-(s^\ast)}&=&-\sqrt{6}\xi_{\Gamma_8^+(\Gamma_5^+),\Gamma_7^-(\Gamma_2^-)}=\sqrt{3}\xi_{\Gamma_7^+(\Gamma_5^+),\Gamma_7^-(\Gamma_2^-)}\\
P_p^{p^*}=\xi_{\Gamma_5^+(p),\Gamma_4^-(p^\ast)}&=&-\sqrt{3}\xi_{\Gamma_8^+(\Gamma_5^+),\Gamma_8^-(\Gamma_4^-)}=-\sqrt{3}\xi_{\Gamma_8^-(\Gamma_4^-),\Gamma_8^+(\Gamma_5^+)}\notag \\
&&=\sqrt{6}\xi_{\Gamma_7^+(\Gamma_5^+),\Gamma_8^-(\Gamma_4^-)}=\sqrt{6}\xi_{\Gamma_8^-(\Gamma_4^-),\Gamma_7^+(\Gamma_5^+)}\notag\\
&&=\sqrt{6}\xi_{\Gamma_8^+(\Gamma_5^+),\Gamma_6^-(\Gamma_4^-)}=\sqrt{6}\xi_{\Gamma_6^-(\Gamma_4^-),\Gamma_8^+(\Gamma_5^+)}\label{eqn:ppinteract}\\
P_p^{f^*}=\xi_{\Gamma_5^+(p),\Gamma_2^-(f^\ast)}&=&-\sqrt{6}\xi_{\Gamma_8^+(\Gamma_5^+),\Gamma_7^-(\Gamma_2^-)}=\sqrt{3}\xi_{\Gamma_7^+(\Gamma_5^+),\Gamma_7^-(\Gamma_2^-)} \\
P_p^{d^*}=\xi_{\Gamma_5^+(p),\Gamma_3^-(d^\ast)}&=&\sqrt{6}\xi_{\Gamma_8^+(\Gamma_5^+),\Gamma_8^-(\Gamma_3^-)}=\sqrt{3}\xi_{\Gamma_7^+(\Gamma_5^+),\Gamma_8^-(\Gamma_3^-)} 
\end{eqnarray}
\end{subequations}
The interaction between states with $\Gamma_5^\pm$ symmetry with respect to the adapted double group bases may be written in block form as,
\begin{eqnarray}
U_{\Gamma_5^+}^\dag (\mathbb{1}_{2\times 2}\otimes K_{\Gamma_5^+,\Gamma_5^-})U_{\Gamma_5^-}&=&\begin{pmatrix}(\mathcal{K}_{\Gamma_8^+,\Gamma_8^-}^1+\mathcal{K}_{\Gamma_8^+,\Gamma_8^-}^2) & \mathcal{K}_{\Gamma_8^+,\Gamma_7^-} \\
\mathcal{K}_{\Gamma_7`^+,\Gamma_8^-} & \mathcal{K}_{\Gamma_7^+,\Gamma_7^-} \end{pmatrix} \label{eqn:g5pm}
\end{eqnarray}
with relations between first order parameters given as,
\begin{subequations}
\begin{eqnarray}
P_p^{\prime f^*}=\xi_{\Gamma_5^+(p),\Gamma_5^-(f^\ast)}&=&-\xi_{\Gamma_8^+(\Gamma_5^+),\Gamma_8^-(\Gamma_4^-)}^1=3\xi_{\Gamma_8^+(\Gamma_5^+),\Gamma_8^-(\Gamma_3^-)}^2\\
&=&-3\sqrt{2}\xi_{\Gamma_8^+(\Gamma_5^+),\Gamma_7^-(\Gamma_5^-)}=-3\sqrt{2}\xi_{\Gamma_7^+(\Gamma_5^+),\Gamma_8^-(\Gamma_5^-)}=\frac{3}{2}\xi_{\Gamma_7^+(\Gamma_5^+),\Gamma_7^-(\Gamma_5^-)}
\end{eqnarray}
\end{subequations}
\end{widetext}
Since the first order interaction parameters $\xi$ are all real\cite{Elder2}, these relations permit some of the sign differences between related parameters to be fixed. These relations shall prove useful in determining the sign of second order interaction parameters in the inter-band block shown in Eq.(\ref{eqn:extended_kane_parameters}), within the extended Weiler model. 

Having identified all the possible first order interactions under single group selection rules, second order interaction parameters due to L\"owdin interaction can be defined in the same way $\zeta$'s are under double group formulation of EWZ, or $F, G, H_1, H_2$ are under the single group formulation of DKK.
\begin{widetext}
\begin{subequations}
\label{eqn:extended_kane_parameters}
\begin{eqnarray}
Z_{\Gamma_8^+,\Gamma_8^+}^{\Gamma_6^-}&=&\frac{1}{3m_0}\sum_{\alpha^v=\Gamma_6^-(\Gamma_4^-)}\frac{\xi_{\Gamma_5^+,\Gamma_4^-}^2}{E_{\alpha^v}-E_{\Gamma_8^+}} \\
Z_{\Gamma_8^+,\Gamma_8^+}^{\Gamma_7^-}&=&\frac{1}{3m_0}\sum_{\alpha^v=\Gamma_7^-(\Gamma_2^-)}\frac{\xi_{\Gamma_5^+,\Gamma_2^-}^2}{E_{\alpha^v}-E_{\Gamma_8^+}}\left\{+\frac{1}{9m_0}\sum_{\alpha^v=\Gamma_7^-(\Gamma_5^-)}\frac{\xi_{\Gamma_5^+,\Gamma_5^-}^2}{E_{\alpha^v}-E_{\Gamma_8^+}}\right\} \\
Z_{\Gamma_8^+,\Gamma_8^+}^{\Gamma_8^-;1}&=&\frac{2}{3m_0}\sum_{\alpha^v=\Gamma_8^-(\Gamma_4^-)}\frac{\xi_{\Gamma_5^+,\Gamma_4^-}^2}{E_{\alpha^v}-E_{\Gamma_8^+}}\left\{+\frac{2}{m_0}\sum_{\alpha^v=\Gamma_8^-(\Gamma_5^-)}\frac{\xi_{\Gamma_5^+,\Gamma_5^-}^2}{E_{\alpha^v}-E_{\Gamma_8^+}}\right\}\\
Z_{\Gamma_8^+,\Gamma_8^+}^{\Gamma_8^-;2}&=&\frac{1}{3m_0}\sum_{\alpha^v=\Gamma_8^-(\Gamma_3^-)}\frac{\xi_{\Gamma_5^+,\Gamma_3^-}^2}{E_{\alpha^v}-E_{\Gamma_8^+}}\left\{+\frac{2}{9m_0}\sum_{\alpha^v=\Gamma_8^-(\Gamma_5^-)}\frac{\xi_{\Gamma_5^+,\Gamma_5^-}^2}{E_{\alpha^v}-E_{\Gamma_8^+}}\right\} \\
Z_{\Gamma_8^+,\Gamma_8^+}^{\Gamma_8^-;3}&=&\left\{-\frac{2}{3m_0}\sum_{\alpha^v=\Gamma_8^-(\Gamma_5^-)}\frac{\xi_{\Gamma_5^+,\Gamma_5^-}^2}{E_{\alpha^v}-E_{\Gamma_8^+}}\right\}\\
Z_{\Gamma_8^+,\Gamma_7^+}^{\Gamma_7^-}&=&-\frac{\sqrt{2}}{3m_0}\sum_{\alpha^v=\Gamma_7^-(\Gamma_2^-)}\frac{\xi_{\Gamma_5^+,\Gamma_2^-}^2(2E_{\alpha^v}-E_{\Gamma_8^+}-E_{\Gamma_7^+})}{2(E_{\alpha^v}-E_{\Gamma_8^+})(E_{\alpha^v}-E_{\Gamma_7^+})} \notag \\
&&\left\{-\frac{2\sqrt{2}}{9m_0}\sum_{\alpha^v=\Gamma_7^-(\Gamma_5^-)}\frac{\xi_{\Gamma_5^+,\Gamma_5^-}^2(2E_{\alpha^v}-E_{\Gamma_8^+}-E_{\Gamma_7^+})}{2(E_{\alpha^v}-E_{\Gamma_8^+})(E_{\alpha^v}-E_{\Gamma_7^+})}\right\} \\
Z_{\Gamma_8^+,\Gamma_7^+}^{\Gamma_8^-;1}&=&-\frac{\sqrt{2}}{3m_0}\sum_{\alpha^v=\Gamma_8^-(\Gamma_4^-)}\frac{\xi_{\Gamma_5^+,\Gamma_4^-}^2(2E_{\alpha^v}-E_{\Gamma_8^+}-E_{\Gamma_7^+})}{2(E_{\alpha^v}-E_{\Gamma_8^+})(E_{\alpha^v}-E_{\Gamma_7^+})}\notag\\
&&\left\{+\frac{\sqrt{2}}{3m_0}\sum_{\alpha^v=\Gamma_8^-(\Gamma_5^-)}\frac{\xi_{\Gamma_5^+,\Gamma_5^-}^2(2E_{\alpha^v}-E_{\Gamma_8^+}-E_{\Gamma_7^+})}{2(E_{\alpha^v}-E_{\Gamma_8^+})(E_{\alpha^v}-E_{\Gamma_7^+})}\right\} \\
Z_{\Gamma_8^+,\Gamma_7^+}^{\Gamma_8^-;2}&=&\frac{\sqrt{2}}{3m_0}\sum_{\alpha^v=\Gamma_8^-(\Gamma_3^-)}\frac{\xi_{\Gamma_5^+,\Gamma_3^-}^2(2E_{\alpha^v}-E_{\Gamma_8^+}-E_{\Gamma_7^+})}{2(E_{\alpha^v}-E_{\Gamma_8^+})(E_{\alpha^v}-E_{\Gamma_7^+})}\notag \\
&&\left\{-\frac{\sqrt{2}}{9m_0}\sum_{\alpha^v=\Gamma_8^-(\Gamma_5^-)}\frac{\xi_{\Gamma_5^+,\Gamma_5^-}^2(2E_{\alpha^v}-E_{\Gamma_8^+}-E_{\Gamma_7^+})}{2(E_{\alpha^v}-E_{\Gamma_8^+})(E_{\alpha^v}-E_{\Gamma_7^+})}\right\} \\
Z_{\Gamma_7^+,\Gamma_7^+}^{\Gamma_7^-}&=&\frac{2}{3m_0}\sum_{\alpha^v=\Gamma_7^-(\Gamma_2^-)}\frac{\xi_{\Gamma_5^+,\Gamma_2^-}^2}{E_{\alpha^v}-E_{\Gamma_7^+}}\left\{+\frac{8}{9m_0}\sum_{\alpha^v=\Gamma_7^-(\Gamma_5^-)}\frac{\xi_{\Gamma_5^+,\Gamma_5^-}^2}{E_{\alpha^v}-E_{\Gamma_7^+}}\right\}   \\
Z_{\Gamma_7^+,\Gamma_7^+}^{\Gamma_8^-}&=&\frac{1}{3m_0}\sum_{\alpha^v=\Gamma_8^-(\Gamma_4^-)}\frac{\xi_{\Gamma_5^+,\Gamma_4^-}^2}{E_{\alpha^v}-E_{\Gamma_7^+}}+\frac{2}{3m_0}\sum_{\alpha^v=\Gamma_8^-(\Gamma_3^-)}\frac{\xi_{\Gamma_5^+,\Gamma_3^-(\alpha^v)}^2}{E_{\alpha^v}-E_{\Gamma_7^+}} \notag \\
&&\left\{+\frac{1}{9m_0}\sum_{\alpha^v=\Gamma_8^-(\Gamma_5^-)}\frac{\xi_{\Gamma_5^+,\Gamma_5^-}^2}{E_{\alpha^v}-E_{\Gamma_7^+}}\right\}
\end{eqnarray}
\end{subequations}
\end{widetext}
where the single group interaction parameters are defined in Eq.(\ref{eqn:xi_relations},\ref{eqn:g5pm}). The appropriate negative sign in front of terms such as $Z_{\Gamma_8^+,\Gamma_7^+}^{\Gamma_7^-}$ and $Z_{\Gamma_8^+,\Gamma_7^+}^{\Gamma_8^-,1}$ are derived from the signs given in Eq.(\ref{eqn:xi_relations}).  

With these definitions, symmetric and anti-symmetric L\"owdin terms are constructed using Eq.(\ref{eqn:Lowdin}). The resulting  Luttinger parameters in the 6 band model may be expressed in terms of these 10 independent second order interaction parameters as follows,
\begin{subequations}
\label{eqn:extended_weiler_gamma}
\begin{widetext}
\begin{eqnarray}
\gamma_1&=&2Z_{\Gamma_8^+,\Gamma_8^+}^{\Gamma_6^-}+2Z_{\Gamma_8^+,\Gamma_8^+}^{\Gamma_7^-}+Z_{\Gamma_8^+,\Gamma_8^+}^{\Gamma_8^-,1}+8Z_{\Gamma_8^+,\Gamma_8^+}^{\Gamma_8^-,2}\left\{+4Z_{\Gamma_8^+,\Gamma_8^+}^{\Gamma_8^-,3}\right\}-1 \\
\gamma_2&=&Z_{\Gamma_8^+,\Gamma_8^+}^{\Gamma_6^-}-Z_{\Gamma_8^+,\Gamma_8^+}^{\Gamma_7^-}-4Z_{\Gamma_8^+,\Gamma_8^+}^{\Gamma_8^-,2}\left\{-2Z_{\Gamma_8^+,\Gamma_8^+}^{\Gamma_8^-,3}\right\}\\
\gamma_3&=&Z_{\Gamma_8^+,\Gamma_8^+}^{\Gamma_6^-}+Z_{\Gamma_8^+,\Gamma_8^+}^{\Gamma_7^-}-2Z_{\Gamma_8^+,\Gamma_8^+}^{\Gamma_8^-,2}\\
\kappa&=&Z_{\Gamma_8^+,\Gamma_8^+}^{\Gamma_7^-}-\frac{1}{2}Z_{\Gamma_8^+,\Gamma_8^+}^{\Gamma_8^-,1}-2Z_{\Gamma_8^+,\Gamma_8^+}^{\Gamma_8^-,2}\left\{-\frac{5}{2}Z_{\Gamma_8^+,\Gamma_8^+}^{\Gamma_8^-,3}\right\}-\frac{g_0}{6}\\
q&=&Z_{\Gamma_8^+,\Gamma_8^+}^{\Gamma_6^-}-\frac{1}{2}Z_{\Gamma_8^+,\Gamma_8^+}^{\Gamma_8^-,1}\left\{-\frac{3}{2}Z_{\Gamma_8^+,\Gamma_8^+}^{\Gamma_8^-,3}\right\} \label{eqn:single_group_q2}\\
\gamma_1^{ss}&=&Z_{\Gamma_7^+,\Gamma_7^+}^{\Gamma_7^-}+4Z_{\Gamma_7^+,\Gamma_7^+}^{\Gamma_8^-}-1\\
\kappa^{ss}&=&2Z_{\Gamma_7^+,\Gamma_7^+}^{\Gamma_7^-}-4Z_{\Gamma_7^+,\Gamma_7^+}^{\Gamma_8^-}+\frac{g_0}{3}\\
\gamma_{2}^{vs}&=&\frac{1}{\sqrt{2}}\left(Z_{\Gamma_8^+,\Gamma_7^+}^{\Gamma_7^-}-Z_{\Gamma_8^+,\Gamma_7^+}^{\Gamma_8^-,1}-4Z_{\Gamma_8^+,\Gamma_7^+}^{\Gamma_8^-,2}\right)\\
\gamma_{3}^{vs}&=& \frac{1}{\sqrt{2}}\left(Z_{\Gamma_8^+,\Gamma_7^+}^{\Gamma_7^-}+Z_{\Gamma_8^+,\Gamma_7^+}^{\Gamma_8^-,1}+2Z_{\Gamma_8^+,\Gamma_7^+}^{\Gamma_8^-,2}\right) \\
\kappa^{vs}&=&\frac{1}{\sqrt{2}}\left(Z_{\Gamma_8^+,\Gamma_7^+}^{\Gamma_7^-}-Z_{\Gamma_8^+,\Gamma_7^+}^{\Gamma_8^-,1}+2Z_{\Gamma_8^+,\Gamma_7^+}^{\Gamma_8^-,2}\right)-\frac{g_0}{3}
\end{eqnarray}
\end{widetext}
In addition, the conduction band Luttinger invariants may be written as
\begin{eqnarray}
\gamma_1^c&=&Z_{\Gamma_7^-,\Gamma_7^-}^{\Gamma_7^+}+4Z_{\Gamma_7^-,\Gamma_7^-}^{\Gamma_8^+}-1\\
g^c&=&2Z_{\Gamma_7^-,\Gamma_7^-}^{\Gamma_7^+}-4Z_{\Gamma_7^-,\Gamma_7^-}^{\Gamma_8^+}-g_0
\end{eqnarray}
\end{subequations}

In Eq.(\ref{eqn:extended_kane_parameters}), contributions from remote states with $\Gamma_5^-$ symmetry are included in curly braces. The first order interaction $\mathcal{K}_{\Gamma_8^+(\Gamma_5^+):\Gamma_8^-(\Gamma_5^-)}$ is expressed as an linear combination of  $\mathcal{K}_{\Gamma_8^+,\Gamma_8^-}^1$ and $\mathcal{K}_{\Gamma_8^+,\Gamma_8^-}^2$, as shown in Eq.(\ref{eqn:g5pm}). In this respect, we may label states of $\Gamma_8^-$ symmetry derived from $\Gamma_5^-$ as `mixed' with respect to the bases of $\Gamma_8^-$ derived from $\Gamma_4^-$ and $\Gamma_3^-$. Such contributions add the additional parameter of $Z_{\Gamma_8^+,\Gamma_8^+}^{\Gamma_8^-;3}$ to the valence band block. It is clear that these parameters have the same labelling scheme as the $\zeta$ parameters defined in EWZ, and there is a one-to-one correspondence provided that remote states with $\Gamma_5^-$ symmetry are included. Thus, the presence of $Z_{\Gamma_8^+,\Gamma_8^+}^{\Gamma_8^-,3}$ in the extended Weiler model is dependent on the inclusion of remote states with $\Gamma_5^-$ symmetry. Similar arguments can be made about single group states with $\Gamma_3^+$ and $\Gamma_4^+$ symmetry which give rise to double group bases of $\Gamma_8^+$. In the implementation of Weiler model or 30 band model in the literature, remote states with $\Gamma_5^-$ symmetry are always excluded. Hence, there are only four independent second order interaction parameters in the valence band related to the five Luttinger invariants. As one expects the invariants to be linearly independent of each other, the presence of only four parameters in Eq.(\ref{eqn:extended_weiler_gamma}) remove such linear independence. In addition, the Weiler model has difficulty in explaining the effective mass in the spin orbit band of Si and conduction band of Ge as shown below.

The relations shown in Eq.(\ref{eqn:extended_weiler_gamma}) derived from the extended Weiler model may be compared with Eq.(\ref{eqn:H_dkk},\ref{eqn:lutt_sdpf}) derived from the DKK model with relativistic corrections, and those in Eq.(\ref{eqn:invariantzeta}) derived from the double group formulation. The Luttinger invariants in each of the blocks are now different from each other. The magnitude of these difference is determined by the different energy scaling in the corresponding $Z$ parameters. The $q$ parameter is now non-zero, provided spin orbit splitting exists in the conduction band state derived from the antibonding $p^\ast$ state. It is therefore clear that extended Weiler model overcomes some of the key difficulties encountered in the DKK model with relativistic corrections.

The impacts of classifying the energetic positions of the zone centre states according to double group under the Weiler model, is best seen in the contraction of the full zone 30-band model\cite{Richard:2004wo,Rideau:2006gu} into models with fewer bands. The relations between the 30-band single group interaction parameters and Luttinger invariants, have been obtained by Richards et al.\cite{Richard:2004wo} and are given here for crystals with diamond lattice. These relations have been obtained by transforming the Hamiltonian referring to product bases\cite{Rideau:2006gu} into adapted double group bases, using the diagonalization matrices shown in appendix \ref{app:unitary_transform}. The L\"owdin terms are then constructed using the resulting interaction matrices and parameters, and Eq.(\ref{eqn:Lowdin}) giving the following relations, 
\begin{widetext}
\begin{subequations}
\label{eqn:30band}
\begin{eqnarray}
\gamma_1&=&\frac{2}{3m_0}\left[\frac{{P_p^{s^*}}^2}{E_{\Gamma_7^-}^{s^*}-E_{\Gamma_8^+}^p}+\frac{{P_p^{p^*}}^2}{E_{\Gamma_6^-}^{p^*}-E_{\Gamma_8^+}^p}+\frac{{P_p^{p^*}}^2}{E_{\Gamma_8^-}^{p^*}-E_{\Gamma_8^+}^p}+\frac{4{P_p^{d^*}}^2}{E_{\Gamma_8^-}^{d^*}-E_{\Gamma_8^+}^p}+\frac{{P_p^{f^*}}^2}{E_{\Gamma_7^-}^{f^*}-E_{\Gamma_8^+}^p}\right] -1\\
\gamma_2&=&\frac{1}{3m_0}\left[-\frac{{P_p^{s^*}}^2}{E_{\Gamma_7^-}^{s^*}-E_{\Gamma_8^+}^p}+\frac{{P_p^{p^*}}^2}{E_{\Gamma_6^-}^{p^*}-E_{\Gamma_8^+}^p}-\frac{4{P_p^{d^*}}^2}{E_{\Gamma_8^-}^{d^*}-E_{\Gamma_8^+}^p}-\frac{{P_p^{f^*}}^2}{E_{\Gamma_7^-}^{f^*}-E_{\Gamma_8^+}^p}\right] \\
\gamma_3&=&\frac{1}{3m_0}\left[\frac{{P_p^{s^*}}^2}{E_{\Gamma_7^-}^{s^*}-E_{\Gamma_8^+}^p}+\frac{{P_p^{p^*}}^2}{E_{\Gamma_6^-}^{p^*}-E_{\Gamma_8^+}^p}-\frac{2{P_p^{d^*}}^2}{E_{\Gamma_8^-}^{d^*}-E_{\Gamma_8^+}^p}+\frac{{P_p^{f^*}}^2}{E_{\Gamma_7^-}^{f^*}-E_{\Gamma_8^+}^p}\right]\\
\kappa&=&\frac{1}{3m_0}\left[\frac{{P_p^{s^*}}^2}{E_{\Gamma_7^-}^{s^*}-E_{\Gamma_8^+}^p}-\frac{{P_p^{p^*}}^2}{E_{\Gamma_8^-}^{p^*}-E_{\Gamma_8^+}^p}-\frac{2{P_p^{d^*}}^2}{E_{\Gamma_8^-}^{d^*}-E_{\Gamma_8^+}^p}+\frac{{P_p^{f^*}}^2}{E_{\Gamma_7^-}^{f^*}-E_{\Gamma_8^+}^p}\right]-\frac{g_0}{6}\\
q&=&\frac{1}{3m_0}\left[\frac{{P_p^{p^*}}^2}{E_{\Gamma_6^-}^{p^*}-E_{\Gamma_8^+}^p}-\frac{{P_p^{p^*}}^2}{E_{\Gamma_8^-}^{p^*}-E_{\Gamma_8^+}^p}\right] \label{eqn:single_group_q1}
\end{eqnarray}
\end{subequations}
\end{widetext}
where the definition of the matrix elements (reduced tensor element) are shown in Fig.\ref{fig:schematics}b, and $g_0\simeq 2$. The effect of $\Delta_{p*}^{d*}$ is not included, in accordance with the single group formulation in the literature. The original atomic orbital has been shown as superscript in the energies of the relevant zone centre states. The order of basis for valence band used in the original work of Richard\cite{Richard:2004wo}, is the same as those in Eq.(\ref{Chuang}), and hence there is a difference in sign of the $\gamma_2$ invariant. The contribution from each remote state symmetry is described by Winkler\cite{fizik:2003wv}. The number of parameters, including the zone centre state energies, are large, but they can be reduced to the set of independent $Z$ parameters defined in Eq.(\ref{eqn:extended_kane_parameters}).

Referring to the 30-band model and interactions designated in Figure \ref{fig:schematics}, the $Z$ parameters can be related to single group first order interaction parameters as follows.
\begin{subequations}
\label{eqn:zeta_single}
For the valence band under 4-band model,
\begin{eqnarray}
Z^{\Gamma_7^-}_{\Gamma_8^+,\Gamma_8^+}&=&\frac{1}{3m_0}\left[\frac{{P_p^{s^*}}^2}{E_{\Gamma_7^-}^{s^*}-E_{\Gamma_8^+}^p}+\frac{{P_p^{f^*}}^2}{E_{\Gamma_7^-}^{f^*}-E_{\Gamma_8^+}^p}\right]\simeq\sigma\\
Z^{\Gamma_6^-}_{\Gamma_8^+,\Gamma_8^+}&=&\frac{1}{3m_0}\frac{{P_p^{p^*}}^2}{E_{\Gamma_6^-}^{p^*}-E_{\Gamma_8^+}^p}\simeq\frac{\pi}{2}\\
Z^{\Gamma_8^-,1}_{\Gamma_8^+,\Gamma_8^+}&=&\frac{2}{3m_0}\frac{{P_p^{p^*}}^2}{E_{\Gamma_8^-}^{p^*}-E_{\Gamma_8^+}^p}\simeq\pi\\
Z^{\Gamma_8^-,2}_{\Gamma_8^+,\Gamma_8^+}&=&\frac{1}{3m_0}\frac{{P_p^{d^*}}^2}{E_{\Gamma_8^-}^{d^*}-E_{\Gamma_8^+}^p}\simeq\frac{\delta}{4}\\
Z^{\Gamma_8^-,3}_{\Gamma_8^+,\Gamma_8^+}&=&0
\end{eqnarray}
For the spin split off band under 2-band or 6-band model,
\begin{eqnarray}
Z^{\Gamma_7^-}_{\Gamma_7^+,\Gamma_7^+}&=&\frac{1}{3m_0}\left[\frac{2{P_p^{s^*}}^2}{E_{\Gamma_7^-}^{s^*}-E_{\Gamma_7^+}^p}+\frac{2{P_p^{f^*}}^2}{E_{\Gamma_7^-}^{f^*}-E_{\Gamma_7^+}^p}\right]\\
Z^{\Gamma_8^-}_{\Gamma_7^+,\Gamma_7^+}&=&\frac{1}{3m_0}\left[\frac{{P_p^{p^*}}^2}{E_{\Gamma_8^-}^{p^*}-E_{\Gamma_7^+}^p}+\frac{2{P_p^{d^*}}^2}{E_{\Gamma_8^-}^{d^*}-E_{\Gamma_7^+}^p}\right]
\end{eqnarray}
For the conduction band with $\Gamma_7^-$ symmetry under 2-band model,
\begin{eqnarray}
Z^{\Gamma_7^+}_{\Gamma_7^-,\Gamma_7^-}&=&\frac{1}{3m_0}\left[\frac{2{P_p^{s^*}}^2}{E_{\Gamma_7^+}^p-E_{\Gamma_7^-}^s}+\frac{2{P_{s^*}^d}^2}{E_{\Gamma_7^+}^d-E_{\Gamma_7^-}^{s^*}}\right]\\
Z^{\Gamma_8^+}_{\Gamma_7^-,\Gamma_7^-}&=&\frac{1}{3m_0}\left[\frac{{P_p^{s^*}}^2}{E_{\Gamma_8^+}^p-E_{\Gamma_7^-}^{s^*}}+\frac{{P_{s^*}^d}^2}{E_{\Gamma_8^+}^d-E_{\Gamma_7^-}^{s^*}}\right].
\end{eqnarray}
For the inter band block under 6-band model,
\begin{eqnarray}
Z^{\Gamma_7^-}_{\Gamma_8^+,\Gamma_7^+}&=&-\frac{1}{3m_0}\left[\frac{\sqrt{2}{P_p^{s^*}}^2}{2}\left(\frac{1}{E_{\Gamma_7^-}^{s^*}-E_{\Gamma_8^+}^p}+\frac{1}{E_{\Gamma_7^-}^{s^*}-E_{\Gamma_7^+}^p}\right)\right.\notag \\
&&\left.+\frac{\sqrt{2}{P_p^{f^*}}^2}{2}\left(\frac{1}{E_{\Gamma_7^-}^{f^*}-E_{\Gamma_8^+}^p}+\frac{1}{E_{\Gamma_7^-}^{f^*}-E_{\Gamma_7^+}^p}\right)\right]\\
Z^{\Gamma_8^-,1}_{\Gamma_8^+,\Gamma_7^+}&=&-\frac{1}{3m_0}\frac{\sqrt{2}{P_p^{p^*}}^2}{2}\left(\frac{1}{E_{\Gamma_8^-}^{p^*}-E_{\Gamma_8^+}^p}+\frac{1}{E_{\Gamma_8^-}^{p^*}-E_{\Gamma_7^+}^p}\right)\\
Z^{\Gamma_8^-,2}_{\Gamma_8^+,\Gamma_7^+}&=&\frac{1}{3m_0}\frac{\sqrt{2}{P_p^{d^*}}^2}{2}\left(\frac{1}{E_{\Gamma_8^-}^{d^*}-E_{\Gamma_8^+}^p}+\frac{1}{E_{\Gamma_8^-}^{d^*}-E_{\Gamma_7^+}^p}\right)
\end{eqnarray}
\end{subequations}
There are no states derived from $\Gamma_5^-$ symmetry in the 30 band model and therefore the parameter $Z^{\Gamma_8^-,3}_{\Gamma_8^+,\Gamma_8^+}$ is absent.

The implementation of the 30-band model\cite{Richard:2004wo,Rideau:2006gu} was intended to provide both an accurate description of dispersion over the whole of the Brillouin zone, and the extraction of energy gaps at the other high symmetry points.  It is however the case that the Luttinger parameters, which are derived from the first order parameters in Ref.\onlinecite{Richard:2004wo} using Eq.(\ref{eqn:30band}), deviate substantially from the experimentally measured parameters. This means that the dispersion relations calculated near the zone centre are not accurate. Utilising the Weiler model, It is possible to extract a set of single group first order interaction parameters from the experimentally determined Luttinger parameters of the valence band. This then permits the extraction of a full set of Luttinger parameters for the inter-band, and spin split-off band blocks, and places a lower limit to the Luttinger parameter in the conduction band ($\Gamma_7^-$) block. These derived parameters can be compared with  experimentally determined values, where available, to ascertain the veracity of assumptions in the Weiler model.

The single group first order interaction parameters can be extracted from experimentally determined Luttinger parameters $\gamma_1, \gamma_2$ and $\gamma_3$ with the aid of Eq.(\ref{eqn:zeta_single},\ref{eqn:extended_weiler_gamma}) and some additional assumptions. Ignoring the remote anti-bonding $f^*$ state with $\Gamma_7^-(\Gamma_2^-)$ symmetry\footnote{Neglecting this state will only lower the limit of conduction band effective mass.}, the first order interaction  parameters $P_p^{s*}(\xi_{\Gamma_5^+,\Gamma_2^-}), P_p^{p*}(\xi_{\Gamma_5^+,\Gamma_4^-})$ and $P_p^{d*}(\xi_{\Gamma_5^+,\Gamma_3^-})$ can be obtained from Eq.(\ref{eqn:zeta_single},\ref{eqn:extended_weiler_gamma}) given the zone centre state energies. Then a full set of Luttinger parameters in the inter-band block, spin split off band block can be obtained. The lower limit to Luttinger parameter in the conduction band ($\Gamma_7^-$) block can also be evaluated by ignoring the contribution of $d$ states with $\Gamma_8^+\oplus\Gamma_7^+$ symmetry in the conduction band. These parameters are listed in Table \ref{tbl:g1_frm_30band} for Si and Ge. 

\begin{table}[h]
\caption{Luttinger parameters of different blocks derived from $\gamma_1, \gamma_2, \gamma_3$ of the valence band under the Weiler model}
\begin{center}
\begin{ruledtabular}
\begin{tabular}{cccccccc}
&$\gamma_1$ & $\gamma_2$ & $\gamma_3$ & $\gamma_2^{vs}$ & $\gamma_3^{vs}$ & $\gamma_1^{ss}$ & $\gamma_1^c(\Gamma_7^-)$\\ \hline
Si &5.26 & -0.38 & 1.56 & -0.39 & -1.55 & 5.18 & $>-5.76$ \\
Ge & 13.38 & -4.24 & 5.69 & -3.73 & -4.98 & 10.47 & $>-21.91$ 
\end{tabular}
\end{ruledtabular}
\end{center}
\label{tbl:g1_frm_30band}
\end{table}%

Under the 2 band model, the effective mass of the spin orbit band is smaller compared with experimentally measured value in Si, and the lower limit of conduction band effective mass is larger compared with experimentally measured value in Ge.  In addition, the dispersion along the $\left<110\right>$ direction in Si, under the 6-band model is un-physical (see Figure \ref{fig:si_dispersion}). These discrepancies are generally attributed to remote states that have not been included in the model. For example, the introduction of the Hermann-Weisbuch\cite{Hermann:1977fc} parameters accounts for a similar over estimation of the conduction band effective mass in compound semiconductors. It is however clear, that all of the valence band states are well identified by experimental methods, and any excluded remote states reside above the lowest conduction band. Such states would include  those in the conduction band with $\Gamma_7^-(\Gamma_2^-)$ symmetry from $f$ atomic orbital, and states with $\Gamma_8^+\oplus\Gamma_7^+$ symmetry from $d$ atomic orbital. The contribution from such un-known, or excluded states, to the L\"owdin interaction would only increase the effective mass of the lowest conduction band, and reduce the effective mass of the spin split off band. These discrepancies are thus intrinsic to the Weiler model.

Under the extended Weiler model, it is possible to provide a set of second order interaction parameters to predict the appropriate effective mass of conduction band in Ge and spin-split-off band in Si. The extended Weiler model includes remote states with $\Gamma_8^-\oplus\Gamma_7^-(\Gamma_5^-)$ symmetry, and the second order interaction parameters include $Z_{\Gamma_8^+,\Gamma_8^+}^{\Gamma_8^-,3}$. However, the magnitude of $Z_{\Gamma_8^+,\Gamma_8^+}^{\Gamma_8^-,3}$ would exceed that of $Z_{\Gamma_8^+,\Gamma_8^+}^{\Gamma_8^-,2}$ (see section \ref{sec:double_group_parameters}), which would not be consistent with known energies and symmetry of the zone centre states.

\section{L\"owdin interaction in multiband models}
\label{sec:lowdin_bases}
In both the double group formulation of EWZ and single group formulation in the general literature, the L\"owdin interaction terms can be constructed from first order interaction matrices and have exactly the same form. This is the case even though the basis functions for the perturbative expansion differ. This section examines the way in which the L\"owdin interaction is derived, and shows that it is only permissible to use the double group formulation to derive the solution of one electron Schr\"odinger equation shown in Eq.(\ref{eqn:H0}) when a limited set of bases is stipulated.

It is well known that the $\bm{k}$ independent spin orbit interaction term, $H_{so1}^s$, has non-zero matrix elements as an intra-band term. Within the single group formulation, such matrix elements are also permitted as inter-band terms between states with $\Gamma_3^\pm$, $\Gamma_4^\pm$, or $\Gamma_5^\pm$ symmetry with the same spatial parity. With respect to the single group product basis, the matrix representation of this interaction was obtained in Eq.(\ref{eqn:so1_single_grp}) and shown {\em not} to be relativistically small. The Hamiltonian in Eq.(\ref{eqn:H0}) can be divided into three terms,
\[
H(\bm{k})=H_0^s+H_{so1}^s(\bm{\sigma})+H_1(\bm{k})
\]
with respect to a complete set of bases. To obtain the Hamiltonian with limited multiband model (e.g. 8-band), the complete set of zone centre states are divided into near and remote sets, which have relatively weak interactions between them. The interaction between states in the near set are included explicitly, and the effects of remote set on the near set are included as perturbation to the desired order. The L\"owdin method calls for a unitary transformation which leaves the inter-band block between the near and remote sets to be zero within the desired order in the relevant perturbation\cite{LowdinPO:1951io,Wagner:1986wm,fizik:2003wv}. 

Here, we wish to include all the terms quadratic in $\bm{k}$ components. To achieve this, the resulting bases must diagonalise $H_0=H_0^s+H_{so1}^s(\bm{\sigma})$, rather than just $H_0^s$. In absence of the inter-band $\bm{k}$ independent spin orbit interaction terms, there is no mixing between states derived from different single group symmetry. The unitary transformation described in appendix \ref{app:unitary_transform}, which is frequently used in the literature, would then diagonalise $H_0^s$ and $H_{so1}^s(\bm{\sigma})$ with respect to the adapted double group bases. This is clearly not true if inter-band $\bm{k}$ independent spin orbit interaction is present. Such interactions can not be made diagonal in the Hamiltonian by the unitary transformation within each IR. Diagonalization of  $H_0$ leads to mixing between single group zone centre states which necessitate the double group selection rules and formulation. In this case, the bases for the L\"owdin term diagonalise $H_0=H_0^s+H_{so1}^s(\bm{\sigma})$ and differ from the adapted double group bases. They are the bases relevant to the double group formulation. Hence, the L\"owdin interaction obtained from perturbation theory in EWZ and {\em any} method of invariant\footnote{The method of Luttinger\cite{LuttingerJM:1956hi} and Bir and Pikus\cite{BirPikus} focuses on obtaining linearly independent generators with the relevant transformation properties under point group from one set of bases. Where more than one linearly independent generators are present for a given IR, different generating operators are used. For example, $\hat{\bm{J}}$ and $\hat{\bm{J}}^3$. The method of EWZ focuses on generating operators with the correct transformation properties. Where more than one linearly independent generators are present for a given IR, different independent bases are used with the same generating operator. For example $\bm{J}$ and $\bm{\mathcal{J}}$ used in Eq.(\ref{eqn:j_mu_double}) are all obtained using the angular momentum operator $\hat{\bm{J}}$, but evaluated with respect to different bases.} obtained using any form of double group bases\footnote{While adapted double group bases, restricted to one single group IR, are incomplete and it is not possible to enumerate required linearly independent matrices from the same generating operator when multiplicity in decomposition of product representation is greater than one, it does form the representation of the double group. Using the method of Luttinger and Bir and Pikus, the incompleteness of such bases is overcome by using different generating operator.}  actually refers to the bases of double group formulation in the presence of inter-band $\bm{k}$ independent spin orbit interaction.

The relativistic corrections to the DKK Hamiltonian, discussed in the section \ref{sec:relativistic_dkk}, refers to the adapted double group bases. The distinction between the relativistically corrected single group formulation and the double group formulation lies in the zone centre bases of both the near set and remote set of states. The effect of mixing due to $\bm{k}$ independent inter-band spin orbit interaction can not be accounted using the perturbation method employed in section \ref{sec:relativistic_dkk} unless such interaction is between states of the near set. Under the 30-band model, the relevant inter-band $\bm{k}$ independent spin orbit interaction can be identified as the $\Delta_{p^*}^{d^*}$ term (which correspond to $\Delta_{3C}$ and is set to zero in Ref.\onlinecite{Richard:2004wo}). Terms such as $\Delta_{p}^{d}$, also cause inter-band mixing but do not change the symmetry of the bases. Hence, it has no direct effect on the form of first order interaction matrices. The limitation of the relativistic corrected DKK model, based on adapted double group bases, is therefore attributable to the failure in dealing with the effect of inter-band $\bm{k}$ independent spin orbit interactions. 

In the extended Weiler model, the ambiguity between the double group classification of energetic position of zone centre states and the use of single group selection rule and single group derived interactions, means that the zone centre states are not well specified. In essence, it lacks a clearly defined un-perturbed Hamiltonian\footnote{It actually corresponds to the unperturbed Hamiltonian $H_0$, shown in Eq.(\ref{eqn:H0}), with the inter-band $\bm{k}$ independent spin orbit interaction selectively removed.}. The use of experimentally observed zone centre states under a double group classification means the model closely resembles the double group formulation, as evident in the identical form of Eq.(\ref{eqn:extended_weiler_gamma}) and Eq.(\ref{eqn:invariantzeta_ccss}). However, the `mixing' of states with the $\Gamma_8^\pm$ symmetry is incorporated in the extended Weiler model only when remote states with $\Gamma_5^-$ symmetry are included. If these remote states with $\Gamma_5^-$ symmetry are considered, parametric fitting of experimentally measured Luttinger parameters should return the same results as the double group formulation. Nevertheless, the $Z$ parameters may not make physical sense from perturbation theory perspective\footnote{This refers to the large magnitude of $Z_{\Gamma_8^+,\Gamma_8^+}^{\Gamma_8^-,3}$ in comparison to $Z_{\Gamma_8^+,\Gamma_8^+}^{\Gamma_8^-,2}$}. However, remote states with $\Gamma_5^-$ symmetry are normally excluded under the Weiler model or 30 band model. In this case, there are only four independent second order interaction parameters for the valence band ($Z_{\Gamma_8^+,\Gamma_8^+}^{\Gamma_6^-},Z_{\Gamma_8^+,\Gamma_8^+}^{\Gamma_7^-},Z_{\Gamma_8^+,\Gamma_8^+}^{\Gamma_8^-,1}$ and $Z_{\Gamma_8^+,\Gamma_8^+}^{\Gamma_8^-,2}$) but five Luttinger invariants.  The mismatch between the four independent second order interaction parameters and the Luttinger invariants, leads to linear dependence among the latter in contradiction to symmetry argument. 

The existence of a finite inter-band $\bm{k}$ independent spin orbit interaction, means that they must be treated under the unperturbed system if quasi-degenerate perturbation theory is to be used to incorporate effects of remote states in a limited multiband model. Hence the double group formulation, in which the mixing is automatically permitted under double group selection rules and material parameters, should be used. The additional mixing mechanism arising from formation of hybridised orbitals with consideration of spin means that the use of double group bases is essential in order to obtain the correct description of electronic dispersion under the $\bm{k\cdot p}$ method. The Hamiltonian obtained from method of invariant does not specify the bases used for $\bm{k}$ expansion other than its transformation properties in terms of IR of the symmetry group and the order of the basis function. However, the $\bm{k}$ dependence is consistent with that of the first order $\bm{k\cdot\pi}$ perturbation and L\"owdin interactions under the quasi-degenerate perturbation theory.  As a result, the basis must be those of the double group formulation instead of the adapted double group bases. 

It is clear from the discussion that the decoupling of states in near and remote set in a limited bases set model should be performed in such a way that all $\bm{k}$ independent perturbation is diagonalised by the unitary transformation. Apart from spin orbit interaction, strain is another perturbation that is $\bm{k}$ independent. In principle, the method of dealing with strain perturbation should be different from those prescribed in the literature\cite{BirPikus}. Upon the application of homogeneous stress, the symmetry of the crystal is lowered. The appropriate selection rules to treat the $\bm{k\cdot\pi}$ perturbation should be those of the prevailing sub-group $H$ derived from the point group $G$ of the unstrained crystal. The IR of the point group $G$ then generally becomes reducible under the compatibility relation between $H$ and $G$. Issues may arise from `mixing' caused by the strain perturbation, as this modifies the first order interaction parameter or Luttinger invariants derived from the unstrained crystal. One can see some analogy between mixing due to $\bm{k}$ independent inter-band spin orbit interaction with respect to the the adapted double group bases and mixing due to strain perturbation. The treatment of stress as described by Bir and Pikus\cite{BirPikus} may be justifiable if the terms in the Hamiltonian due to strain and causing mixing is smaller compared to other having similar mixing effects. However, the mixing would not be negligible in the infinite stress limit employed by Hensel and Suzuki\cite{Hensel:1974da,Suzuki:1974gd} when extracting the Luttinger parameters from experimental data. The assumption holds for stress applied the the $\left<001\right>$ direction, since all the terms first order in strain are diagonal, but fails in the $\left<111\right>$ and $\left<110\right>$ directions. This may cast some doubt on the particular method used.
 
\section{Material parameters for Si and Ge in double group formulation}
\label{sec:double_group_parameters}
\begin{figure*}
\begin{centering}
\includegraphics[width=6.0cm]{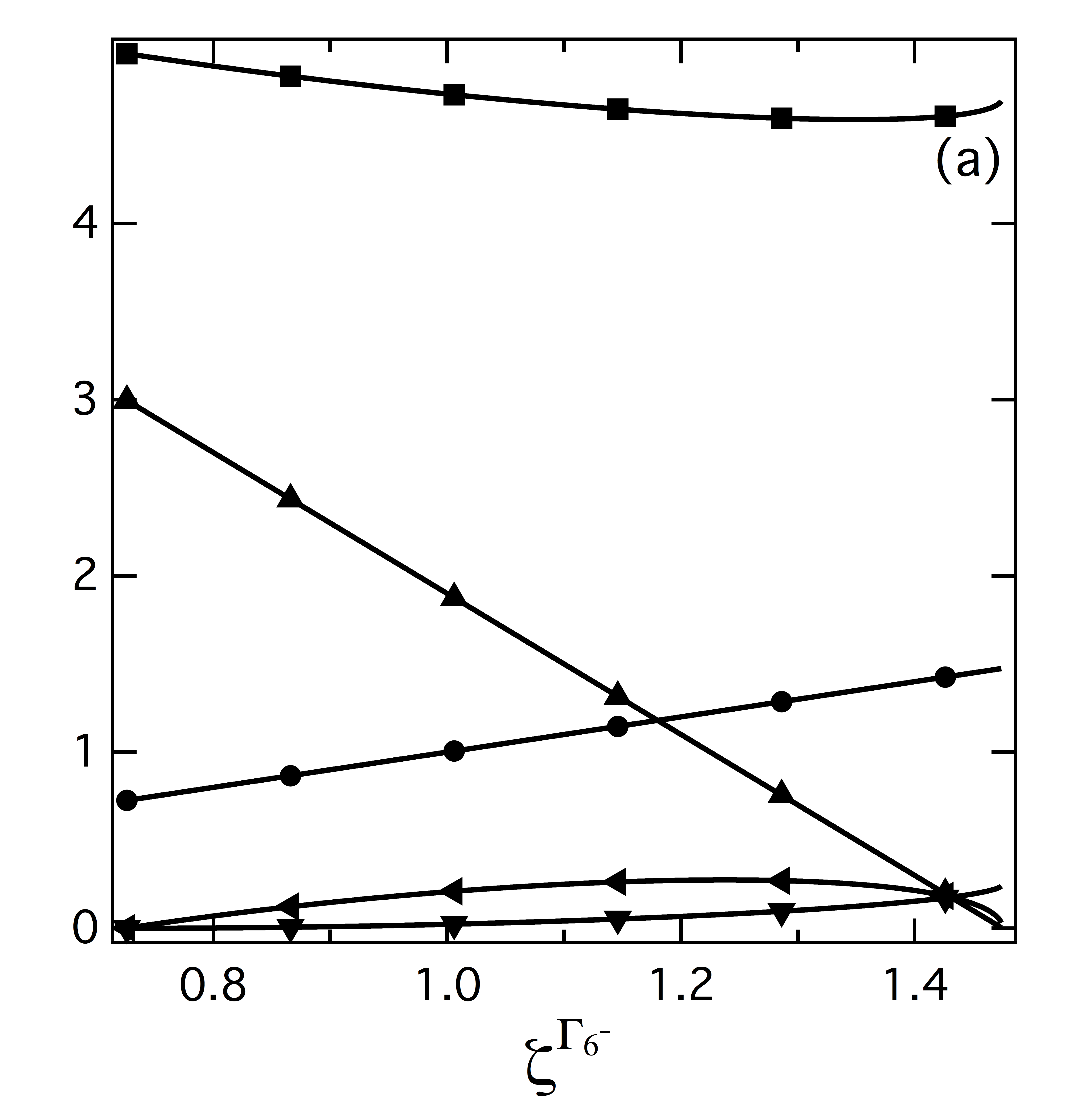}
\includegraphics[width=6.0cm]{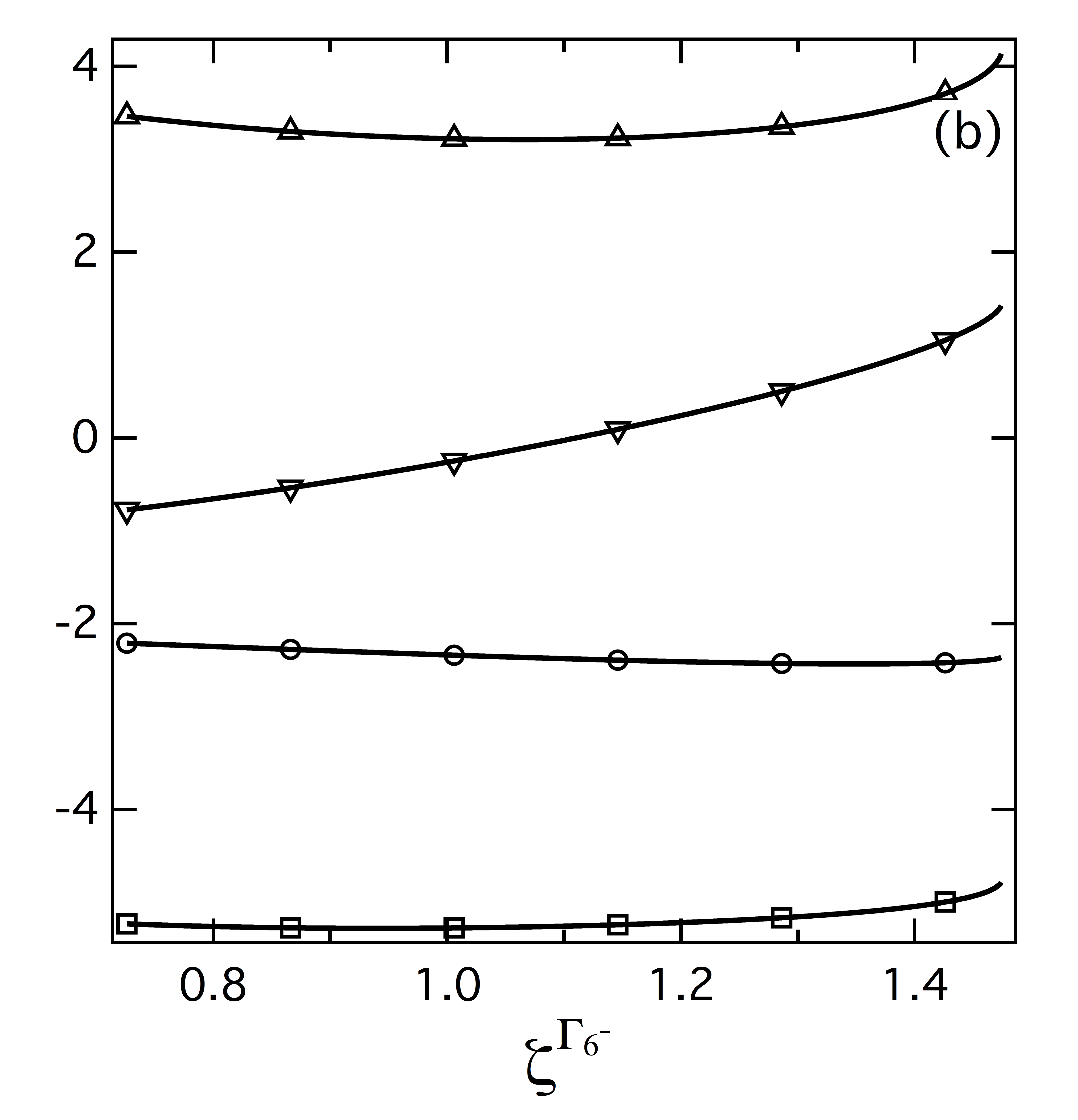}\\
\includegraphics[width=6.0cm]{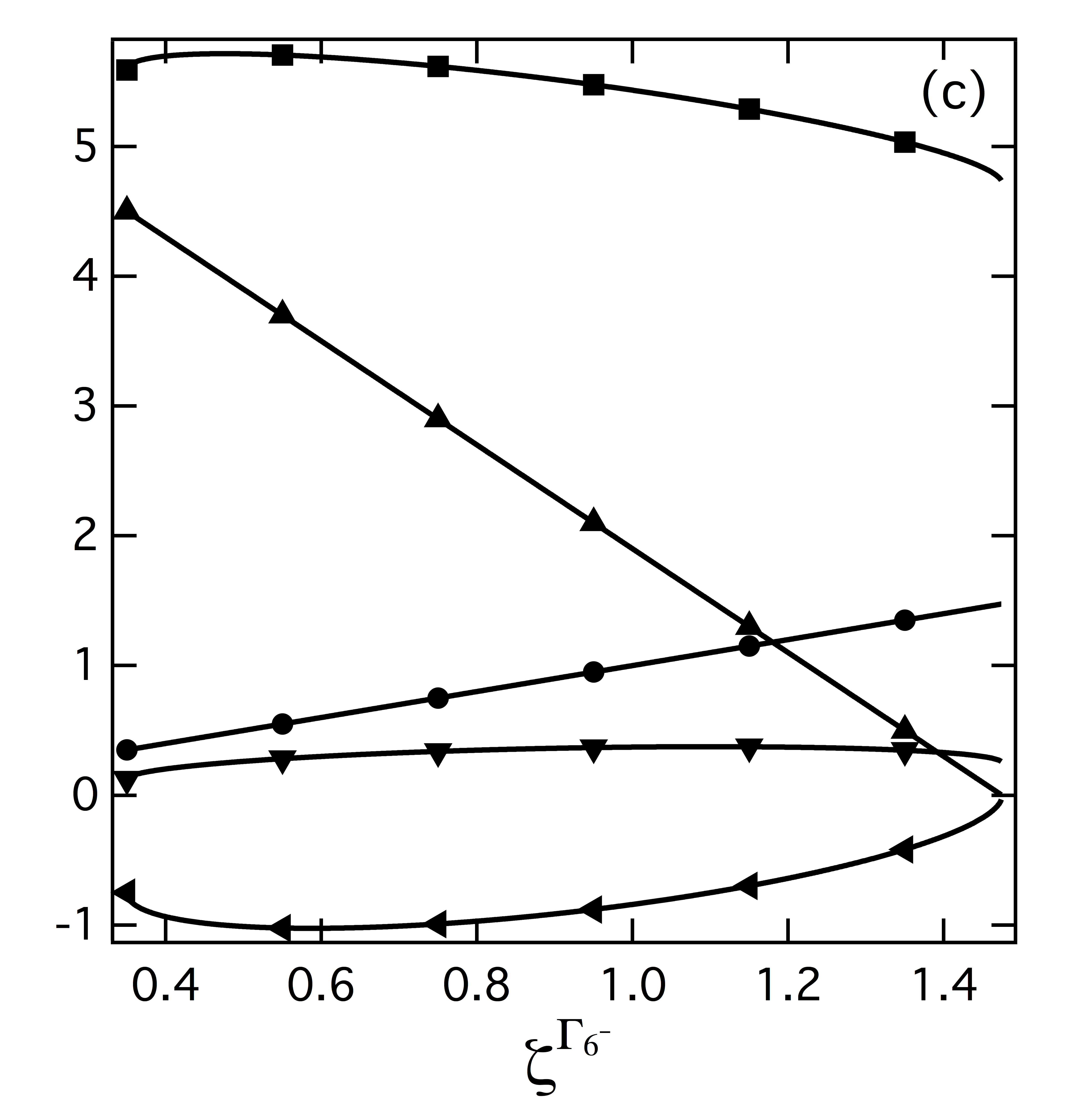}
\includegraphics[width=6.0cm]{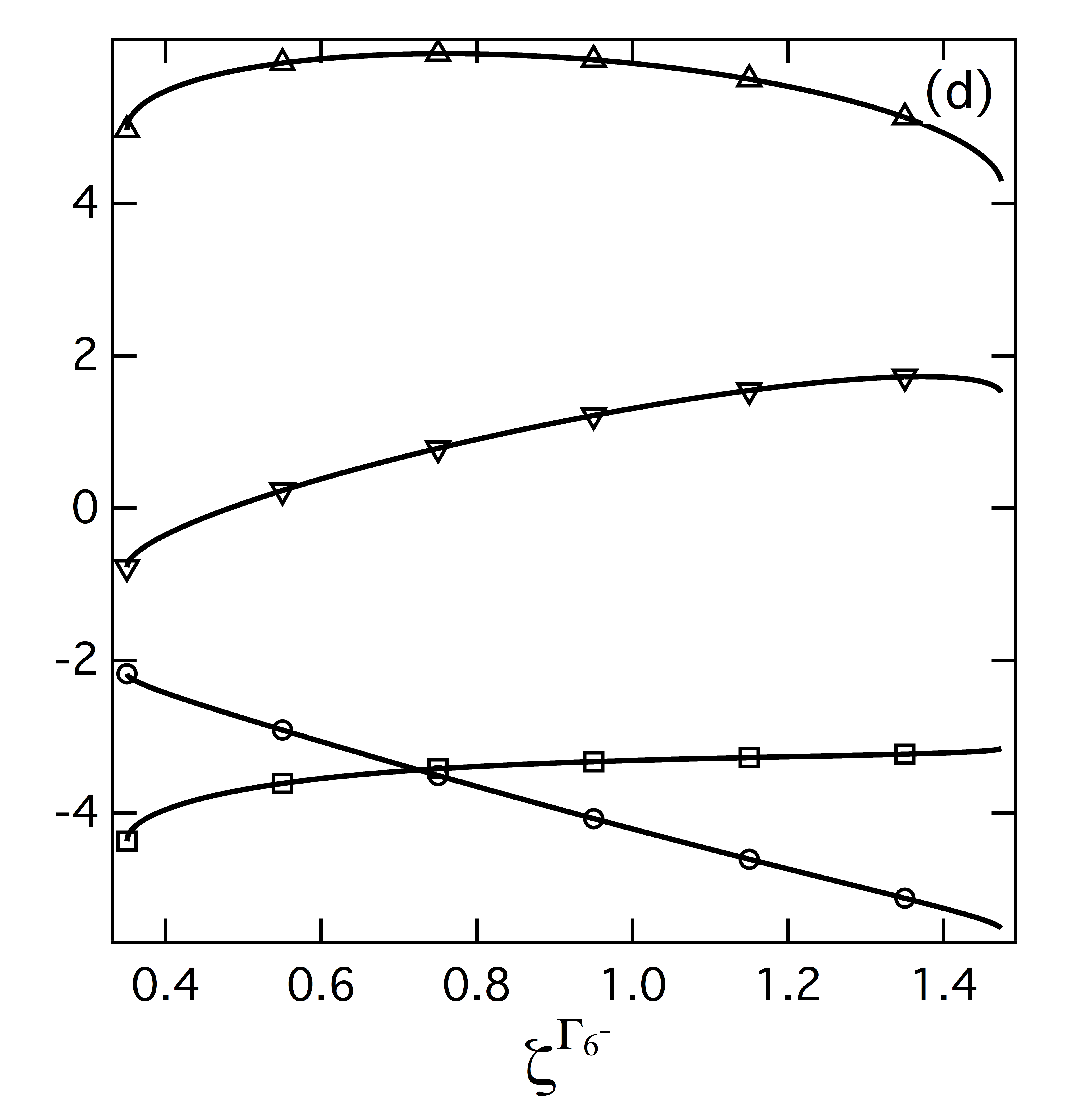}
\caption{Parametric dependence of dimensionless double group material parameters of the valence band, Luttinger parameters $\kappa, q, \gamma_2^{vs}$, and $\gamma_3^{vs}$ on $\zeta^{\Gamma_6^-}_{\Gamma_8^+,\Gamma_8^+}$ compliant to $\gamma_1=13.38$, $\gamma_2=-4.24$, and $\gamma_3=5.69$ for Ge. The symbols associated with the parameters are $\zeta^{\Gamma_6^-}_{\Gamma_8^+,\Gamma_8^+} (\bullet)$, $\zeta^{\Gamma_7^-}_{\Gamma_8^+,\Gamma_8^+} (\blacksquare)$, $\zeta^{\Gamma_8^-,1}_{\Gamma_8^+,\Gamma_8^+} (\blacktriangle)$, $\zeta^{\Gamma_8^-,2}_{\Gamma_8^+,\Gamma_8^+} (\blacktriangledown)$, $\zeta^{\Gamma_8^-,3}_{\Gamma_8^+,\Gamma_8^+} (\blacktriangleleft)$, $\gamma_2^{vs} ( \circ )$, $\gamma_3^{vs} (\square)$, contribution to $\kappa (\vartriangle)$ from anti-symmetric L\"owdin interaction, and contribution to $q (\triangledown)$ from anti-symmetric L\"owdin interaction. Panel (a) and (b) are for $\zeta^{\Gamma_8^-,3}_{\Gamma_8^+,\Gamma_8^+}>0$ and panels (c) and (d) are for $\zeta^{\Gamma_8^-,3}_{\Gamma_8^+,\Gamma_8^+}<0$. See text for methods used to obtain these graphs.}\label{fig1}
\end{centering}
\end{figure*}
The single group formulation provides an efficient way of calculating dispersion relations, but has some deficiencies which were identified and discussed in the last few sections. These deficiencies are generally ignored in the literature but can be overcome by using the double group formulation. To implement double group formulation with a limited set of near states, both linear and quadratic k terms must be considered in the construction of the k-dependent Hamiltonian. These interaction terms should be defined by first order interaction parameters appropriate for the zone centre states included in the near set, and Luttinger invariants that describe interactions mediated by remote states. Given the one-to-one relations between the Luttinger invariants and corresponding second order interaction parameters, as defined in Eq.(\ref{eqn:invariantzeta}), the L\"owdin contribution to the k-dependent Hamiltonian may also be described in terms of the second order interaction parameters $\zeta$.

The extraction of Luttinger parameters from experimental data generally relies on magneto-optical measurements with or without an externally applied stress. The works of Pidgeon and Brown\cite{Pidgeon:1966jn}, Suzuki and Hensel\cite{Suzuki:1974gd,Hensel:1974da}, and Trebin et al.\cite{Trebin:1979jl,Ranvaud:1979cz} provide the methodology and description of such measurements. Analysis of data relies on the use of $k\cdot p$ models with fitting parameters that are specified under perturbation theory and derived under a single group formulation. As the single group formulation utilises relations given in Eq.(\ref{eqn:single_grp_relations}), the extracted Luttinger parameters are model dependent. Moreover, the linear dependence between the Luttinger invariants under the single group formulation is problematic, as there are insufficient second order interaction parameters to describe all of the interactions. In this section, we examine the experimentally derived Luttinger parameters for both Ge and Si, and obtain a set of first order ($\xi$) and second order ($\zeta$) parameters under the double group fromulation. Data is available for the valence band ($\Gamma_8^+$), and spin split-off band ($\Gamma_7^+$) for both Si and Ge, and the lowest conduction band ($\Gamma_7^-$) for Ge. Second order interaction parameters ($\zeta$) are extracted from Eq.(\ref{eqn:invariantzeta}), and dispersion relations are calculated under the 6-, 8-, 14-band models. The dispersion relation should be physical, and the derived parameters should satisfy Eq.(\ref{eqn:invariantzeta}), with the appropriate second order parameters remaining positive as required by their definition.

\begin{table*}
\centering 
\caption{Double group parameters for valence band of silicon and germanium}\label{tbl:zeta_vb}
\begin{ruledtabular}
\begin{tabular}{ccccccccccccccc} 
  &$\zeta^{\Gamma_6^-}_{\Gamma_8^+,\Gamma_8^+}$ & $\zeta^{\Gamma_7^-}_{\Gamma_8^+,\Gamma_8^+}$ & $\zeta^{\Gamma_8^-,1}_{\Gamma_8^+,\Gamma_8^+}$ & $\zeta^{\Gamma_8^-,2}_{\Gamma_8^+,\Gamma_8^+}$ & $\zeta^{\Gamma_8^-,3}_{\Gamma_8^+,\Gamma_8^+}$ & $\zeta^{\Gamma_7^-}_{\Gamma_7^+,\Gamma_7^+}$ & $\zeta^{\Gamma_8^-}_{\Gamma_7^+,\Gamma_7^+}$ & $\zeta^{\Gamma_7^-}_{\Gamma_8^+,\Gamma_7^+}$ & $\zeta^{\Gamma_8^-,1}_{\Gamma_8^+,\Gamma_7^+}$  & $\zeta^{\Gamma_8^-,2}_{\Gamma_8^+,\Gamma_7^+}$ & $\gamma_1^{ss}$ & $\gamma_1^c(\Gamma_7^-)$ & $\gamma_2^{vs}$ & $\gamma_3^{vs}$ \\ \hline
Ge & 0.48 & 5.715 & 3.980 & 0.253 & -1.002 & 2.456 & 2.268 & -3.786 & -3.007 & 0.757& 10.53 &-26.32 & -2.693 & -3.732	\\ Si & 0.375 & 1.500 & 3.00 & 0.157 & -0.687 & 0.0288 & 1.105 & -0.208 & -1.821 & 0.417 & 3.448 & -7.028 & -0.039 & -0.845
\end{tabular}
\end{ruledtabular}
\end{table*}

The first concern is the second order interaction parameters, derivable from Luttinger parameters, that describe the main valence band block. The Luttinger parameters\cite{Anonymous:ba} for Ge are; $\gamma_1=13.35$, $\gamma_2=-4.25$, $\gamma_3=5.69$, $\kappa=3.59$, and $q=0.05$\footnote{The values of $\kappa$ and $q$ are obtained from $\kappa^\prime=3.41,\:q^\prime=0.07$ using Eq.(\ref{eqn:kappaqprime}).}. In principle, the second order parameters can be obtained by solving Eq.(\ref{eqn:invariantzeta}), using all the experimentally determined Luttinger parameters. However, the $\zeta$ parameters obtained from these values do not have the appropriate signs as required by their definition. Assuming the Luttinger parameters associated with the symmetric L\"owdin term are more reliable, it is possible to obtain the parametric dependence of the $\zeta$ parameters with the assumption that there are only 16 zone centre states corresponding to $\Gamma_6^+$, $\Gamma_7^+$, $\Gamma_8^+$, $\Gamma_6^-$, $\Gamma_7^-$ and $\Gamma_8^-$ IRs. Thus, the $\zeta$ parameters describing the interaction with the $\Gamma_8^-$ remote states have the relation $\zeta^{\Gamma_8^-,3}_{\Gamma_8^+,\Gamma_8^+}=\pm\sqrt{\zeta^{\Gamma_8^-,1}_{\Gamma_8^+,\Gamma_8^+}\zeta^{\Gamma_8^-,2}_{\Gamma_8^+,\Gamma_8^+}}$. The $\pm$ sign depends on whether the first order interaction parameters $\xi_{\Gamma_8^+,\Gamma_8^-}^1$ and $\xi_{\Gamma_8^+,\Gamma_8^-}^2$, have the same or opposite sign. Choosing $\zeta^{\Gamma_6^-}_{\Gamma_8^+,\Gamma_8^+}$ as the parameter to vary, two sets of parametric relations can be obtained depending on the choice of sign of $\zeta^{\Gamma_8^-,3}_{\Gamma_8^+,\Gamma_8^+}$. A choice on the sign of $\zeta^{\Gamma_8^-,3}_{\Gamma_8^+,\Gamma_8^+}$ and values for the second order parameters is deferred, until discussions are made on the parameters concerning the spin split-off, and inter-band blocks. These relations are shown in Figure \ref{fig1}(a) and (c), and calculated values of $\kappa$ and $q$ are shown in Figure \ref{fig1}(b) and (d). It is clear that the values of $\kappa$ and $q$ obtained in the literature do not provide a consistent solution to Eq.(\ref{eqn:invariantzeta}). The importance of double group formulation can be seen in the larger magnitude of $\zeta^{\Gamma_8^-,3}_{\Gamma_8^+,\Gamma_8^+}$ compared with $\zeta^{\Gamma_8^-,2}_{\Gamma_8^+,\Gamma_8^+}$. Under the extended Weiler model, it would be difficult to envisage such a large $Z^{\Gamma_8^-,3}_{\Gamma_8^+,\Gamma_8^+}$ compared to $Z^{\Gamma_8^-,2}_{\Gamma_8^+,\Gamma_8^+}$ given the remoteness of states with $\Gamma_5^-$ symmetry. 

The Luttinger invariants in the inter-band block can not be measured directly in experiment. While it is recognised that such parameters should be different from those in the main valence band block\cite{Weiler:1978da}, all previous models employed in the extraction of Luttinger parameters makes the assumption of Eq.(\ref{eqn:single_grp_relations}). To complete the 6-, and 8-band models, a systematic way of obtaining these parameters under the double group formulation is required. A 14-band model is constructed for calculating the bulk dispersion using double group selection rules and first order interaction matrices given in Eq.(\ref{eqn:double_first_order}), with the constraint that the Luttinger parameters, obtained by contracting a such model, should yield the experimentally determined values of $\gamma_1, \gamma_2$, and $\gamma_3$. In other words, the second order interaction parameters describing the valence band block should follow those prescribed by the parametric relations described in the previous paragraph and shown in Figure\ref{fig1} (a) or (c). The Luttinger parameters in the lower band models can then be evaluated by appropriate calculation of the L\"owdin terms from the 14-band model, or more simply by using Eq.(\ref{eqn:invariantzeta}) and removing relevant terms corresponding to states included in the near set. This model differs from the corresponding single group formulation\cite{Rossler:1984fd, Mayer:1991ef, Zawadzki:1985je, Pfeffer:1990cq} model in terms of the description of the first order interaction between states with $\Gamma_8^\pm$ symmetry, the order of the basis in $\Gamma_8^+$ IR, and the absence of $k$ independent inter-band spin orbit interaction. The Hamiltonian, in absence of other external perturbations, is then given by,
\begin{widetext}
\begin{eqnarray}
H(\bm{k})&=&H_0+H_{\bm{k\cdot \pi}}(\bm{k})+H_{L\ddot{o}dwin}(\bm{k}) \\
H_{\bm{k\cdot \pi}}&=&\begin{pmatrix}
\bm{0} & \bm{0} & \bm{0} & K_{\Gamma_8^-,\Gamma_8^+}^{cv}(\bm{k}) & K_{\Gamma_8^-,\Gamma_7^+}^{cv}(\bm{k}) \\
\bm{0} & \bm{0} & \bm{0} & K_{\Gamma_6^-,\Gamma_8^+}^{cv}(\bm{k}) & \bm{0} \\
\bm{0} & \bm{0} & \bm{0} & K_{\Gamma_7^-,\Gamma_8^+}^{cv}(\bm{k}) & K_{\Gamma_7^-,\Gamma_7^+}^{cv}(\bm{k}) \\
K_{\Gamma_8^+,\Gamma_8^-}^{vc}(\bm{k}) & K_{\Gamma_8^+,\Gamma_6^-}^{vc}(\bm{k}) & K_{\Gamma_8^+,\Gamma_7^-}^{vc}(\bm{k}) & \bm{0} & \bm{0} \\
K_{\Gamma_7^+,\Gamma_8^-}^{vc}(\bm{k}) &\bm{0} & K_{\Gamma_7^+,\Gamma_7^-}^{vc}(\bm{k}) & \bm{0} & \bm{0} 
\end{pmatrix} \\
H_{L\ddot{o}dwin}(\bm{k})&=&\begin{pmatrix}
L_{c8c8} & L_{c8c6} & \bm{0} & \bm{0} &\bm{0} \\
L_{c8c6}^\prime & L_{c6c6} & \bm{0} & \bm{0} & \bm{0} \\
\bm{0} & \bm{0} & \bm{0} & \bm{0} & \bm{0} \\
\bm{0} & \bm{0} & \bm{0} & \bm{0} & \bm{0} \\
\bm{0} & \bm{0} & \bm{0} & \bm{0} & \bm{0} 
\end{pmatrix}
\end{eqnarray}
\end{widetext}
where $H_0$ is diagonal and contains the energies of the zone centre states, and the first order interaction matrices are given in Eq.(\ref{eqn:double_first_order}). The matrix is ordered in states with $\Gamma_8^-, \Gamma_6^-, \Gamma_7^-, \Gamma_8^+$, and $\Gamma_7^+$ symmetry with increasing row and column numbers. The lowest valence band zone centre states with $\Gamma_6^+$ symmetry is considered as a remote state. The L\"owdin term only exists between the conduction band states with $\Gamma_8^-$ and $\Gamma_6^-$ symmetry due to double group selection rules. However, the inclusion of L\"owdin terms, that arise from interactions with the remote state with $\Gamma_6^+$ symmetry, have no impact on the dispersion of lowest conduction band, valence band, and spin split off band. As we are primarily interested in the dispersion of these bands, no L\"owdin terms are considered in the 14-band model. Several relations can now be derived from these assertions,  $\zeta^{\Gamma_7^-}_{\Gamma_8^+,\Gamma_8^+}=-\zeta^{\Gamma_8^+}_{\Gamma_7^-,\Gamma_7^-}$, and $\zeta^{\Gamma_7^-}_{\Gamma_7^+,\Gamma_7^+}=-\zeta^{\Gamma_7^+}_{\Gamma_7^-,\Gamma_7^-}$. Utilising the relations shown in Eq.(\ref{eqn:invariantzeta}) we have,
\begin{subequations}
\begin{eqnarray}
\zeta^{\Gamma_7^-}_{\Gamma_7^+,\Gamma_7^+}&=&-\left(\gamma_1^e+1+4\zeta^{\Gamma_7^-,}_{\Gamma_8^+,\Gamma_8^+}\right) \\
\zeta^{\Gamma_8^-}_{\Gamma_7^+,\Gamma_7^+}&=&\frac{1}{4}\left(\gamma_1^{ss}+1-\zeta^{\Gamma_7^-}_{\Gamma_7^+,\Gamma_7^+}\right)
\end{eqnarray}
\end{subequations}
There are now only three adjustable parameters in the model, $\zeta^{\Gamma_6^-}_{\Gamma_8^+,\Gamma_8^+},\:\xi_{\Gamma_7^+,\Gamma_7^-},\:\xi_{\Gamma_7^+,\Gamma_8^-}$ given that the energies of the zone centre states are found to be $E_{\Gamma_7^-}-E_{\Gamma_8^+}=0.887eV$, $E_{\Gamma_6^-}-E_{\Gamma_8^+}=3.006eV$, $E_{\Gamma_8^-}-E_{\Gamma_8^+}=3.206eV$, and $E_{\Gamma_7^+}-E_{\Gamma_8^+}=-0.297eV$. The magnitude of the first order interaction parameters, such as $\xi_{\Gamma_8^+,\Gamma_7^-}$ may be calculated from the corresponding second order interaction parameter such as $\zeta^{\Gamma_7^-}_{\Gamma_8^+,\Gamma_8^+}$. If one choose the condition that $\zeta^{\Gamma_8^-,3}_{\Gamma_8^+,\Gamma_8^+}>0$, it can be easily verified that the dispersion in the $\left<110\right>$ direction is un-physical (valence band maximum not at zone centre) for all the possible range of $\zeta^{\Gamma_6^-}_{\Gamma_8^+,\Gamma_8^+}$. Thus, it is the case that $\zeta^{\Gamma_8^-,3}_{\Gamma_8^+,\Gamma_8^+}<0$, and consequently $\xi_{\Gamma_8^+,\Gamma_8^-}^1,\:\xi_{\Gamma_8^+,\Gamma_8^-}^2$ have opposite signs, and therefore $\zeta^{\Gamma_8^-,1}_{\Gamma_8^+,\Gamma_7^+}$, and $\zeta^{\Gamma_8^-,2}_{\Gamma_8^+,\Gamma_7^+}$ also have opposite signs. 
Using the relations shown in Eq.(\ref{eqn:xi_relations}) as a guide, one can further assume that $\zeta^{\Gamma_7^-}_{\Gamma_8^+,\Gamma_7^+}<0$ and $\zeta^{\Gamma_8^-,1}_{\Gamma_8^+,\Gamma_7^+}<0$.

\begin{figure*}
\begin{center}
\includegraphics[width=6.5in]{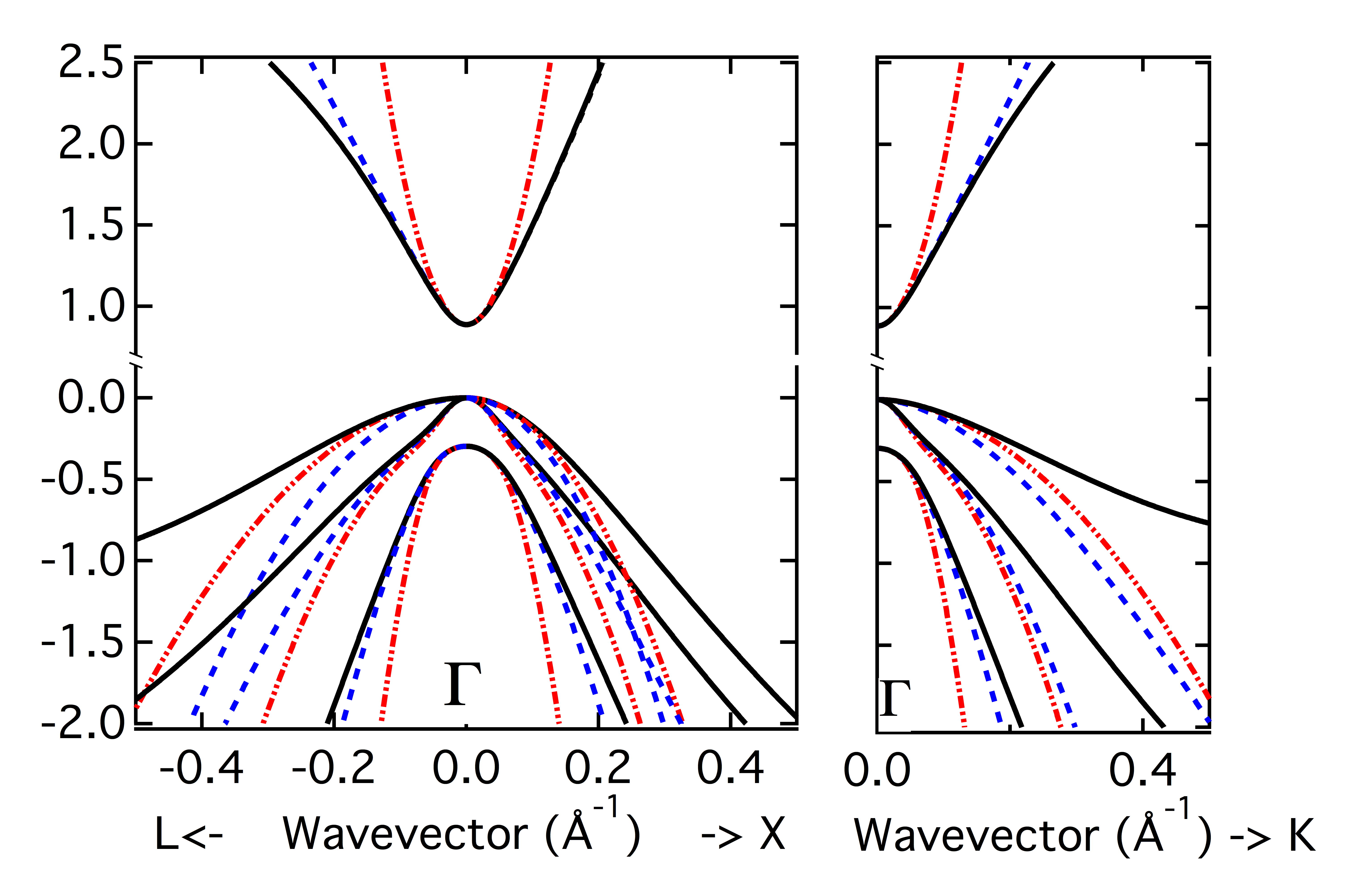}
\caption{Dispersion in Ge calculated using double group formulation, for 14-band (solid line), 8-band (dash line), and 6-band (dot-dash line) models.} 
\label{fig:ge_dispersion}
\end{center}
\end{figure*}

There is still a range of allowed values obtained from the adjustable parameters of $\zeta^{\Gamma_6^-}_{\Gamma_8^+,\Gamma_8^+}$, $\xi_{\Gamma_7^+,\Gamma_7^-}$, $\xi_{\Gamma_7^+,\Gamma_8^-}$ that provide physical dispersion relations in the valence band. The magneto-reflectance measurements of Aggarwal\cite{Aggarwal:1970jh} provide values of Luttinger parameters $\gamma_1^c(\Gamma_7^-)=-26.32\:(m_c=0.038m_0)$, and $\gamma_1^{ss}=10.52\:(m_{ss}=0.095m_0)$. 
These additional constraints reduce the number of adjustable parameters to just one-$\zeta^{\Gamma_6^-}_{\Gamma_8^+,\Gamma_8^+}$. The resulting dispersion in the $X, L,$ and $K$ directions are shown in Figure \ref{fig:ge_dispersion} with the corresponding second order parameters given in Table \ref{tbl:zeta_vb}. Given the way the fitting parameters are chosen, the dispersion at zone centre necessarily reflects the experimentally determined Luttinger parameters for the valence, conduction and spin-split off bands. 

Under the Weiler model (which exclude the remote states with $\Gamma_5^-$ symmetry), the lower limit of $\Gamma_7^-$ conduction band effective mass can be extracted solely from the Luttinger parameters of the $\Gamma_8^+$ valence band using relations in Eq.(\ref{eqn:extended_weiler_gamma}). This still exceed the experimentally measured value of Aggarwal, and requires the inclusion of the Herman-Weisbuch parameter. Under double group formulation the correct $\Gamma_7^-$ conduction band effective mass can be extracted from the Luttinger parameters of the $\Gamma_8^+$ valence band using only relations in Eq.(\ref{eqn:invariantzeta}). The reasons for this can be demonstrated by comparing the relations between the $\gamma_1$ and second order interaction parameters given in Eq.(\ref{eqn:invariantzeta}), Eq.(\ref{eqn:extended_weiler_gamma}), and Eq.(\ref{eqn:lutt_sdpf}). For the Weiler model, the $Z_{\Gamma_8^+,\Gamma_8^+}^{\Gamma_8^-,3}=0$. Thus $\zeta_{\Gamma_8^+,\Gamma_8^+}^{\Gamma_8^-,3}$ is the term that differentiate the double group formulation of EWZ, Weiler model, and relativistically corrected DKK model. A smaller conduction band effective mass requires a larger magnitude of $\zeta_{\Gamma_7^-,\Gamma_7^-}^{\Gamma_8^+}=-\zeta_{\Gamma_8^+,\Gamma_8^+}^{\Gamma_7^-}$, $Z_{\Gamma_7^-,\Gamma_7^-}^{\Gamma_8^+}=-Z_{\Gamma_8^+,\Gamma_8^+}^{\Gamma_7^-}$ or $\sigma$ (in Eq.(\ref{eqn:lutt_sdpf})).  Because of the negative contribution of $\zeta_{\Gamma_8^+,\Gamma_8^+}^{\Gamma_8^+,3}$ to $\gamma_1$ in Eq.(\ref{eqn:invariantzeta}), a larger magnitude of $\zeta_{\Gamma_8^+,\Gamma_8^+}^{\Gamma_7^-}$ is possible under EWZ compared to $Z_{\Gamma_8^+,\Gamma_8^+}^{\Gamma_7^-}$ or $\sigma$ in Eq.(\ref{eqn:lutt_sdpf}) for a given value of $\gamma_1$. Hence, the mixing term $\zeta_{\Gamma_8^+,\Gamma_8^+}^{\Gamma_8^+,3}$ is the key to the success of the double group formulation of EWZ in correctly evaluating the interaction between the zone centre valence and conduction band states. Similar effects means that inter-band Luttinger parameter $\gamma_2^{vs}$ and $\gamma_3^{vs}$ under EWZ are smaller than their counter parts in the valence band. This allows the dispersion in the $\left<110\right>$ direction to remain physical. The importance of zone centre states in the conduction band with $\Gamma_8^-$ symmetry is further demonstrated by the differences observed between the 6-, 8-, and 14-band models. Thus, the dispersion in the valence band and conduction band away from zone centre obtained in 6-, or 8-band models differ from that of the 14-band model.

\begin{figure*}
\begin{center}
\includegraphics[width=6.5in]{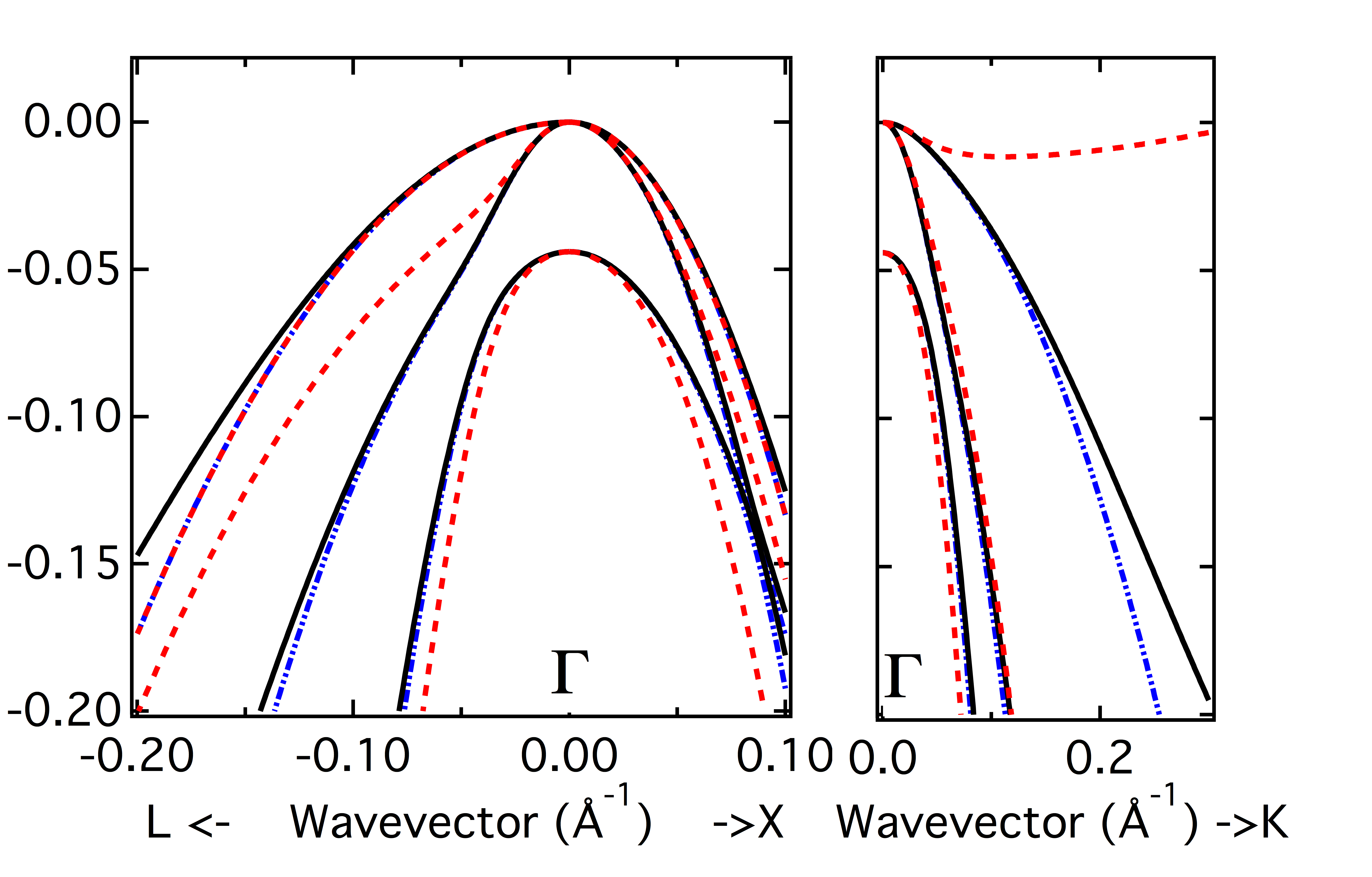}
\caption{Dispersion relation in the valence and spin split off bands of Si calculated using double group formulation, for 14-band (solid line), 6-band (dot-dash line) and using 6-band Weiler model (dash line).}
\label{fig:si_dispersion}
\end{center}
\end{figure*}

For Si, there is no experimentally determined value for the $\Gamma_7^-$ conduction band effective mass. However, the effective mass of the spin split-off band is known\cite{OHYAMA:1970kx} to be $m_{s}=0.29m_0$. Parameters may still be extracted for Si using a similar approach to the one employed to extract parameters in Ge. A suitable set of material parameters for Si are shown in Table \ref{tbl:zeta_vb}. The dispersion relation for the valence and spin split off bands are calculated using double group parameters under 6-, and 14-band models, and compared with results from parameters obtained under the 6-band Weiler model (Table \ref{tbl:g1_frm_30band}) in Figure \ref{fig:si_dispersion}. The parameters shown in Table \ref{tbl:g1_frm_30band} are used in the 6-band Weiler model. The dispersion relation in the $\left<110\right>$ direction shows the inadequacy of the single group formulation, both in terms of the position of valence band maximum, and value of zone centre effective mass for the spin split-off band.

The use of the double group formulation of the $\bm{k\cdot p}$ method can provide a good agreement to a range of experimentally measured effective masses and thus provide an adequate description of dispersion. However, one must be aware that experimentally extracted Luttinger parameters are obtained using models based on the single group formulation. Since all the valence band parameters for Ge were obtained using single group formulation\cite{Hensel:1974da}, there is no reason to give preference to the accuracy of Luttinger parameters associated with symmetric L\"owdin term over those associated with anti-symmetric L\"owdin terms. The discrepancy between the predicted Luttinger invariants ($\kappa$ and $q$) associated with anti-symmetric L\"owdin term using Eq.(\ref{eqn:invariantzeta}) from their experimentally determined values is then due to the use of single group formulation in the analysis of data and uncertainty in the contribution from spin degree of freedom for mixed zone centre states. Nevertheless, this is a significant improvement on the Weiler model, which has limited parameter set, and has difficulty in explaining effective masses measured in different bands simultaneously.

\section{Discussion}
\label{sec:discussion}
The treatment of spin orbit interaction is central to the differences between the double and single group formulations of the $\bm{k\cdot p}$ method within the framework of one electron theory. The two formulations differ in the selection rules applied to the construction of matrix representation of $\bm{k\cdot\pi}$ perturbation. They are equivalent when the zone centre bases for perturbative expansion in $\bm{k}$ are complete (no L\"owdin interaction), as the only difference is in the choice of bases that describe the system. In this case, the spin orbit interaction terms, $H_{so1}^s$ and $H_{so2}^s$, appear explicitly in the Hamiltonian. In constructing the Hamiltonian with a limited bases set, unitary transformations are applied to all zone centre bases to eliminate the $\bm{k\cdot\pi}$ perturbation between states in the near and remote set to the desired order, and diagonalise the un-perturbed Hamiltonian $H_0$. The double group formulation of EWZ treats $H_0$ as its un-perturbed Hamiltonian, with the effects of $H_{so1}^s$ incorporated within the zone centre states and reflected by their energies. Since these eigenstates form the bases of representation of the double group, double group selection rules are required in the description of $\bm{k\cdot\pi}$ interaction between zone centre states. In contrast, the single group formulation retains the single group selection rules when evaluating the $\bm{k\cdot\pi}$ perturbation. As such, the adapted double group bases are derived from unitary transformations of bases {\em within} the single group product bases associated with a {\em single occurrence} of a particular single group IR. In other words, no mixing is permitted between states arising from different single group IRs in the construction of double group zone centre states. This situation occurs only if there is {\em no} $\bm{k}$ independent inter-band spin orbit interaction included within the formulation. This situation corresponds to an unperturbed Hamiltonian equal to $H_0$ with the $\bm{k}$ independent inter-band spin orbit interaction selectively removed. Thus, a single group formulation with a limited bases set represents a system that differs from those described by the Hamiltonian indicated in Eq.(\ref{eqn:H0}). 

More importantly, crystals are systems that contain many electrons. Hybridisation is a well known many electron effect in the formation of covalent bond in cubic semiconductor crystals. With consideration of spin degree of freedom, hybridised orbitals introduces additional mechanism of mixing within the zone centre states. Mixing due to many electron effects permits zone centre states to form bases of representation of the double group, which are drawn from multiple single group IRs that are compatible with the double group representation. This mixing mechanism can not be incorporated directly within the one electron theory, and therefore requires the use of double group selection rules in the $\bm{k\cdot p}$ method. In particular, such mixing allows the zone centre states with $\Gamma_7^-$ and $\Gamma_6^+$ symmetry to experience the effects of spin orbit interaction. This provides an additional mechanism that can modify the ordering of anti-bonding states which is absent under single group consideration. In particular, this mechanism provides an explanation to the observed trend that in heavier elemental semiconductors the $\Gamma_7^-$ conduction band shifts towards the valence band. This mechanism may explain why $\alpha$-Sn has an inverted band structure. Given that there are eight valence electrons per primitive cell, there are eight bonding and eight anti-bonding states under the consideration of hybridised orbitals. This means that each of the double group IRs of the $O_h$ group occurs only once, with the positive parity IRs corresponding to valence band zone centre states, and the negative parity IRs corresponding to the conduction band zone centre states. Any other states that may exist, and are available for occupancy, are un-bound states above the vacuum level. The effect of these states on the dispersion relation tends to increase the lowest zone centre conduction band effective mass, but reduces the valence band effective masses. Therefore, they can not be relied on to explain the difference between predicted mass under single group formulation and that of experimentally observed values. From group theoretical perspective, mixing in zone centre states requires the use of double group selection rules. The multiplicity in the decomposition of direct product of IRs $\Gamma_8^\pm$ lead to difference in the description of $\bm{k\cdot\pi}$ interaction between the single and double group formulations.

The treatment of Zeeman interaction under the double group formulation `de-couples' the link between these terms and the angular momentum. In deriving the Hamiltonian with states quantised in a given orientation, the unitary transformation required to diagonalise the Zeeman interaction differs from that generated by angular momentum operator. Such transformation becomes material dependent on the $\kappa$ and $q$ invariant for the valence band. The unitary transformation introduced by Luttinger only diagonalises one of the generators, but is nevertheless a valid unitary transformation. Due to mixed nature of the zone centre states, the $\hat{\bm{S}}\cdot \bm{B}$ term derived from Dirac equation has to be dealt with using group theoretical method. The matrix representation of $\hat{\bm{S}}$ are no longer the Pauli matrices projected into the adapted double group bases but requires the general description of an operator which transforms according to $\Gamma_4^+$ IR of the $O_h$ group. As generators for transformation of spin states under point group operation, reduced tensor elements for matrix representation of $\hat{\bm{S}}$ is subject the same constraints as that of angular momentum operator as shown in Eq.(\ref{eqn:J_constraint}). The decoupling of Zeeman interaction from total angular momentum and impact of mixed zone centre states on the matrix representation of the spin operator makes the theoretical prediction of the associated Luttinger invariants difficult.

Despite the prevalence of single group formulation, group theoretical methods are employed to determine the number of invariants for a given interaction using double group selection rules. This introduces some disparity between the number of independent second order interaction parameters in single group formulation and number of invariants obtained using the double group selection rules. Perturbation theory allows these invariants to be related to symmetric and anti-symmetric L\"owdin interactions. The linear dependence among the invariants introduced by such relations is an undesirable consequence of the the assumption of single group formulation. The use of perturbation theory leads to two well defined generators for anti-symmetric L\"owdin interactions. The underlying physics of such interactions are made clearer by the use of the two generators given by $\mathcal{J}$ and $J$, as apposed to $\mathcal{J}$ and $\mathcal{J}^3$ used in the literature, since they remove the inter-dependence of invariants $\kappa^\prime$ and $q^\prime$, and highlight the contributions from remote states. The mixed nature of zone centre states also requires different treatment of $\bm{\sigma}\cdot\bm{B}$ term in the Hamiltonian in the same way as the anti-symmetric L\"owdin terms. Symmetry demands that the Luttinger invariants are linearly independent from each other. The double group formulation ensures there are the same number of independent interaction parameters as invariants. Therefore, the relations between the Luttinger invariants and second order interaction parameters preserves the linear independence between the invariants. In contrast, the relations between second order interaction parameters and Luttinger invariants frequently lead to linear dependence among the latter under single group formulation. 

\begingroup
\squeezetable 
\begin{table*}
\caption{Comparison of a number of different single group formulations of the $\mathbf{k\cdot p}$ method with the double group formulation.}
\begin{center}
\begin{ruledtabular}
\begin{tabular}{lcccc}
\multirow{2}{*}{Model}&\multirow{2}{*}{DKK\cite{Dresselhaus:1955da}} & DKK with & \multirow{2}{*}{Extended Weiler Model\footnote{The model was originally described in Ref. \onlinecite{Weiler:1978da} and extended in this manuscript.}} & \multirow{2}{*}{EWZ\cite{Elder:2011kg}} \\ 
& & Relativistic Correction\cite{Suzuki:1974gd} &  & \\ \hline
Selection Rules & Single group & Single group & Single group & Double group \\
Unperturbed zonecentre energy & $E(\Gamma_5^+)$ & $E(\Gamma_5^+)$ & $E(\Gamma_8^+); E(\Gamma_7^+)$ & $E(\Gamma_8^+); E(\Gamma_7^+)$ \\
Predicted zone centre energy & $E(\Gamma_5^+)$ & $E(\Gamma_8^+); E(\Gamma_7^+)$ & $E(\Gamma_8^+); E(\Gamma_7^+)$ & $E(\Gamma_8^+); E(\Gamma_7^+)$\\
Invariants & L, M, N, P & $\left\{\begin{array}{l}\gamma_1,\gamma_2, \gamma_3, \kappa, q=0\\ \gamma_2^{vs}=\gamma_2, \gamma_3^{vs}=-\gamma_3,\\ \kappa^{vs}=\kappa-\frac{g_0}{6}\\
\gamma_1^{ss}=\gamma_1, \kappa^{s}=4\kappa+g_0\end{array}\right.$ & $\left\{\begin{array}{l}\gamma_1,\gamma_2, \gamma_3, \kappa, q\\ \gamma_2^{vs}, \gamma_3^{vs}, \kappa^{vs} \\
\gamma_1^{ss}, \kappa^{ss}\end{array}\right.$ & $\left\{\begin{array}{l}\gamma_1,\gamma_2, \gamma_3, \kappa, q\\ \gamma_2^{vs}, \gamma_3^{vs}, \kappa^{vs} \\
\gamma_1^{ss}, \kappa^{ss}\end{array}\right.$ \\
\begin{tabular}{l}
Second order\\Interaction Parameters\end{tabular} & $F, G, H_1, H_2$ & $F, G, H_1, H_2$ &  $\left\{\begin{array}{l}Z_{\Gamma_8^+,\Gamma_8^+}^{\Gamma_6^-}, Z_{\Gamma_8^+,\Gamma_8^+}^{\Gamma_7^-}, \\ Z_{\Gamma_8^+,\Gamma_8^+}^{\Gamma_8^-;1}, Z_{\Gamma_8^+,\Gamma_8^+}^{\Gamma_8^-;2}, \left\{ Z_{\Gamma_8^+,\Gamma_8^+}^{\Gamma_8^-;3} \right\}\footnote{This parameter is only present if remote state with $\Gamma_5^-$ symmetry is included.} \\ Z_{\Gamma_8^+,\Gamma_7^+}^{\Gamma_7^-}, Z_{\Gamma_8^+,\Gamma_7^+}^{\Gamma_8^-;1}, Z_{\Gamma_8^+,\Gamma_7^+}^{\Gamma_8^-;2} \\
Z_{\Gamma_7^+,\Gamma_8^+}^{\Gamma_7^-}, Z_{\Gamma_7^+,\Gamma_7^+}^{\Gamma_8^-}\end{array}\right.$ & $\left\{\begin{array}{l}\zeta_{\Gamma_8^+,\Gamma_8^+}^{\Gamma_6^-}, \zeta_{\Gamma_8^+,\Gamma_8^+}^{\Gamma_7^-}, \\ \zeta_{\Gamma_8^+,\Gamma_8^+}^{\Gamma_8^-;1}, \zeta_{\Gamma_8^+,\Gamma_8^+}^{\Gamma_8^-;2}, \zeta_{\Gamma_8^+,\Gamma_8^+}^{\Gamma_8^-;3} \\ \zeta_{\Gamma_8^+,\Gamma_7^+}^{\Gamma_7^-}, \zeta_{\Gamma_8^+,\Gamma_7^+}^{\Gamma_8^-;1}, \zeta_{\Gamma_8^+,\Gamma_7^+}^{\Gamma_8^-;2} \\
\zeta_{\Gamma_7^+,\Gamma_8^+}^{\Gamma_7^-}, \zeta_{\Gamma_7^+,\Gamma_7^+}^{\Gamma_8^-}\end{array}\right.$ \\
Bases & Single group & \multicolumn{2}{c}{Adapted double group bases or single group product bases } & Double group bases \\

\end{tabular}
\end{ruledtabular}
\end{center}
\label{tbl:model_comparison}
\end{table*}
\endgroup

A number of models discussed in the literature have been compared and contrasted with the double group formulation of EWZ. The key features of the models, when treating 6-bands are summarised in table \ref{tbl:model_comparison}. The use of selection rules which define the $\bm{k\cdot \pi}$ interaction, assigns each model to either the single or double group formulation. In addition, the models are differentiated according to the way zone centre states are classified. The double group formulation of EWZ on the right and the DKK model on the left represent the two extremes in the treatment of spin effects. 

The DKK model does not account for spin orbit interaction, and as such the spin degree of freedom does not arise, meaning single group selection rules apply. There are four invariants ($L, M, N,$ and $P$) associated with the degenerate states with $\Gamma_5^+$ symmetry, and four second order interaction parameters ($F, G, H_1$ and $H_2$). The model is self consistent in terms of group theoretical description, but does not account for any of the effects of spin observed in experimental measurements. When the bases are transformed into the adapted double group bases, this model gives equality among the Luttinger invariants in different blocks and the $q$ invariant, permitted under double group selection rules, is absent. The relativistically corrected DKK model uses the same unperturbed Hamiltonian as the DKK model, but treats the spin orbit interactions $H_{so1}^s$ and $H_{so2}^s$, together with $\bm{k\cdot p}$ term as perturbations. The resulting bases correspond to those of the adapted double group bases, and the Hamiltonian is partitioned according to the double group classification of zone centre states. While symmetry arguments show that the Luttinger invariants differ between blocks of the Hamiltonian, and $q$ invariant is permitted to exist, the evaluation of the $\bm{k}$ dependent relativistic corrections indicate that such differences are negligible. Thus, the relations between the Luttinger invariants in different blocks still hold, and the $q$ invariant is predicted to be approximately zero. Thus, the relativistically corrected DKK model fails to explain the finite $q$ parameter observed experimentally, the effective mass of the spin orbit band, and the effective mass of the lowest conduction band. Under the relativistically corrected DKK model, the intra-band $\bm{k}$ independent spin orbit interaction is not relativistically small, and generally assumed to give rise to the finite splitting due to spin orbit interaction observed experimentally. However, the small factor at the front of Eq.(\ref{eqn:so1_single_grp}) means a substantial number of states far away from the relevant states are required to obtain a finite result. This is perhaps unphysical. The most obvious reason we can call upon to explain this is the discrepancy between the adapted double group bases used in the relativistically correct DKK model and that of the zone centre state under the double group formulation as discussed in section \ref{sec:lowdin_bases}.

The Kane and extended Kane models are utilised most frequently in the implementation of single group formulation in the literature. The models assume double group classified zone centre state energies for the near set, but retain the single group selection rules to evaluate the $\bm{k\cdot\pi}$ perturbation. It is also assumed that there is equality among Luttinger invariants in blocks associated by spin splitting (Eq.(\ref{eqn:single_grp_relations})), but the $q$ invariant is allowed to be finite. It is difficult to attribute these assumptions to a consistent set of group theoretical rules or a definitive un-perturbed Hamiltonian. The selective use of Eq.(\ref{eqn:single_grp_relations}) and exclusion of Eq.(\ref{eqn:single_grp_qrelations}), both derivable under the relativistically corrected DKK model, suggests that a different classification of zone centre state energies have been used for the evaluation of $q$ invariant compared with other Luttinger invariants. The finite $q$ invariant can arise by the incorporation of states with $\Gamma_8^-\oplus\Gamma_6^-(\Gamma_4^-)$ symmetry into the near set in the extended Kane model. If the same energy scaling is applied to the evaluation of other Luttinger invariants, it is clear that relations in Eq.(\ref{eqn:single_grp_relations}) are also invalid. Therefore this calls for double group classification of all zone centre states, as in the Weiler model.

The Weiler model classifies all zone centre states according to the double group, but retains the single group selection rules when evaluating the $\bm{k\cdot\pi}$ perturbation. While there is no mixing of states between different single group IR, as in all single group formulations, this model constitutes an improvement over the relativistically corrected DKK model. This is principally due to the double group classification of zone centre states in both the near \emph{and} remote set which ensures that Luttinger invariants differ between blocks of the Hamiltonian. In the extended Weiler model, discussed in this manuscript, there are ten independent second order interaction parameters and ten invariants describing a 6-band model. Ten independent second order interaction parameters exist only if $\Gamma_8^-$ remote states derived from $\Gamma_5^-$ IR are included in the near set interactions. In the original Weiler model, the $\Gamma_8^-$ remote states derived from $\Gamma_5^-$ IR are excluded. Parametric fitting of the {\em full} set of second order interaction parameter will also lead to un-physical values for some of the interaction parameters ($\left|Z_{\Gamma_8^+,\Gamma_8^+}^{\Gamma_8^-,3}\right|>\left|Z_{\Gamma_8^+,\Gamma_8^+}^{\Gamma_8^-,2}\right|$). The inclusion of additional conduction band zone centre states will continue to have difficulty in explaining the conduction band effective mass in this model. 

The double group formulation of EWZ follows naturally the prescription of L\"owdin method in partition of states into near and remote sets. Compared with the Weiler model, mixing of double group bases derivable from the compatible single group IRs is permitted in addition to the double group classification of the zone centre states. From a one electron theory perspective, this takes into account the $\bm{k}$ independent inter-band spin orbit interactions among the remote set of states. The double group selection rules also enable the semi-empirical method to account for the effect of mixing due to formation of hybridised orbitals with consideration of spin. This model has the same number of independent second order interaction parameters as invariants in each block. Furthermore, the same set of second order interaction parameters are compatible with effective masses observed experimentally in the lowest zone centre conduction band, valence band, and spin split-off band. Compared with the extended Weiler model, the un-physical values of the second order parameter can be explained through mixing, which occurs in all states with $\Gamma_8^\pm$ symmetry. Therefore, the number of independent second order interaction parameters in each block is already the same as number of invariants without the need to specifically include remote state derived from $\Gamma_5^-$ symmetry. All the other positive features of the Weiler model are retained, enabling good description of the electron dispersion near the zone centre.

The $M_j$ quantum number and its correlation to the $z$ component of angular momentum have been used extensively in the explanation of many physical observations such as circular dichroism. However, the historically chosen unitary transformation, which diagonalises the $\bm{k}$ independent intra-band spin orbit interaction, leads to an order of basis in the valence band zone centre states which is not the same as $D_{\frac{3}{2}}^+$ IR of the $O(3)$ group. Hence, one can not make a correlation between the $M_j$ number of the historically chosen basis and the $z$ component of angular momentum. The difference between the historically chosen basis and those bases with the same order as $D_{\frac{3}{2}}^+$ IR of the $O(3)$ group, can easily be accounted for by a similarity transformation. As a result of this transformation, discussed in section \ref{sec:unitary_transform}, the form of the inter-band Hamiltonian in a 6-band model differs. Also there are some sign differences between the Luttinger parameters, most notably in the change in sign of $\gamma_2$ invariant in the valence band. Under the double group formulation, the matrix representation of the angular momentum operator consists of two linearly independent matrices, as shown in Eq.(\ref{eqn:j_mu_double}). The explanation of phenomena, such as circular dichroism, can still be made provided appropriate momentum matrix elements are evaluated between relevant states using group theoretical techniques. A prime example of this, is their applicability in deriving the selection rules on polarisation dependence of optical transitions between valence and conduction band zone centre states, as discussed in section \ref{sec:double_group}.

Experimentally determined Luttinger parameters are dependent on the model used to describe the interactions between the electrons and the external perturbation such as stress or magnetic field\cite{Hensel:1974da,Suzuki:1974gd}. Specifically, they are dependent on the model derived under the single group formulation, which utilises the relations given in Eq.(\ref{eqn:single_grp_relations}). The experimentally obtained Luttinger parameters are then generally not in agreement with the relations derived between the second order interaction parameters and Luttinger invariant under the double group formulation. The inter-band Luttinger parameters have not been determined independently either. There is a need to re-analyse the raw data and obtain new Luttingers parameter using models which follow double group selection rules. The difficulty in obtaining inter-band Luttinger parameters is then partially overcome, if there are measurements of Luttinger parameters describing relevant intra-band interactions. Such parameters, have a strong influence upon the dispersion away from the zone centre. 
With the advent of modern photoemission instruments\cite{Koralek:2007dd,Liu:2008kl}, a case can be made to re-study these old material systems in order to obtain appropriate Luttinger parameters. For the same reason, the parameters shown in Table \ref{tbl:zeta_vb} and dispersion relation obtained under the double group formulation, shown in Figure \ref{fig:ge_dispersion},\ref{fig:si_dispersion} may not be reliable due to its dependence on Luttinger parameters extracted using single group formulation.
\begin{acknowledgements}
The authors acknowledge Prof. Lee Chang of Tsinghua University for valuable discussion on the Foldy-Wouthuysen transformation. EW acknowledge EPSRC,UK, for the provision of a studentship.
\end{acknowledgements}
\appendix
\section{Basis functions used in calculation of interaction matrice}
\label{app:bases}
\begin{table*}
\caption{The single group bases are expressed in terms of real harmonics, whereas the double group bases are expressed in terms of real harmonics and spinor states $\{\alpha, \beta\}$. The double group bases are compliant to the time reversal condition stated in Eq.(\ref{eqn:time_reversal}).} \label{tab:basis}
\begin{ruledtabular}
\begin{tabular}{cllcl}
IR & basis functions &&IR & basis functions\\ \hline
$\Gamma_1^+$ & $(l=0):\:1$ & &$\Gamma_2^-$ &$(l=3):\:xyz$  \\
$\Gamma_3^+$ & $(l=2):\left\{\begin{array}{l}3z^2-r^2,\\ \sqrt{3}(x^2-y^2)\end{array}\right.$& &$\Gamma_3^-$ & $(l=5):\left\{\begin{array}{l}\sqrt{3}xyz(x^2-y^2),\\ xyz(r^2-3z^2)\end{array}\right.$\\
$\Gamma_5^+$ & $(l=2):\left\{\begin{array}{l}yz\\ zx\\ xy\end{array}\right. $&$(l=4):\left\{\begin{array}{l}yz(6x^2-y^2-z^2)\\ zx(6y^2-z^2-x^2)\\ xy(6z^2-x^2-y^2)\end{array}\right.$&$\Gamma_4^-$ & $(l=1):\left\{\begin{array}{l}x,\\ y,\\ z\end{array}\right.$\\
$\Gamma_4^-$&$(l=3):\:\left\{\begin{array}{l}x(3y^2+3z^2-2x^2)\\y(3z^2+3x^2-2y^2)\\z(3x^2+3y^2-2z^2)\end{array}\right.$&&$\Gamma_5^-$ &$(l=3):\left\{\begin{array}{l}x(y^2-z^2),\\ y(z^2-x^2),\\ z(x^2-y^2)\end{array}\right.$ \\
$\Gamma_6^+$ & $(l=0):\left\{\begin{array}{l}i\alpha\\ i\beta\end{array}\right.$ &&$\Gamma_7^-$ &$(l=3):\left\{\begin{array}{l}ixyz\alpha,\\ ixyz\beta\end{array}\right.$\\
$\Gamma_7^+$ & $(l=2):\left\{\begin{array}{l}xy\alpha+(yz+izx)\beta,\\ -xy\beta+(yz-izx)\alpha\end{array}\right.$ &&$\Gamma_6^-$ &$(l=1):\left\{\begin{array}{l}-z\alpha-(x+iy)\beta,\\ z\beta-(x-iy)\alpha\end{array}\right.$ \\
$\Gamma_8^+$ &$(l=2):\left\{\begin{array}{l}-(yz-izx)\alpha-2xy\beta\\ \sqrt{3}(yz-izx)\beta\\ -\sqrt{3}(yz+izx)\alpha\\ (yz+izx)\beta-2xy\alpha\end{array}\right.$ &$(l=2):\left\{\begin{array}{l}i(x^2-y^2)\beta\\ \frac{i}{\sqrt{3}}(r^2-3z^2)\alpha\\-\frac{i}{\sqrt{3}}(r^2-3z^2)\beta\\-i(x^2-y^2)\alpha\}\end{array}\right.$&$\Gamma_8^-$ & $(l=1):\left\{\begin{array}{l}-(x+iy)\alpha,\\ \sqrt{\frac{1}{3}}\left[2z\alpha-(x+iy)\beta\right],\\ \sqrt{\frac{1}{3}}\left[2z\beta+(x-iy)\alpha\right],\\ (x-iy)\beta\end{array}\right.$ 
\end{tabular}
\label{tbl:basis_fn}
\end{ruledtabular}
\end{table*}
The solution of un-perturbed Schr\"odinger equation (zone centre states) form the basis of irreducible representations of the symmetry group of the crystal\cite{Inui:1996wk}. The first order interaction matrices between states belonging to irreducible representations may be calculated using the extension of Wigner-Eckart theorem for the point group\cite{Elder:2011kg}. This reduces the interaction matrices to a sum of products between a scaling parameter (reduced tensor element), and an angular dependent part. As states belonging to a particular IR with dimension larger than one are degenerate, linear combinations of such states are also energy eigenstates, and form basis of the IR. One is free to choose such basis, but the basis with the same ordering within in the IR must be used for calculation of all the relevant interaction matrices. This applies to first  and higher order terms in perturbation theory. 

It is customary to generate representation matrices by considering the point group as a subgroup of $O(3)$. Using the total angular momentum operator $\hat{\bm{J}}$ as generators of the infinitesimal rotation in the $O(3)$ group, representation matrices for proper rotations can be constructed, and effects of improper rotations in the $O_h$ or $T_d$ groups can be introduced afterwards\cite{Inui:1996wk}. For the $O_h$ point group, the effect of improper rotations differentiate representations that have different parity.  These representation matrices fix the order of states within an IR. This aspect is essential in deriving the necessary similarity transformation required in combining the basis of $\Gamma_8^\pm$ in the $O_h$ group, into basis of $\Gamma_8$ in $T_d$ group under compatibility relations. If a different basis is used, then the underlying unitary transformation must be considered. It is preferable to work with system with higher symmetry ($O_h$ group in this case) and consider the effect of lower symmetry ($T_d$ group) via the sub group relationship and compatibility relations. For a Hamiltonian constructed using the method of invariant\cite{LuttingerJM:1956hi}, the form of the matrix representation of components of the total angular momentum operator with respect to the $J=\frac{3}{2}$ valence band bases, implies an order of the states same as those used in EWZ\footnote{Much of the literature claims to use $J=\frac{3}{2}$ bases of $\Gamma_8^+$ IR to obtain the matrix representation of angular momentum and construct the valence band Hamiltonian from method of invariant. The form inter-band block of the Hamiltonian suggests that the bases used are ordered according to the $D_{\frac{3}{2}}^-$ IR of the $O(3)$ group. The use of these bases ensures agreement with the 6 band Hamiltonian shown in Eq.(\ref{Chuang}).}.

It is also important to consider the time reversal properties of basis when considering intra-band interactions. This is of particular importance when obtaining first order, intra-band interactions, allowed under $T_d$ group symmetry. To facilitate this, we use the relation stated by Koster\cite{Koster:1963um} and used in EWZ,
\begin{equation}
\label{eqn:time_reversal}
T\left|J,M_j\right>=(-1)^{J-M_j}\left|J,-M_j\right>
\end{equation}
where $T$ is the time reversal operator, $\left|J,M_j\right>$ is a basis of IR of dimension $2J+1$ and $M_j=-J, -J+1, \cdots, J$ index the basis within the IR.  We ensure the double group bases follow this convention, but retain the real harmonic format generally used for single group bases. However, a necessary scaling factor of $i$ is included in the even single group bases (expressed as real spherical harmonics) to ensure compliance to ordering of basis prescribed by Eq,(\ref{eqn:time_reversal}). This also ensure the reduced tensor element of momentum operator is real.

Table \ref{tab:basis} lists the basis functions of representation of the $O_h$ group used in the calculation of interaction matrices in appendix \ref{app:first_order}. The single group bases are expressed in terms of real harmonics whereas the double group bases are expressed in terms of real harmonics and spinor states $\{\alpha, \beta\}$. The double group bases are compliant to the time reversal condition stated in Eq.(\ref{eqn:time_reversal}).
\section{First order interaction matrices}
\label{app:first_order}
The perturbation operator, $\hat{\bm{\pi}}=\hat{\bm{p}}+\frac{\hbar}{2m_0^2c^2}\left(\hat{\bm{S}}\times\nabla V(r)\right)$ (or $\hat{\bm{p}}$ under the single group) transforms according to $\Gamma_4^-$ IR of the $O_h$ group, and is time reversal odd. A general irreducible tensor operator is time reversal even. Thus, the equivalent operator of $\hat{\bm{\pi}}$ constructed from irreducible tensor operator must be scaled by a factor of ``$i$''. The angular dependence of $\hat{\bm{\pi}}$ with respect to given bases, may be evaluated using the constructed equivalent operator, Wigner-Eckart theorem. Therefore,
\begin{eqnarray}
\hat{\pi}_x&=&\frac{T_1^{-1}-T_1^{+1}}{\sqrt{2}i} \notag \\
\hat{\pi}_y&=&\frac{T_1^{-1}+T_1^{+1}}{\sqrt{2}} \\
\hat{\pi}_z&=&\frac{T_1^0}{i} \notag
\end{eqnarray}
where $T_1^q$ are irreducible spherical tensor operators of rank one. 
Given the set of zone centre states associated with the $\alpha^u,\alpha^v$ IR of the symmetry group, the first order interaction may be written in matrix form as,
\begin{subequations}
\begin{eqnarray}
K_{\alpha^u,\alpha^v}&=&\frac{\hbar}{m_0}\bm{k\cdot\pi}_{\alpha^u,\alpha^v}=\sum_{\mu\in\{x,y,z\}}k_\mu \bm{M}_{\alpha^u,\alpha^v}^\mu \\
\bm{M}_{\alpha^u,\alpha^v:ij}^\mu&=&\frac{\hbar}{m_0}\left<\phi_i^{\alpha^u}|\pi_\mu|\phi_j^{\alpha^v}\right> \label{eqn:M_matrix}
\end{eqnarray}
\end{subequations}
Where selection rules indicate that there are multiple linearly independent angular parts, they can be generated by enumeration of basis of the IRs. 

With the single group bases given in appendix \ref{app:bases}, the matrix representation of the $\bm{k\cdot p}$ interaction can be written as,
\begin{widetext}
\begin{equation}
\begin{array}{ll}
K_{\Gamma_5^+,\Gamma_2^-}=\frac{\hbar}{m_0}\xi_{\Gamma_5^+,\Gamma_2^-}\begin{pmatrix}k_x & k_y & k_z\end{pmatrix}^T &K_{\Gamma_4^-,\Gamma_1^+}=\frac{\hbar}{m_0}\xi_{\Gamma_4^-,\Gamma_1^+}\begin{pmatrix}k_x & k_y & k_z\end{pmatrix}^T \\
K_{\Gamma_5^+,\Gamma_3^-}=\frac{\hbar}{m_0}\xi_{\Gamma_5^+,\Gamma_3^-}\begin{pmatrix}\sqrt{3}k_x & k_x\\-\sqrt{3}k_y & k_y\\0 & -2k_z\end{pmatrix} &K_{\Gamma_4^-,\Gamma_3^+}=\frac{\hbar}{m_0}\xi_{\Gamma_4^-,\Gamma_3^+}\begin{pmatrix}-k_x & \sqrt{3}k_x\\-k_y & -\sqrt{3}k_y\\2k_z & 0\end{pmatrix} \\
K_{\Gamma_5^+,\Gamma_4^-}=\frac{\hbar}{m_0}\xi_{\Gamma_5^+,\Gamma_4^-}\begin{pmatrix}0 & k_z & k_y\\ k_z & 0 & k_x\\k_y & k_x & 0\end{pmatrix} &
K_{\Gamma_5^+,\Gamma_5^-}=\frac{\hbar}{m_0}\xi_{\Gamma_5^+,\Gamma_5^-}\begin{pmatrix}0 & -k_z & k_y\\ -k_z & 0 & -k_x\\k_y & -k_x & 0\end{pmatrix}
\end{array}
\label{eqn:single_first_order}
\end{equation}
With the double group bases given in appendix \ref{app:bases}, the matrix representation of the $\bm{k\cdot\pi}$ interaction can be written as,
\begin{eqnarray}
\label{eqn:double_first_order}
&&
\begin{array}{ll}
\mathcal{K}_{\Gamma_6^\pm,\Gamma_6^\mp}=\frac{\hbar}{m_0}\xi_{\Gamma_6^\pm,\Gamma_6^\mp}\begin{pmatrix}
k_z & k_- \\ k_+ & -k_z\end{pmatrix} &\mathcal{K}_{\Gamma_7^\pm,\Gamma_7^\mp}=\frac{\hbar}{m_0}\xi_{\Gamma_7^\pm,\Gamma_7^\pm}\begin{pmatrix}k_z & k_- \\ k_+ & -k_z\end{pmatrix}\\
\mathcal{K}_{\Gamma_8^\pm,\Gamma_6^\mp}=\frac{\hbar}{m_0}\xi_{\Gamma_8^\pm,\Gamma_6^\mp}\begin{pmatrix}
-\sqrt{3}k_- & 0 \\ 2k_z & -k_- \\ k_+ & 2k_z \\ 0 & \sqrt{3}k_+ \end{pmatrix} & \mathcal{K}_{\Gamma_8^\pm,\Gamma_7^\mp}=\frac{\hbar}{m_0}\xi_{\Gamma_8^\pm,\Gamma_7^\mp}\begin{pmatrix}
k_+ & 2k_z \\ 0 & -\sqrt{3}k_+ \\ \sqrt{3}k_- & 0 \\ 2k_z & -k_-\end{pmatrix} 
\end{array}\\
&&
\mathcal{K}_{\Gamma_8^\pm,\Gamma_8^\mp}=\mathcal{K}_{\Gamma_8^\pm,\Gamma_8^\mp}^1+\mathcal{K}_{\Gamma_8^\pm,\Gamma_8^\mp}^2=\frac{\hbar}{m_0}\xi_{\Gamma_8^\pm,\Gamma_8^\mp}^1\begin{pmatrix}
k_z & 0 & 0 & k_+\\ 0 & -k_z & k_- & 0 \\ 0 & k_+ & k_z & 0 \\ k_- & 0 & 0  & k_z\end{pmatrix} +\frac{\hbar}{m_0}\xi_{\Gamma_8^\pm,\Gamma_8^\mp}^2\begin{pmatrix}
4k_z & \sqrt{3}k_- & 0 & k_+\\ \sqrt{3}k_+ & 0 & 3k_- & 0 \\ 0 & 3k_+ & 0 & \sqrt{3}k_- \\ k_- & 0 & \sqrt{3}k_+ & -4k_z\end{pmatrix}  \label{eqn:double_first_order_G8pm} 
\end{eqnarray}
\end{widetext}
To obtain the individual component matrix $\bm{M}^\mu_{\alpha^u,\alpha^v}$, one takes the corresponding $K_{\alpha^u,\alpha^v}$ and set $k_\mu=1$ and $k_\nu=0\:\: \forall\: \nu\ne\mu$. The first order interaction with respect to the double group bases may also be obtained by transforming the first order interaction matrix with respect to the product bases into adapted double group bases, using the transformation matrices given in the following section. This is illustrated in appendix \ref{app:xis}. In applying Winger-Eckart theorem to obtain these interaction matrices, the first order interaction parameters $\xi$'s are real, but their signs are determined by the radial part of the wave functions.

\section{Transformation matrices from single group to adapted double group bases}
\label{app:unitary_transform}
The $O_h$ point group character table indicates that $\Gamma_6^+\otimes\Gamma_4^-=\Gamma_8^-\oplus\Gamma_6^-$, and $\Gamma_6^+\otimes\Gamma_5^+=\Gamma_8^+\oplus\Gamma_7^+$. Therefore, the adapted double group basis for $\Gamma_8^-\oplus\Gamma_6^-$ and $\Gamma_8^+\oplus\Gamma_7^+$, may be obtained from a direct product of spinor states with single group basis of $\Gamma_4^-$ and $\Gamma_5^+$, followed by an appropriate unitary transformation. All bases transform as row vectors. The unitary transformation \footnote{The resulting adapted double group bases comply with time reversal requirement stated by Eq.(2) of EWZ.} may be written as,
\begin{widetext}
\begin{subequations}
\label{single2double}
\begin{eqnarray}
\left(\psi_1^{\Gamma_8^-}, \psi_2^{\Gamma_8^-}, \psi_3^{\Gamma_8^-}, \psi_4^{\Gamma_8^-}, \psi_1^{\Gamma_6^-}, \psi_2^{\Gamma_6^-} \right)&=&\left(\phi_1^{\Gamma_4^-}\alpha, \phi_2^{\Gamma_4^-}\alpha, \cdots, \phi_3^{\Gamma_4^-}\beta\right)\begin{pmatrix}
-\frac{1}{\sqrt{2}} & 0 & \frac{1}{\sqrt{6}} & 0 & 0 & -\frac{1}{\sqrt{3}} \\
-\frac{i}{\sqrt{2}} & 0 & -\frac{i}{\sqrt{6}} & 0 & 0 & \frac{i}{\sqrt{3}} \\
0 & \frac{2}{\sqrt{6}} & 0 & 0 & -\frac{1}{\sqrt{3}} & 0 \\
0 & -\frac{1}{\sqrt{6}} & 0 & \frac{1}{\sqrt{2}} & -\frac{1}{\sqrt{3}} & 0 \\
0 & -\frac{i}{\sqrt{6}} & 0 & -\frac{i}{\sqrt{2}} & -\frac{i}{\sqrt{3}} & 0 \\
0 & 0 & \frac{2}{\sqrt{6}} & 0 & 0 & \frac{1}{\sqrt{3}} \\
\end{pmatrix} \notag \\
&=&\left(\phi_1^{\Gamma_4^-}\alpha, \phi_2^{\Gamma_4^-}\alpha, \phi_3^{\Gamma_4^-}\alpha, \phi_1^{\Gamma_4^-}\beta, \phi_2^{\Gamma_4^-}\beta, \phi_3^{\Gamma_4^-}\beta\right)U^{\Gamma_4^-} \\
\left(\psi_1^{\Gamma_8^+}, \psi_2^{\Gamma_8^+}, \psi_3^{\Gamma_8^+}, \psi_4^{\Gamma_8^+}, \psi_1^{\Gamma_7^+}, \psi_2^{\Gamma_7^+} \right)&=&\left(\phi_1^{\Gamma_5^+}\alpha, \phi_2^{\Gamma_5^+}\alpha, \cdots, \phi_3^{\Gamma_5^+}\beta\right)\begin{pmatrix}
-\frac{1}{\sqrt{6}} & 0 & -\frac{1}{\sqrt{2}} & 0 & 0 & \frac{1}{\sqrt{3}}\\ 
\frac{i}{\sqrt{6}} & 0 & -\frac{i}{\sqrt{2}} & 0 & 0 & -\frac{i}{\sqrt{3}}\\ 
0 & 0 & 0 & -\frac{2}{\sqrt{6}} & \frac{1}{\sqrt{3}} & 0\\ 
0 & \frac{1}{\sqrt{2}} & 0 & \frac{1}{\sqrt{6}} & \frac{1}{\sqrt{3}} & 0\\ 
0 & -\frac{i}{\sqrt{2}} & 0 & \frac{i}{\sqrt{6}} & \frac{i}{\sqrt{3}} & 0\\ 
-\frac{2}{\sqrt{6}} & 0 & 0 & 0 & 0 & -\frac{1}{\sqrt{3}} 
\end{pmatrix} \notag \\
&=&\left(\phi_1^{\Gamma_5^+}\alpha, \phi_2^{\Gamma_5^+}\alpha, \phi_3^{\Gamma_5^+}\alpha,\phi_1^{\Gamma_5^+}\beta, \phi_2^{\Gamma_5^+}\beta, \phi_3^{\Gamma_5^+}\beta\right)U^{\Gamma_5^+}  
\end{eqnarray}
\end{subequations}
The two matrices are related by 
\begin{equation}
\label{eqn:MU_matrix}
U^{\Gamma_4^-}=U^{\Gamma_5^+}\begin{pmatrix}U & \mathbb{0}_{4\times 2} \\ \mathbb{0}_{2\times 4} & -\mathbb{1}_{2\times 2}\end{pmatrix}
\end{equation} 
where $U$ is defined by Eq.(\ref{eq:u_matrix}).

\section{Construction of the L\"owdin term and relations of Luttinger invariants to interaction parameters}
\label{app:lowdin}

Without consideration of spin, the DKK hamiltonian ($H_{L\ddot{o}wdin}$) with respect to the $\Gamma_5^+$ basis of $\{yz, zx, xy\}$ in the valence band, are obtained from first order interaction matrices, and Eq.(\ref{eqn:Lowdin_S},\ref{eqn:Lowdin_A}) as,
\begin{subequations}
\label{eqn:H_dkk}
\begin{eqnarray}
H_{DKK}&=&\begin{pmatrix}
Lk_x^2+M(k_y^2+k_z^2) & Nk_xk_y+iP[k_x,k_y] & Nk_xk_z-iP[k_z,k_x] \\
Nk_xk_y-iP[k_x,k_y] & Lk_y^2+M(k_z^2+k_x^2) & Nk_yk_z+Pi[k_y,k_z] \\
Nk_xk_z+iP[k_z,k_x] & Nk_yk_z-iP[k_y,k_z] &Lk_z^2+M(k_x^2+k_y^2)
\end{pmatrix} \\
L&=&F+2G \\
M&=& H_1+H_2\\
N&=&F+H_1-G-H_2 \\
P&=&\frac{1}{2}\left(F-H_1-G+H_2\right)
\end{eqnarray}
\end{subequations}
\end{widetext}
where the second order parameter $F, G, H_1, H_2$ are as defined by Kane\cite{Kane:1966wa}. These second order parameters may be related to first order interaction parameters in appendix \ref{app:first_order} by,

\begin{subequations}
\label{eqn:luttinger_inv_single}
\begin{eqnarray}
F&=&-\frac{\hbar^2}{2m_0}6\sigma=\frac{\hbar^2}{m_0^2}\frac{\left|\xi_{\Gamma_5^+,\Gamma_2^-}\right|^2}{E_{\Gamma_5^+}^v-E_{\Gamma_2^-}^{s^*}}\\
G&=&-\frac{\hbar^2}{2m_0}6\delta=\frac{\hbar^2}{2m_0^2}\frac{\left|\xi_{\Gamma_5^+,\Gamma_3^-}\right|^2}{E_{\Gamma_5^+}^v-E_{\Gamma_3^-}^{d^*}}\\
H_1&=&-\frac{\hbar^2}{2m_0}6\pi=\frac{\hbar^2}{m_0^2}\frac{\left|\xi_{\Gamma_5^+,\Gamma_4^-}\right|^2}{E_{\Gamma_5^+}^v-E_{\Gamma_4^-}^{p^*}}\\
H_2&=&-\frac{\hbar^2}{2m_0}6\phi=\frac{\hbar^2}{m_0^2}\frac{\left|\xi_{\Gamma_5^+,\Gamma_5^-}\right|^2}{E_{\Gamma_5^+}^v-E_{\Gamma_5^-}^{f^*}}
\end{eqnarray}
\end{subequations}
where the sign of each term in the summation is determined by the relative energetic position of the remote states and near states. For diamond, Si, and Ge, all are negative. The parameters, $\sigma,\:\delta,\:\pi$ and $\phi$ are dimensionless versions of the second order interaction parameters for the valence/spin split-off bands defined along the lines of Ref.\onlinecite{Roth:1959kd,Foreman:1993gb}.

The 6 band Hamiltonian with respect to the adapted double group basis can e derived from obtained from 
\begin{equation}
H^{adap}=H_0^s+{U^{\Gamma_5^+}}^\dag(\mathbb{1}_{2\times 2}\otimes H_{DKK})U^{\Gamma_5^+}+g_0\mu_B\hat{\bm{S}}\cdot\bm{B}
\end{equation}
where the $U^{\Gamma_5^+}$ is given in appendix \ref{app:unitary_transform}. This is formally the DKK Hamiltonian, with $\bm{k}$ independent relativistic correction only. The effect of $H_{so3}=g_0\mu_B\hat{\bm{S}}\cdot\bm{B}$ is also included by projecting the matrix representation of $\hat{\bm{S}}$ into the appropriate spaces. It returns the form of Hamiltonian given by Eq.(\ref{EWZ}), with the Luttinger parameters related to the single group invariants and second order interaction parameters as,
\begin{subequations}
\label{eqn:lutt_sdpf}
\begin{eqnarray}
\gamma_1&=&-\frac{2m_0}{\hbar^2}\frac{(L+2M)}{3}-1= -\frac{2m_0}{\hbar^2}\frac{(F+2G+2H_1+2H_2)}{3}-1\notag \\
&=&2\sigma+4\pi+4\delta+4\phi-1\\
\gamma_2&=&\frac{2m_0}{\hbar^2}\frac{(L-M)}{6}=\frac{2m_0}{\hbar^2}\frac{(F+2G-H_1-H_2)}{6} \notag \\
&=&-\sigma+\pi-2\delta+\phi \label{eqn:g2_single}\\
\gamma_3&=&-\frac{2m_0}{\hbar^2}\frac{N}{6}=-\frac{2m_0}{\hbar^2}\frac{(F-G+H_1-H_2)}{6} \notag \\
&=&\sigma+\pi-\delta-\phi  \\
\kappa&=&-\frac{2m_0}{\hbar^2}\frac{P}{3}-\frac{g_0}{6}=-\frac{2m_0}{\hbar^2}\frac{(F-G-H_1+H_2)}{6}-\frac{g_0}{6}\notag \\
&=&\sigma-\pi-\delta+\phi-\frac{g_0}{6}\\
q&=&0 \label{eqn:single_grp_qrelations}\\
\gamma_2^{vs}&=&\gamma_2,\quad
\gamma_3^{vs}=-\gamma_3,\quad\kappa^{vs}+\frac{g_0}{3}=\kappa+\frac{g_0}{6}, \notag\\ 
\gamma_1^{ss}&=&\gamma_1, \quad \kappa^{s}-\frac{g_0}{3}=4\left(\kappa+\frac{g_0}{6}\right).\label{eqn:single_grp_relations}
\end{eqnarray}
\end{subequations}
These results have been obtained previously in the literature\cite{Roth:1959kd}. Within the framework of single group approximation, there are four interaction parameters ($F, G, H_1,$ and $H_2$) and four invariants ($L, M, N,$ and $P$) permitted by single group selection rules.  Referring to the adapted double group bases, the single group selection rules only permit four invariants ($\gamma_1, \gamma_2, \gamma_3,$ and $\kappa$). Note that the $\kappa$ invariant in the spin-split off band differs from that of the valence band. Under this particular approach, the effect of inter-band $\bm{k}$ independent spin orbit interaction, described at end of section \ref{sec:incomplete_bases} and beginning of section \ref{sec:extended_kane}, could not have been included. 

In the double group formulation of EWZ, the second order interaction parameters $\zeta$'s are defined by Eq.(11-13) of EWZ. Using the first order interaction matrices obtained in appendix \ref{app:first_order} and Eq.(\ref{eqn:Lowdin_S}), the symmetric L\"owdin term has the form of Eq.(\ref{EWZ}) with the Luttinger invariants defined by,
\begin{widetext}
\begin{subequations}
\label{eqn:luttinger_inv}
\begin{eqnarray}
\gamma_1&=&2\zeta_{\Gamma_8^+,\Gamma_8^+}^{\Gamma_6^-}+2\zeta_{\Gamma_8^+,\Gamma_8^+}^{\Gamma_7^-}+\zeta_{\Gamma_8^+,\Gamma_8^+}^{\Gamma_8^-,1}+8\zeta_{\Gamma_8^+,\Gamma_8^+}^{\Gamma_8^-,2}+4\zeta_{\Gamma_8^+,\Gamma_8^+}^{\Gamma_8^-,3}-1 \\
\label{eq:g2}
\gamma_2&=&\zeta_{\Gamma_8^+,\Gamma_8^+}^{\Gamma_6^-}-\zeta_{\Gamma_8^+,\Gamma_8^+}^{\Gamma_7^-}-4\zeta_{\Gamma_8^+,\Gamma_8^+}^{\Gamma_8^-,2}-2\zeta_{\Gamma_8^+,\Gamma_8^+}^{\Gamma_8^-,3}  \label{eqn:g2_double}\\
\gamma_3&=&\zeta_{\Gamma_8^+,\Gamma_8^+}^{\Gamma_6^-}+\zeta_{\Gamma_8^+,\Gamma_8^+}^{\Gamma_7^-}-2\zeta_{\Gamma_8^+,\Gamma_8^+}^{\Gamma_8^-,2}\\
\gamma_{2}^{vs}&=&\frac{1}{\sqrt{2}}\left(\zeta_{\Gamma_8^+,\Gamma_7^+}^{\Gamma_7^-}-\zeta_{\Gamma_8^+,\Gamma_7^+}^{\Gamma_8^-,1}-4\zeta_{\Gamma_8^+,\Gamma_7^+}^{\Gamma_8^-,2}\right)\\
\gamma_{3}^{vs}&=& \frac{1}{\sqrt{2}}\left(\zeta_{\Gamma_8^+,\Gamma_7^+}^{\Gamma_7^-}+\zeta_{\Gamma_8^+,\Gamma_7^+}^{\Gamma_8^-,1}+2\zeta_{\Gamma_8^+,\Gamma_7^+}^{\Gamma_8^-,2}\right) \\
\gamma_1^s&=&\zeta_{\Gamma_7^+,\Gamma_7^+}^{\Gamma_7^-}+4\zeta_{\Gamma_7^+,\Gamma_7^+}^{\Gamma_8^-}-1\\
\gamma_1^e(\Gamma_7^-)&=&\zeta_{\Gamma_7^-,\Gamma_7^-}^{\Gamma_7^+}+4\zeta_{\Gamma_7^-,\Gamma_7^-}^{\Gamma_8^+}-1
\end{eqnarray}
\end{subequations}
\end{widetext}
It is clear from Eq.(\ref{eqn:g2_single},\ref{eqn:g2_double}) that the sign of $\gamma_2$ is determined by relative strength of $\sigma$ versus $\pi$ under single group formulation, and $\zeta_{\Gamma_8^+,\Gamma_8^+}^{\Gamma_7^-}$ versus $\zeta_{\Gamma_8^+,\Gamma_8^+}^{\Gamma_6^-}$ under the double group formulation. Given that the definitions of these parameters contain the energy differences between the corresponding zone centre states, the sign of $\gamma_2$ is mostly determined by the symmetry of lowest zone centre conduction band states with the possible  exception of Si. Thus for most materials except diamond, and possibly Si, $\gamma_2<0$ for the form of Hamiltonian given in Eq.(\ref{EWZ}). In Si, the contribution from other parameters in Eq.(\ref{eqn:g2_single},\ref{eqn:g2_double}) may become important in determining the final sign of $\gamma_2$. While intra-band absorption data suggest a positive $\gamma_2$\cite{Braunstein:1963fu}, cyclotron resonance data under uniaxial stress \cite{Hensel:1963dd} suggests a negative $\gamma_2$.

Group theory based methodologies are empolyed to obtain the first order interaction matrices in appendix \ref{app:first_order}, but can not specify the sign of $\xi$ parameters. The second order interaction parameters $\zeta$ are all positive for the main valence band block, except the mixed term of $\zeta_{\Gamma_8^+,\Gamma_8^+}^{\Gamma_8^-,3}$. The sign of Luttinger parameters can, therefore, be inferred from the relations shown in Eq.(\ref{eqn:luttinger_inv},\ref{eqn:invariantzeta}) obtained using perturbation theory.

\section{Relationsip between first order parameters}
\label{app:xis}
The first order interaction between states belonging to any single group IR have been listed in appendix \ref{app:first_order}. Using appropriate transformation matrices between the product bases and adapted double group bases (see appendix \ref{app:unitary_transform}), the interaction between states of $\Gamma_i$ and $\Gamma_j$ referring to adapted double bases is given by $\mathcal{K}_{\Gamma_i\otimes\Gamma_6:\Gamma_j\otimes\Gamma_6}={U^{\Gamma_i}}^\dag(\mathbb{1}_{2\times 2}\otimes K_{\Gamma_i:\Gamma_j})U^{\Gamma_j}$. If we consider the interaction between $\Gamma_5^+$ and $\Gamma_4^-$, we have,
\begin{widetext}
\begin{subequations}
\begin{eqnarray}
\mathcal{K}_{\Gamma_8^+\oplus\Gamma_7^+(\Gamma_5^+),\Gamma_8^-\oplus\Gamma_6^-(\Gamma_4^-)}&=&{U^{\Gamma_5^+}}^\dag\left(\mathbb{1}_{2\times 2}\otimes K_{\Gamma_5^+,\Gamma_4^-}\right)U^{\Gamma_4^-}\\
&=&\frac{\hbar}{m_0}\xi_{\Gamma_5^+,\Gamma_4^-}\begin{pmatrix}
-\frac{k_z}{\sqrt{3}} & 0 & 0 & -\frac{k_+}{\sqrt{3}} &  - \frac{k_+}{\sqrt{2}} & 0\\ 
0 & \frac{k_z}{\sqrt{3}} &  - \frac{k_-}{\sqrt{3}} & 0 & \frac{2k_z}{\sqrt{6}} &  - \frac{k_-}{\sqrt{6}}\\ 
0 & -\frac{k_+}{\sqrt{3}} & -\frac{k_z}{\sqrt{3}} & 0 & \frac{k_+}{\sqrt{6}} & \frac{2k_z}{\sqrt{6}}\\ 
- \frac{k_-}{\sqrt{3}} & 0 & 0 & \frac{k_z}{\sqrt{3}} & 0 & \frac{k_+}{\sqrt{2}}\\ 
\frac{k_-}{\sqrt{6}} & 0 & \frac{k_+}{\sqrt{2}} & \frac{2k_z}{\sqrt{6}} & 0 & 0\\ 
\frac{2k_z}{\sqrt{6}} &  - \frac{k_-}{\sqrt{2}} & 0 & -\frac{k_+}{\sqrt{6}} & 0 & 0 
\end{pmatrix}\\
&=&\frac{\hbar}{m_0}\begin{pmatrix}
\mathcal{K}_{\Gamma_8^+,\Gamma_8^-}^1 & \mathcal{K}_{\Gamma_8^+,\Gamma_6^-} \\
\mathcal{K}_{\Gamma_7^+,\Gamma_8^-} & \mathbb{0}_{2\times 2}
\end{pmatrix}
\end{eqnarray}
where $\xi_{\Gamma_5^+,\Gamma_4^-}=P_p^{p^*}$. Comparing the reduced tenor elements of each block, we obtain:
\begin{eqnarray}
P_p^{p^*}=\xi_{\Gamma_5^+(p),\Gamma_4^-(p^\ast)}&=&-\sqrt{3}\xi_{\Gamma_8^+(\Gamma_5^+),\Gamma_8^-(\Gamma_4^-)}=-\sqrt{3}\xi_{\Gamma_8^-(\Gamma_4^-),\Gamma_8^+(\Gamma_5^+)}\notag \\
&&=\sqrt{6}\xi_{\Gamma_7^+(\Gamma_5^+),\Gamma_8^-(\Gamma_4^-)}=\sqrt{6}\xi_{\Gamma_8^-(\Gamma_4^-),\Gamma_7^+(\Gamma_5^+)}\notag\\
&&=\sqrt{6}\xi_{\Gamma_8^+(\Gamma_5^+),\Gamma_6^-(\Gamma_4^-)}=\sqrt{6}\xi_{\Gamma_6^-(\Gamma_4^-),\Gamma_8^+(\Gamma_5^+)}
\end{eqnarray}
\end{subequations}
\end{widetext}
The interaction matrices refer to the adapted double group basis  given in appendix \ref{app:bases}, with the symmetry of the IRs determined from the character table. It is worth noting that the relation between signs of first order parameters are fixed between different blocks, even though the symmetry does not permit the determination of the actual signs. Particularly, the sign of $\xi$ in $\mathcal{K}_{\Gamma_8^+(\Gamma_5^+),\Gamma_8^-(\Gamma_4^-)}$ is negative relative to $\mathcal{K}_{\Gamma_7^+(\Gamma_5^+),\Gamma_8^-(\Gamma_4^-)}$.

Similar relations in Eq.(\ref{eqn:xi_relations}) may be obtained under the single group formulation by transforming the appropriate interaction matrices that refer to single group product bases.

\section{Group theoretical background behind the unitary transformation of Eq.(\ref{eq:u_matrix})}
\label{app:grp_background}
When applying group theoretical method to problems such as crystals with diamond and zincblende lattices, where a group and subgroup ($T_d\subset O_h$) relation exists, it is advantageous to study the higher symmetry $O_h$ group, and examine the subgroup under compatibility relations.  This approach has been taken by EWZ, and has been used successfully to derive all necessary generators for the $T_d$ group, including those that are spatially odd, such as the linear k term, from results of the $O_h$ group. It is not generally possible to take this process in reverse and infer the properties of  parent group from the structure of the subgroup.

Under the method of invariant\cite{BirPikus}, the generators for irreducible perturbations are obtained from known bases of appropriate IRs associated with the zone centre states. The form of these generators, as well as the form of the representation matrices, are dependent on the choice of bases. The point groups, such as $O_h$ and $T_d$, are subgroups of $O(3)$. As a result, the IR of the $O(3)$ group also form representations of the point group which are generally reducible. For the specific cases of $D^\pm_\frac{1}{2}$ and $D^\pm_\frac{3}{2}$ IRs of the $O(3)$ group, they also form irreducible representation $\Gamma_6^\pm$ and $\Gamma_8^\pm$ IR of the $O_h$ group, respectively. In the literature, the $M_j$ number used to label basis of IR of the point group is frequently equated to $z$ components of angular momentum. It must be recognised that in point group, there is no corresponding  generators of infinitesimal rotations, as in the continuos Lie group such as $O(3)$ or $SO(3)$. Therefore, no link can be established between the $M_j$ number and $z$ component of angular momentum in the same way. If one is to correlate the $M_j$ number as scaled $z$ component of angular momentum, then the order of bases must be derived from the appropriate representation of the $O(3)$ group. In the $O(3)$ group, the infinitesimal generator of rotation about the $z$ axes is scaled from $z$ component of the total angular momentum operator. The order of bases within IR $D^\pm_j$ has the $m_j$ numbers ranging from $2j+1$ to $-(2j+1)$, which labels the row of the representation matrices. The corresponding order of bases in the IRs of the point group are then given by the corresponding $\left|J, M_j\right>$ numbers that correlate to the $\left|j, m_j\right>$ number in the $D^\pm_j$ representation of the $O(3)$ group with $j=\frac{1}{2}$ for $\Gamma_{6}^\pm$, and $j=\frac{3}{2}$ for $\Gamma_8^\pm$ in the $O_h$ group. For the $D^\pm_j$ IR of the $O(3)$ group, the representation matrices are the same for proper rotations but differ in sign for improper rotations. Within the $O_h$ group, such differences in the representation matrices are reflected in the isomorphism between the product of IR representations $\Gamma_1^-\otimes\Gamma_i^-$ and IR $\Gamma_i^+$ (or $\Gamma_i^+=\Gamma_1^-\otimes\Gamma_i^-$). The corresponding representation matrices are linked by,
\begin{equation}
\label{eqn:product_g1m}
D^{\Gamma_i^+}(g)=D^{\Gamma_1^-}(g)\otimes D^{\Gamma_i^-}(g)=D^{\Gamma_1^-}(g)\cdot D^{\Gamma_i^-}(g)
\end{equation}
since $\Gamma_1^-$ is one dimensional and $D^{\Gamma_1^-}(g)$ is just a number. The representation matrices and corresponding bases functions derived by Onodera et al. follows the relations shown in Eq.(\ref{eqn:product_g1m}). 

Similar relations between IRs may be constructed from other one-dimensional representations of the $O_h$ group. One such example is 
\begin{widetext}
\begin{eqnarray}
\Gamma_i^+&=&\Gamma_2^-\otimes\Gamma_j^-\quad\{i,j\}\mbox{~or~}\{j,i\}\:\in\:[\{1,2\},\{3,3\},\{4,5\},\{6,7\},\{8,8\}]\\
D^{\Gamma_i^+\prime}(g)&=&D^{\Gamma_2^-}(g)\otimes D^{\Gamma_j^-}(g)=D^{\Gamma_2^-}(g)\cdot D^{\Gamma_j^-}(g) \label{eqn:product_g2m}
\end{eqnarray}
\end{widetext}
If one makes the same choice of basis, and representation matrices for $\Gamma_8^-$ as in $D^-_\frac{3}{2}$, then there will be some sign differences between the corresponding representation matrices of $D^{\Gamma_8^+}(g)$ (from Eq.(\ref{eqn:product_g1m})) and $D^{\Gamma_8^+\prime}(g)$ (from Eq.(\ref{eqn:product_g2m})) depending on the value of $D^{\Gamma_1^-}(g)$ and $D^{\Gamma_2^-}(g)$. Both clearly form representations of the $\Gamma_8^+$ IR, and must be linked by a similarity transformation $U$, which relate the basis functions, 
\begin{equation}
\label{eqn:similarity_transfrom}
D^{\Gamma_8^+\prime}(g)=U^\dag D^{\Gamma_8^+}(g)U.
\end{equation}
It can be shown that $U$ is defined by Eq.(\ref{eq:u_matrix}).

In the case of a general matrix  of the form 
\begin{equation}
P_{mn}=\left<\phi_{\alpha^u}^m\right|\hat{\bm{P}}_{\alpha^w}\left|\phi_{\alpha^v}^n\right>
\end{equation} 
where the operator $\hat{\bm{P}}$ transforms according to $\Gamma_{\alpha^w}$ IR of the symmetry group, its transformation properties are given by the product representation $\Gamma_{\alpha^u}^*\otimes\Gamma_{\alpha^w}\otimes\Gamma_{\alpha^v}$, which is generally reducible and decomposable into IRs. The matrix representation $\bm{P}$ is represented by a number of linearly independent matrices and associated invariants depending on the decomposition of the product representations. Specifically, the number of linearly independent matrices is given by the multiplicity factor of $\Gamma_{\alpha^w}$ in the decomposition of $\Gamma_{\alpha^u}^*\otimes\Gamma_{\alpha^v}$\cite{Koster:1958kea}. To find these linearly independent matrices, one must utilise irreducible tensor operators transforming according to $\Gamma_{\alpha^w}$ IR and bases of IR $\Gamma_{\alpha^u}$ and $\Gamma_{\alpha^v}$. This procedure has been used in EWZ utilising the bases given in appendices \ref{app:first_order} to calculate the first order $\bm{k}\cdot\bm{\pi}$ interaction matrices. 

Since $\Gamma_2^{-*}\otimes\Gamma_2^-=\Gamma_1^+$ and $\Gamma_2^-$ is one dimensional, we can write,
\begin{subequations}
\begin{eqnarray}
\Gamma_{\alpha^u}^*\otimes\Gamma_{\alpha^w}\otimes\Gamma_{\alpha^v}&=&(\Gamma_{2}^{-*}\otimes\Gamma_{\alpha^u}^*)\otimes\Gamma_{\alpha^w}\otimes(\Gamma_2^-\otimes\Gamma_{\alpha^v}) \\
&=&(\Gamma_{2}^{-*}\otimes\Gamma_{\alpha^u}^*)\otimes(\Gamma_2^-\otimes\Gamma_{\alpha^w})\otimes\Gamma_{\alpha^v}
\end{eqnarray}
\end{subequations}
and the corresponding relation in matrix representation may be written as
\begin{widetext}
\begin{subequations}
\begin{eqnarray}
D^{\Gamma_{\alpha^u}^*}(g)\otimes D^{\Gamma_{\alpha^w}}(g)\otimes D^{\Gamma_{\alpha^v}}(g)&=&\left(D^{\Gamma_{2}^{-*}}(g)\otimes D^{\Gamma_{\alpha^u}^*}(g)\right)\otimes D^{\Gamma_{\alpha^w}}(g)\otimes\left(D^{\Gamma_2^-}(g)\otimes D^{\Gamma_{\alpha^v}}(g)\right) \label{eqn:quadratic_generator}\\
&=&\left(D^{\Gamma_{2}^{-*}}(g)\otimes D^{\Gamma_{\alpha^u}^*}(g)\right)\otimes \left(D^{\Gamma_2^-}(g)\otimes D^{\Gamma_{\alpha^w}}(g)\right)\otimes D^{\Gamma_{\alpha^v}}(g)\label{eqn:linear_generator}
\end{eqnarray}
\end{subequations}
\end{widetext}

Historically, the point group bases are not well known, and calculations are generally made utilising the fact that the bases of $D_{\frac{1}{2}}^\pm$ and $D_{\frac{3}{2}}^-$ IRs of the $O(3)$ group also from representation of the $\Gamma_6^\pm$ and the $\Gamma_8^-$ IRs of the $O_h$ group. Trebin et al. \cite{Trebin:1979jl} has utilised the relation in Eq.(\ref{eqn:quadratic_generator}) to show that the generators between states of any of the $\{\Gamma_7^-, \Gamma_7^+, \Gamma_8^+\}$ triple, are the same as the corresponding generators between states of the $\{\Gamma_6^+,\Gamma_6^-, \Gamma_8^-\}$ triple. Thus, the full 8-band Hamiltonian for the zone centre states transforming according to $\{\Gamma_7^-, \Gamma_7^+, \Gamma_8^+\}$ triple can be constructed from generator obtained from the  $\{\Gamma_6^+,\Gamma_6^-, \Gamma_8^-\}$ triple. This leads to a form of Hamiltonian similar to those shown in Eq.(\ref{Chuang}). Such a process does not require any knowledge of point group bases of the  $\{\Gamma_7^-, \Gamma_7^+, \Gamma_8^+\}$ triple. The methodology is clearly valid, but should be used with the following considerations,
\begin{enumerate}
\item the vector space of $\Gamma_8^+$ IR is orthogonal to that of $\Gamma_8^-$ IR, and the bases of $\Gamma_8^-$ IR are {\em not} the bases of $\Gamma_8^+$ IR.
\item the method is not applicable if either $\Gamma_{\alpha^u}$ or $\Gamma_{\alpha^v}$ does not belong to the  $\{\Gamma_7^-, \Gamma_7^+, \Gamma_8^+\}$ triple.
\item the representation matrices generated by  $D^{\Gamma_8^+\prime}(g)=D^{\Gamma_8^-}(g)\otimes D^{\Gamma_2^-}(g)=D^{\Gamma_8^-}(g)\cdot D^{\Gamma_2^-}(g)$ constitute one of the many equivalent representations. It is equivalent to the one commonly used for symmetry analysis\cite{Onodera:1966bj,Wormer:2001ec,Altmann:1965br} by a similarity transformation shown in Eq.(\ref{eqn:similarity_transfrom}).
\item correlation of the $M_j$ number in the $D^{\Gamma_8^+\prime}(g)$ representation with the $z$ component of angular momentum is not appropriate, since it is not compliant to relation Eq.(\ref{eqn:product_g1m}). A unitary transformation of the basis described by the inverse of Eq.(\ref{eq:u_matrix}) is required before such correlation can be made.
\end{enumerate}
The second point leads to the scarcity of information in the literature on the derivation of the two linearly independent first order interaction matrices, between states of $\Gamma_8^\pm$ IR. This deficiency is overcome by considering the relation shown in Eq.(\ref{eqn:linear_generator}). The generators with $\Gamma_{\alpha^w}$ symmetry between states associated with IRs in the two different triples is the same as that with $\Gamma_2^-\otimes\Gamma^{\alpha^w}$ symmetry between states of compatible IR in the one of the triples.  The compatible IR in the opposing triple is obtained by direct product of the original with $\Gamma_2^-$. The time reversal properties of the operator should be considered with introduction of appropriate factor of ``i''  (general irreducible tensor operator is time reversal even, angular momentum is timer reversal odd).  For example, the generator transforming according to $\Gamma_{\alpha^w}$, between states of the $\Gamma_8^-$ and $\Gamma_8^+$ IR, is the same as the generator of $\Gamma_2^-\otimes\Gamma_{\alpha^w}$ between states of the $\Gamma_8^-$ and $\Gamma_8^-$ IR scaled by a factor of ``i''.  Thus, all the generators may be obtained from the $\{\Gamma_6^+, \Gamma_6^-, \Gamma_8^-\}$ triple. If we consider the generator for the $\bm{\pi}$ operator transforming according to $\Gamma_4^-$ between states of $\Gamma_8^+$ and $\Gamma_8^-$ symmetry, then {\em they} are the same as that of generator for $\Gamma_2^-\otimes\Gamma_4^-=\Gamma_5^+$ between states of $\Gamma_8^-$ and $\Gamma_8^-$. These generators have been obtained as matrix representations of $i\{J_y,J_z\},i\{J_z,J_x\}, i\{J_x,J_y\}$ and $\{J_x, (J_y^2-J_z^2)\},\{J_y, (J_z^2-J_x^2)\},\{J_z, (J_x^2-J_y^2)\}$ with respect to the $D^-_\frac{3}{2}$ basis. These generators refer to the basis set for which $D^{\Gamma_8^+\prime}(g)$ are the representation matrices of the $\Gamma_8^+$ IR and the $M_J$ number does not correspond to the $z$ component of angular momentum. The second set of generators has not been used in the construction of multiband Hamiltonian in the literature\cite{Rossler:1984fd,Pfeffer:1990cq,Mayer:1991ef,Richard:2004wo,Rideau:2006gu}.
    
The unitary transformation $T$ introduced to the basis of $\Gamma_8^-$ in EWZ, ensures that the compatible $\Gamma_8^-$ and $\Gamma_8^+$ IRs have the same representation matrices as $\Gamma_8^+$ for the elements of the $T_d$ group. This corresponds to $\Gamma_8^-=\Gamma_8^+\otimes\Gamma_2^-$. In essence, this is the reverse of the process that infers representation matrices of $\Gamma_8^+$ IR from that of $\Gamma_8^-$ IR. Applying the inverse unitary transformation to the $\Gamma_8^+$ bases, yields the result obtained by Chuang as shown in section \ref{sec:unitary_transform}. Thus, one would expect the transformation matrix in Eq.(\ref{eq:u_transform}) to be given by $M={U^{\Gamma_5^+}}^\dag U^{\Gamma_4^-}$ as defined in Eq.(\ref{eqn:MU_matrix}). There is, however, difference in the sign of the basis for the $\Gamma_7^+$ and $\Gamma_6^-$ IRs used, which accounts for the difference between Eq.(\ref{eq:u_transform}) and Eq.(\ref{eqn:MU_matrix}).

\section{Unitary transformation induced by rotation of coordinate system}
\label{app:orientation}
The diagonalization of the specific magnetic interaction in an arbitrary orientation is always possible for any zone centre states belonging to the $\Gamma_6^\pm, \Gamma_7^\pm$, and $\Gamma_8^\pm$ IRs. The process is relatively simple for states with $\Gamma_6^\pm$ and $\Gamma_7^\pm$ symmetry. This is due to the one linearly independent generator required under double group selection rules for operator or irreducible perturbation that transforms according to $\Gamma_4^+$ IR of the $O_h$ group for states belong to one of these IRs. On the other hand, two linearly independent matrices are required when we consider matrix representation of Zeeman interaction between states with $\Gamma_8^\pm$ symmetry. Thus, the two linearly independent matrices of the third component of angular momentum operator can not be {\em simultaneously} diagonalised in general (they can be simultaneously diagonalised only when referring to the $\{xyz\}$ coordinate system) for zone centre states belonging to the $\Gamma_8^\pm$ IRs.  This is one of the characteristics of the double group basis that can not be represented by a direct product between single group basis from one IR and spinor states. If one really wishes to diagonalise the matrix representation of the Zeeman interaction, in the direction of the field, the unitary transformation should be those that diagonalise the total Zeeman interaction $\kappa\mathcal{J}_3^{xyz}+qJ_3^{xyz}$. In this case, a single diagonal matrix would be obtained. This is in contrast to the suggestion of Luttinger\cite{LuttingerJM:1956hi} where the transformation described by Eq.(47) of Ref.\onlinecite{LuttingerJM:1956hi} was assumed to have re-quantised the zone centre states along the 3 axis. In fact, it was the $\mathcal{J}_3^{xyz}$ generator that was diagonalised but the $J_3^{123}$ generator remain non-diagonal. Here we will continue the treatment by Luttinger, and use $\mathcal{J}$ as the generator for the induced unitary transformation on the basis of $\Gamma_8^+$ IR. This leaves $\mathcal{J}_3^{123}$ diagonal, but $J_3^{123}$ non-diagonal and orientation dependent. 

The unitary transformation which diagonalised $\mathcal{J}_3^{xyz}$ in coordinate system $\{123\}$ are defined by the appropriate matrix representation of the angular momentum operator scaled by $\frac{1}{\hbar}$,
\begin{widetext}
\begin{equation}
\label{u_transform}
U^{\Gamma}=U_1U_2U_3=\exp[-iG_z^\Gamma\alpha]\exp[-iG_{x^\prime}^\Gamma \beta]\exp[-iG_{z^{\prime\prime}}^\Gamma\gamma]=\exp[-i\bm{G}\cdot\bm{\theta}]
\end{equation} 
\end{widetext}
where $G_\mu$ has the form of Pauli matrices scaled by $\frac{1}{2}$ for $\Gamma_6^\pm$, and $\Gamma_7^\pm$ IR, $J_\mu$ for $\Gamma_8^-$ IR, and $\mathcal{J}_\mu$ for $\Gamma_8^+$ IR, $\alpha, \beta$ and $\gamma$ are Euler angles which define the rotation $\bm{\theta}$ that takes the $\{xyz\}$ coordinate system into $\{123\}$ coordinate system. $J_\mu$ and $\mathcal{J}_\mu$ are those used in by Eq.(\ref{eqn:j_mu_double}). The similarity transform for the $\Gamma_8^\pm$ transformation are related by,
\begin{eqnarray}
U^{\Gamma_8^+}(\bm{\theta})&=&T^\dag U^{\Gamma_8^-}(\bm{\theta})T\quad\mbox{where}\\
T&=&\begin{pmatrix} 0 & 0 & 1 & 0\\ 0 & 0 & 0 & -1\\ -1 & 0 & 0 & 0 \\ 0 & 1 & 0 & 0\end{pmatrix}\label{eqn:tmatrix}
\end{eqnarray}

For magnetic field aligned in the $[111]$ direction  (the $\{1,2,3\}$ axes are in the $[\overline{1}10], [11\overline{2}]$ and $[111]$ directions), $\alpha=-\pi/4,\:\beta=-\arctan(\sqrt{2})$ and $\gamma=0$. The unitary transformation are given by,
\begin{widetext}
\begin{subequations}
\begin{eqnarray}
U^{\Gamma_{6,7}^\pm}([111])&=&\frac{1}{2}\begin{pmatrix}
\left(a+bi\right)\cos\!\left(\frac{\theta}{2}\right) & \left(ai - b\right)\sin\!\left(\frac{\theta}{2}\right)\\ 
-\left(ai + b\right)\sin\!\left(\frac{\theta}{2}\right) & \left(a-bi\right)\cos\!\left(\frac{\theta}{2}\right)
\end{pmatrix}\\
U^{\Gamma_8^-}([111])&=&\frac{1}{8}\left(\begin{array}{cc}
 4\, {\cos\!\left(\frac{\theta}{2}\right)}^3\, \left(a \mathrm{i} + b\right) & \sqrt{3}\, \left(\sin\!\left(\frac{\theta}{2}\right) + \sin\!\left(\frac{3\, \theta}{2}\right)\right)\, \left(b \mathrm{i} - a\right) \\ 
 \sqrt{3}\, \left(b \mathrm{i} + a\right)\, \left(\sin\!\left(\frac{\theta}{2}\right) + \sin\!\left(\frac{3\, \theta}{2}\right)\right)\, \mathrm{i} & \left(\cos\!\left(\frac{\theta}{2}\right) + 3\, \cos\!\left(\frac{3\,\theta}{2}\right)\right)\, \left(b \mathrm{i} + a\right) \\
 \sqrt{3}\, \left(\cos\!\left(\frac{\theta}{2}\right) - \cos\!\left(\frac{3\, \theta}{2}\right)\right)\, \left(b \mathrm{i} - a\right) & \left(\sin\!\left(\frac{\theta}{2}\right) - 3\, \sin\!\left(\frac{3\, \theta}{2}\right)\right)\, \left(b \mathrm{i} - a\right)\, \mathrm{i} \\
 - 4\, {\sin\!\left(\frac{\theta}{2}\right)}^3\, \left(b \mathrm{i} + a\right) & \sqrt{3}\, \left(\cos\!\left(\frac{\theta}{2}\right) - \cos\!\left(\frac{3\, \theta}{2}\right)\right)\, \left(a \mathrm{i} - b\right)\, \\
\end{array}\right.\hookleftarrow\notag\\
 &&\quad\hookrightarrow\left.\begin{array}{cc}
- \sqrt{3}\, \left( \cos\!\left(\frac{\theta}{2}\right) - \cos\!\left(\frac{3\, \theta}{2}\right)\right)\, \left(a \mathrm{i} + b\right) & - 4\, {\sin\!\left(\frac{\theta}{2}\right)}^3\, \left(b \mathrm{i} - a\right)\\
- \left(\sin\!\left(\frac{\theta}{2}\right) - 3\, \sin\!\left(\frac{3\, \theta}{2}\right)\right)\, \left(b \mathrm{i} + a\right)\, \mathrm{i} & - \sqrt{3}\, \left(\cos\!\left(\frac{\theta}{2}\right) - \cos\!\left(\frac{3\, \theta}{2}\right)\right)\, \left(b \mathrm{i} + a\right)\\ 
- \left(\cos\!\left(\frac{\theta}{2}\right) + 3\, \cos\!\left(\frac{3\, \theta}{2}\right)\right)\, \left(b \mathrm{i} - a\right) & - \sqrt{3}\, \left(\sin\!\left(\frac{\theta}{2}\right) + \sin\!\left(\frac{3\, \theta}{2}\right)\right)\, \left(b \mathrm{i} - a\right)\, \mathrm{i}\\ 
\sqrt{3}\, \left(b \mathrm{i} + a\right)\, \left(\sin\!\left(\frac{\theta}{2}\right) + \sin\!\left(\frac{3\, \theta}{2}\right)\right) & - 4\, {\cos\!\left(\frac{\theta}{2}\right)}^3\, \left(a \mathrm{i} - b\right)
 \end{array}\right) \\
U^{\Gamma_8^+}([111])&=&\frac{1}{8}\left(\begin{array}{cc}
- \left(\cos\!\left(\frac{\theta}{2}\right) + 3\, \cos\!\left(\frac{3\, \theta}{2}\right)\right)\, \left(b \mathrm{i} - a\right) & 
  	- \sqrt{3}\, \left(\sin\!\left(\frac{\theta}{2}\right) + \sin\!\left(\frac{3\, \theta}{2}\right)\right)\, \left(a \mathrm{i} + b\right) \\
- \sqrt{3}\, \left(b \mathrm{i} + a\right)\, \left(\sin\!\left(\frac{\theta}{2}\right) + \sin\!\left(\frac{3\, \theta}{2}\right)\right) & - 4\, {\cos\!\left(\frac{\theta}{2}\right)}^3\, \left(a \mathrm{i} - b\right) \\
\sqrt{3}\, \left(\cos\!\left(\frac{\theta}{2}\right) - \cos\!\left(\frac{3\, \theta}{2}\right)\right)\, \left(a \mathrm{i} + b\right) & - 4\, {\sin\!\left(\frac{\theta}{2}\right)}^3\, \left(b \mathrm{i} - a\right) \\
- \left(\sin\!\left(\frac{\theta}{2}\right) - 3\, \sin\!\left(\frac{3\, \theta}{2}\right)\right)\, \left(a \mathrm{i} - b\right) & \sqrt{3}\, \left(\cos\!\left(\frac{\theta}{2}\right) - \cos\!\left(\frac{3\, \theta}{2}\right)\right)\, \left(b \mathrm{i} + a\right) 
\end{array}\right.\hookleftarrow\notag\\
 &&\quad\hookrightarrow\left.\begin{array}{cc}
 - \sqrt{3}\, \left(\cos\!\left(\frac{\theta}{2}\right) - \cos\!\left(\frac{3\, \theta}{2}\right)\right)\, \left(b \mathrm{i} - a\right) & - \left(\sin\!\left(\frac{\theta}{2}\right) - 3\, \sin\!\left(\frac{3\, \theta}{2}\right)\right)\, \left(a \mathrm{i} + b\right)\\ 
 - 4\, {\sin\!\left(\frac{\theta}{2}\right)}^3\, \left(b \mathrm{i} + a\right) & - \sqrt{3}\, \left(\cos\!\left(\frac{\theta}{2}\right) - \cos\!\left(\frac{3\, \theta}{2}\right)\right)\, \left(a \mathrm{i} - b\right)\\ 
4\, {\cos\!\left(\frac{\theta}{2}\right)}^3\, \left(a \mathrm{i} + b\right) & - \sqrt{3}\, \left(\sin\!\left(\frac{\theta}{2}\right) + \sin\!\left(\frac{3\, \theta}{2}\right)\right)\, \left(b \mathrm{i} - a\right)\\
- \sqrt{3}\, \left(a \mathrm{i} - b\right)\, \left(\sin\!\left(\frac{\theta}{2}\right) + \sin\!\left(\frac{3\, \theta}{2}\right)\right) & \left(\cos\!\left(\frac{\theta}{2}\right) + 3\, \cos\!\left(\frac{3\, \theta}{2}\right)\right)\, \left(b \mathrm{i} + a\right) 
 \end{array}\right) 
\end{eqnarray}
\end{subequations}
where $a=\sqrt{2+\sqrt{2}},\quad b=\sqrt{2-\sqrt{2}},\quad\theta=-\beta,\quad\mbox{and~}\tan(\theta)=\sqrt{2}$. 

For magnetic field aligned in the $[110]$ direction  (the $\{1,2,3\}$ axes are in the $[001], [\overline{1}10]$ and $[110]$ directions), $\alpha=-\pi/4,\:\beta=-\pi/2$ and $\gamma=-\pi/2$. The unitary transformation are given by:
\begin{subequations}
\begin{eqnarray}
U^{\Gamma_{6,7}^\pm}([110])&=&\frac{1}{2(1+\mathrm{i})}\begin{pmatrix}
a \mathrm{i} - b & a \mathrm{i} - b\\ b \mathrm{i} - a & a - b\mathrm{i}
\end{pmatrix}\\
U^{\Gamma_8^-}([110])&=&\frac{1-\mathrm{i}}{8}\left(\begin{array}{cccc}
  - a \mathrm{i} - b &  - \sqrt{3} (a \mathrm{i}\, + b)\, &  - \sqrt{3} a \mathrm{i}\, - \sqrt{3} b\, &  - a \mathrm{i} - b\\
   - \sqrt{3}\, \left(- b + a\, \mathrm{i}\right) & b - a \mathrm{i} & a \mathrm{i} - b & \sqrt{3}\, \left(b \mathrm{i}\, \mathrm{i} + a\, \mathrm{i}\right)\\
    \sqrt{3} a\, - \sqrt{3} b \mathrm{i}\, & b \mathrm{i} - a & b \mathrm{i} - a & \sqrt{3} a\, - \sqrt{3} b \mathrm{i}\,\\ b \mathrm{i} + a &  - \sqrt{3} b \mathrm{i}\, - \sqrt{3} a\, & \sqrt{3} b \mathrm{i}\, + \sqrt{3} a\, &  - b \mathrm{i} - a  \end{array}\right) \\
U^{\Gamma_8^+}([110])&=&\frac{1-i}{8}\left(\begin{array}{cccc}
b \mathrm{i} - a & \sqrt{3} b \mathrm{i}\, - \sqrt{3} a\, & \sqrt{3} b \mathrm{i}\, - \sqrt{3} a\, & b \mathrm{i} - a\\  - \sqrt{3} b \mathrm{i}\, - \sqrt{3} a\, &  - b \mathrm{i} - a & b \mathrm{i} + a & \sqrt{3} b \mathrm{i}\, + \sqrt{3} a\,\\ \sqrt{3} a \mathrm{i}\, + \sqrt{3} b\, &  - a \mathrm{i} - b &  - a \mathrm{i} - b & \sqrt{3} a \mathrm{i}\, + \sqrt{3} b\,\\ a \mathrm{i} - b & - \sqrt{3}\, \left(b \mathrm{i}\, \mathrm{i} + a\, \mathrm{i}\right) & \sqrt{3}\, \left(b \mathrm{i}\, \mathrm{i} + a\, \mathrm{i}\right) & b - a \mathrm{i}
 \end{array}\right) 
\end{eqnarray}
\end{subequations}
\end{widetext}
%
\end{document}